%% file: ms.tex
\documentclass{emulateapj}

\slugcomment{{\sc ApJS, in press}}
\begin{document}

\shortauthors{Moustakas et~al.}
\shorttitle{SINGS Oxygen Abundances}

\newcommand{\lz}{\ensuremath{L-Z}}
\newcommand{\mz}{\ensuremath{M-Z}}
\newcommand{\dbpt}{\ensuremath{D_{\rm BPT}}}
\newcommand{\hi}{\textrm{H}~\textsc{i}}
\newcommand{\hii}{\textrm{H}~\textsc{ii}}
\newcommand{\mb}{\ensuremath{M_{B}}}
\newcommand{\cms}{\ensuremath{\textrm{cm~s}^{-1}}}
\newcommand{\kms}{\ensuremath{\textrm{km~s}^{-1}}}
\newcommand{\msun}{\ensuremath{\mathcal{M}_{\sun}}}
\newcommand{\flunits}{\textrm{erg~s\ensuremath{^{-1}}~cm\ensuremath{^{-2}}~\AA\ensuremath{^{-1}}}}
\newcommand{\lunits}{\ensuremath{\textrm{erg~s}^{-1}}}
\newcommand{\fluxunits}{\textrm{erg~s\ensuremath{^{-1}}~cm\ensuremath{^{-2}}}}
\newcommand{\lsun}{\ensuremath{L_{\sun}}}

\newcommand{\ha}{\textrm{H}\ensuremath{\alpha}}
\newcommand{\hb}{\textrm{H}\ensuremath{\beta}}
\newcommand{\hg}{\textrm{H}\ensuremath{\gamma}}
\newcommand{\oii}{[\textrm{O}~\textsc{ii}]}
\newcommand{\oiii}{[\textrm{O}~\textsc{iii}]}
\newcommand{\oi}{[\textrm{O}~\textsc{i}]}
\newcommand{\sii}{[\textrm{S}~\textsc{ii}]}
\newcommand{\siii}{[\textrm{S}~\textsc{iii}]}
\newcommand{\nii}{[\textrm{N}~\textsc{ii}]}
\newcommand{\niii}{[\textrm{N}~\textsc{iii}]}
\newcommand{\cii}{[\textrm{C}~\textsc{ii}]}
\newcommand{\mgii}{Mg~\textsc{ii}}
\newcommand{\neon}{[\textrm{Ne}~\textsc{iii}]}
\newcommand{\ariii}{[\textrm{Ar}~\textsc{iii}]}

\newcommand{\halam}{\textrm{H}\ensuremath{\alpha~\lambda6563}}
\newcommand{\hblam}{\textrm{H}\ensuremath{\beta~\lambda4861}}
\newcommand{\hglam}{\textrm{H}\ensuremath{\gamma~\lambda4340}}
\newcommand{\hdlam}{\textrm{H}\ensuremath{\delta~\lambda4101}}
\newcommand{\oiilam}{\oii~\ensuremath{\lambda3727}}
\newcommand{\oiiilam}{[\textrm{O}~\textsc{iii}]~\ensuremath{\lambda5007}}
\newcommand{\niilam}{[\textrm{N}~\textsc{ii}]~\ensuremath{\lambda6584}} 
\newcommand{\oilam}{[\textrm{O}~\textsc{i}]~\ensuremath{\lambda6300}}
\newcommand{\mgiilam}{\textrm{Mg}~\textsc{ii}~\ensuremath{\lambda2800}}
\newcommand{\nevlam}{[\textrm{Ne}~\textsc{v}]~\ensuremath{\lambda3426}} 

\newcommand{\oiiobs}{\oii\ensuremath{_{\rm obs}}}
\newcommand{\oiicor}{\oii\ensuremath{_{\rm cor}}}
\newcommand{\oiiicor}{\oiii\ensuremath{_{\rm cor}}}
\newcommand{\oiiiobs}{\oiii\ensuremath{_{\rm obs}}}
\newcommand{\haobs}{\ha\ensuremath{_{\rm obs}}}
\newcommand{\hacor}{\ha\ensuremath{_{\rm cor}}}
\newcommand{\hbobs}{\hb\ensuremath{_{\rm obs}}}
\newcommand{\hbcor}{\hb\ensuremath{_{\rm cor}}}

\newcommand{\oiidoublet}{[\textrm{O}~\textsc{ii}]~\ensuremath{\lambda\lambda3726,3729}}
\newcommand{\oiiidoublet}{[\textrm{O}~\textsc{iii}]~\ensuremath{\lambda\lambda4959,5007}}
\newcommand{\niidoublet}{[\textrm{N}~\textsc{ii}]~\ensuremath{\lambda\lambda6548,6584}}
\newcommand{\siidoublet}{[\textrm{S}~\textsc{ii}]~\ensuremath{\lambda\lambda6716,6731}}

\newcommand{\ewoii}{\textrm{EW}(\oii)}
\newcommand{\ewoiii}{\textrm{EW}(\oiii)}
\newcommand{\ewoiilam}{\textrm{EW}(\oii~\ensuremath{\lambda3727})}
\newcommand{\ewoiiilam}{\textrm{EW}(\oiii~\ensuremath{\lambda5007})}
\newcommand{\ewoiiidoublet}{\textrm{EW}(\oiii~\ensuremath{\lambda4959,5007})}
\newcommand{\ewha}{\textrm{EW}(\textrm{H}\ensuremath{\alpha})}
\newcommand{\ewhanii}{\textrm{EW}(\ha+\nii)}
\newcommand{\ewhb}{\textrm{EW}(\textrm{H}\ensuremath{\beta})}

\newcommand{\ioniz}{\ensuremath{O_{32}}}
\newcommand{\pagel}{\ensuremath{R_{23}}}
\newcommand{\logu}{\ensuremath{\log\,(U)}}
\newcommand{\logq}{\ensuremath{\log\,(q)}}
\newcommand{\ewioniz}{\textrm{EW}(\ensuremath{O_{32}})}
\newcommand{\ewpagel}{\textrm{EW}(\ensuremath{R_{23}})}
\newcommand{\pagelcor}{(\pagel)\ensuremath{_{\rm cor}}}
\newcommand{\ionizcor}{(\ioniz)\ensuremath{_{\rm cor}}}
\newcommand{\pagelobs}{(\pagel)\ensuremath{_{\rm obs}}}
\newcommand{\ionizobs}{(\ioniz)\ensuremath{_{\rm obs}}}

\newcommand{\logsunoh}{\ensuremath{12+\log\,(\textrm{O}/\textrm{H})_{\sun}}}
\newcommand{\logoh}{\ensuremath{12+\log\,(\textrm{O}/\textrm{H})}}
\newcommand{\oh}{\ensuremath{\log\,(\textrm{O}/\textrm{H})}}
\newcommand{\logohcor}{\ensuremath{12+\log\,(\textrm{O}/\textrm{H})_{{\rm cor}}}}
\newcommand{\logohchar}{\ensuremath{12+\log\,(\textrm{O}/\textrm{H})_{\rm char}}}
\newcommand{\logohavg}{\ensuremath{12+\log\,(\textrm{O}/\textrm{H})_{\rm avg}}}
\newcommand{\logohfinal}{\ensuremath{12+\log\,(\textrm{O}/\textrm{H})_{\rm final}}}
\newcommand{\logohnuc}{\ensuremath{12+\log\,(\textrm{O}/\textrm{H})_{\rm nuclear}}}
\newcommand{\logohcen}{\ensuremath{12+\log\,(\textrm{O}/\textrm{H})_{\rm central}}}
\newcommand{\logohstrip}{\ensuremath{12+\log\,(\textrm{O}/\textrm{H})_{\rm strip}}}
\newcommand{\logohcircum}{\ensuremath{12+\log\,(\textrm{O}/\textrm{H})_{\rm circum}}}
\newcommand{\ewlogoh}{\ensuremath{12+\log\,(\textrm{O}/\textrm{H})_{\rm EW}}}
\newcommand{\dlogoh}{\ensuremath{\Delta\log\,(\textrm{O}/\textrm{H})}}
\newcommand{\debv}{\ensuremath{\Delta\,E(B-V)}}

\newcommand{\sfr}{\ensuremath{\psi}}
\newcommand{\sfrunits}{\ensuremath{\mathcal{M}_{\sun}~\textrm{yr}^{-1}}}
\newcommand{\rv}{\ensuremath{R_{V}}}

\title{Optical Spectroscopy and Nebular Oxygen Abundances of the \\
  \emph{Spitzer}/SINGS Galaxies}

\author{John Moustakas\altaffilmark{1}, Robert~C. Kennicutt,
  Jr.\altaffilmark{2,3}, Christy~A. Tremonti\altaffilmark{4},
  Daniel~A. Dale\altaffilmark{5},
  \\ John-David~T. Smith\altaffilmark{6}, Daniela
  Calzetti\altaffilmark{7}}  

\altaffiltext{1}{Center for Astrophysics and Space Sciences,
  University of California, San Diego, 9500 Gilman Drive, La Jolla,
  California, 92093} 
\altaffiltext{2}{Institute of Astronomy, University of Cambridge,
  Madingley Road, Cambridge CB3 0HA, UK} 
\altaffiltext{3}{Steward Observatory, University of Arizona, 933 N 
  Cherry Ave., Tucson, AZ 85721} 
\altaffiltext{4}{Department of Astronomy, University of
  Wisconsin-Madison, 4524 Sterling Hall, 475 N. Charter Street,
  Madison, WI 53706} 
\altaffiltext{5}{Department of Physics and Astronomy, University of
  Wyoming, Laramie, WY, 82071} 
\altaffiltext{6}{Department of Physics and Astronomy, University of
  Toledo, Ritter Obs., MS \#113, Toledo, OH 43606}
\altaffiltext{7}{Department of Astronomy, University of Massachusetts, 
  710 N. Pleasant Street, Amherst, MA 01003}

\begin{abstract}
We present intermediate-resolution optical spectrophotometry of $65$
galaxies obtained in support of the \emph{Spitzer} Infrared Nearby
Galaxies Survey (SINGS).  For each galaxy we obtain a nuclear,
circumnuclear, and semi-integrated optical spectrum designed to
coincide spatially with mid- and far-infrared spectroscopy from the
{\em Spitzer Space Telescope}.  We make the reduced,
spectrophotometrically calibrated one-dimensional spectra, as well as
measurements of the fluxes and equivalent widths of the strong nebular
emission lines, publically available.  We use optical emission-line
ratios measured on all three spatial scales to classify the sample
into star-forming, active galactic nuclei (AGN), and galaxies with a
mixture of star formation and nuclear activity.  We find that the
relative fraction of the sample classified as star-forming versus AGN
is a strong function of the integrated light enclosed by the
spectroscopic aperture.  We supplement our observations with a large
database of nebular emission-line measurements of individual \hii{}
regions in the SINGS galaxies culled from the literature.  We use
these ancillary data to conduct a detailed analysis of the radial
abundance gradients and average \hii-region abundances of a large
fraction of the sample.  We combine these results with our new
integrated spectra to estimate the central and characteristic
(globally-averaged) gas-phase oxygen abundances of all $75$ SINGS
galaxies.  We conclude with an in-depth discussion of the absolute
uncertainty in the nebular oxygen abundance scale.
\end{abstract}


\section{Introduction}\label{sec:intro}

The \emph{Spitzer} Infrared Nearby Galaxies Survey (SINGS) was
designed to investigate the star formation and dust emission
properties of $75$ nearby galaxies using the full complement of
instruments available on the {\em Spitzer Space Telescope}
\citep{kenn03b}.  With the survey now complete, SINGS has delivered to
the astrophysics community among the most detailed mid- and
far-infrared wide-field images and spectral maps of nearby galaxies
ever obtained.\footnote{Publically available at http://sings.stsci.edu
  and http://irsa.ipac.caltech.edu/data/SPITZER/SINGS.}  

Supplemented with ancillary multi-wavelength observations from the
ultraviolet (UV) to the radio, these data have facilitated a wide
range of studies of the global and small-scale interstellar medium
(ISM) properties of galaxies and active galactic nuclei (AGN),
including: analyses of the mid-infrared nebular, aromatic, and
molecular hydrogen spectra of galactic nuclei and extranuclear \hii{}
regions \citep{smith04a, smith07a, roussel07a, prescott07a, dale06a,
  dale09a}; the construction of broadband $0.15-850$~\micron{} galaxy
spectral energy distributions \citep[SEDs;][]{dale05a, dale07a}, and
their interpretation using physical dust models \citep{draine07a,
  munoz09a, munoz09b}; detailed studies of the mid- and far-infrared
morphologies of both low-mass dwarfs \citep{cannon06a, cannon06b,
  walter07a} and massive early- and late-type galaxies
\citep{regan04a, murphy06a, bendo07a}; and the development of robust
global and spatially resolved optical and infrared (IR) star formation
rate (SFR) diagnostics \citep{calzetti05a, calzetti07a, calzetti10a,
  kenn07a, kenn09a, boquien09a, boquien10a}, among others.

In addition to imaging and spectroscopy with \emph{Spitzer}, SINGS has
also assembled a large, homogeneous database of multi-wavelength
observations designed to maximize the scientific impact and legacy
value of the survey \citep{kenn03b, meurer06a, dale05a, daigle06a,
  dale07a, calzetti07a, braun07a, dicaire08a, walter08a}.  As part of
this larger effort, this paper presents intermediate-resolution,
($\sim8$~\AA{} FWHM), high signal-to-noise ratio (${\rm
  S/N}=5-100$~pixel$^{-1}$) optical ($3600-6900$~\AA)
spectrophotometry of the SINGS galaxies on several spatial scales
designed to match the coverage of the \emph{Spitzer} spectroscopy,
ranging from the inner nucleus, to spectra that enclose a significant
fraction ($30\%-100\%$) of the integrated optical light.  In addition
to making the reduced one-dimensional spectra publically available, we
also provide measurements of the strong nebular emission lines
corrected for underlying stellar absorption using state-of-the-art
stellar population synthesis models.

These optical spectra provide a valuable complement to the SINGS
multi-wavelength dataset in several respects.  First, the calibrated
emission-line spectra provide measures of the instantaneous SFR and
dust reddening that can be compared to independent measures of SFRs
and extinctions derived from the UV and IR \citep[e.g.,][]{kenn09a,
  calzetti07a, calzetti10a, rieke09a}.  The forbidden-line spectra
provide gas-phase metal abundances and constrain the nature of the
primary ionizing radiation sources (i.e., AGN vs.~star formation),
both of which are important for interpreting the infrared line spectra
and SEDs, and for probing dependences of the dust properties and star
formation on metallicity \citep[e.g.,][]{smith07a, draine07a}.
Finally the optical stellar continuum provides valuable constraints on
the stellar populations, and the properties of the stars that are
responsible for heating the dust \citep[e.g.,][]{gordon00a,
  cortese08a}.

In this paper we combine our optical spectra with spectroscopy of
individual \hii{} regions culled from the literature to derive the
nebular (gas-phase) metallicities\footnote{Unless otherwise indicated,
  in this paper we use the term \emph{metallicity} to mean the
  gas-phase oxygen-to-hydrogen abundance ratio, \logoh.  For
  reference, the currently favored solar oxygen abundance in these
  units is $\logsunoh=8.69\pm0.05$ \citep{asplund09a}.} of the SINGS
galaxies.  Dust grains, which absorb and reradiate a significant
fraction of the bolometric luminosity of a galaxy into the IR, are
composed of heavy elements such as C, O, Mg, Si, and Fe
\citep{draine03a}.  Therefore, the infrared SED of a galaxy is
fundamentally related to its chemical composition.  For example,
numerous studies have reported a link between the observed deficit of
polycyclic aromatic hydrocarbon (PAH) emission in the mid-IR of
galaxies more metal-poor than a threshold gas-phase oxygen abundance
of $\logoh\approx8.1$ \citep[e.g.,][]{madden00a, engelbracht05a,
  engelbracht08a}, which may be due, in part, to a paucity of metals
from which to form these complex molecules \citep[e.g.,][]{draine07a}.
Integrated and monochromatic infrared SFR calibrations must also
carefully consider metallicity effects, as galaxies at UV/optical
wavelengths become increasingly transparent with decreasing
metallicity and dust content \citep[e.g.,][]{draine07a,
  calzetti07a, calzetti10a, zhu08a}.

We organize the remainder of the paper as follows.  In
\S\ref{sec:data} we briefly describe the SINGS sample, present our
optical spectrophotometry and measurements of the fluxes and
equivalent widths of the strong optical emission lines, and describe
our spectroscopic database of \hii{} regions in the SINGS galaxies.
We classify the sample into star-forming galaxies and AGN in
\S\ref{sec:class}, and present a detailed analysis of the nebular
oxygen abundances of the sample based on an analysis of our optical
spectra and the \hii-region observations in \S\ref{sec:analysis}.  In
\S\ref{sec:discussion} we provide a detailed discussion of the
discrepancy among abundances derived using different strong-line
methods.  Finally, we summarize our results in \S\ref{sec:summary}.

\section{Observations}\label{sec:data}

\subsection{SINGS Sample and Properties}\label{sec:sample}   

The SINGS galaxies were selected using three principal criteria ---
morphology, optical luminosity, and infrared-to-optical luminosity
ratio --- such that the sample would span the broadest possible range
of star formation and dust properties exhibited by nearby, normal
galaxies \citep{kenn03b}.  An approximate distance limit,
$D\lesssim30$~Mpc, was also imposed so that the interstellar medium
could be studied with adequate spatial resolution with \emph{Spitzer}.
Galaxies hosting powerful AGN were intentionally excluded from the
sample as a more complete sample of AGNs have been observed as part of
complementary \emph{Spitzer} Guaranteed Time Observer (GTO) programs
\citep[e.g.,][]{weedman05a, wu09a}.

\begin{figure}[hb]
\epsscale{1.2}
\plotone{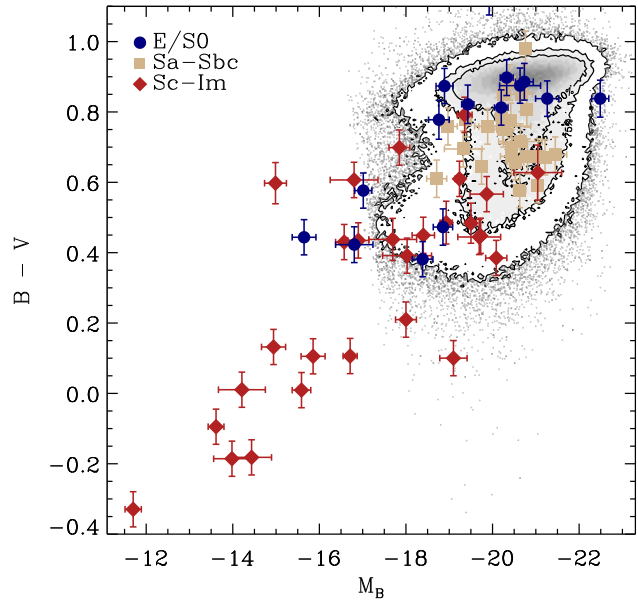}
\caption{Optical color-magnitude diagram for the SINGS galaxies coded
  by morphological type: E/S0 ({\em dark blue points}), Sa-Sbc ({\em
    tan squares}), and Sc-Im ({\em dark red diamonds}).  For
  comparison, the contoured greyscale image shows the color-magnitude
  diagram of a flux-limited sample of $\sim5\times10^{5}$ SDSS
  galaxies.  This figure demonstrates the broad range of optical
  luminosity and color spanned by the SINGS sample. \label{fig:cmd}} 
\end{figure}

Table~\ref{table:properties} presents some basic properties of the
sample used in the current analysis.  The table includes an updated
set of distances based on a careful search of the literature.  For
galaxies without a direct distance estimate (e.g., from observations
of Cepheid variables), we use their Hubble distance corrected for
peculiar motions \citep{masters05a} assuming $H_{0} =
70$~\kms~Mpc$^{-1}$.  Optical $B$- and $V$-band photometry for the
SINGS sample has been taken either from \citet{munoz09a} or
\citet{dale07a}, listed in order of preference.  The photometry in
\citet{munoz09a} for $54/75$ galaxies is based on a careful
curve-of-growth analysis and recalibration of imaging available in the
literature \citep{prugniel98a, dale07a}.  They provide asymptotic
(integrated) $B$- and $V$-band magnitudes for $22$ of these objects,
and integrated $ugriz$ magnitudes from the SDSS \citep{york00a} for
the other $32$ galaxies.  The SDSS magnitudes were transformed to $B$-
and $V$-band magnitudes using the statistical linear transformations
derived by \citet{blanton07a} based on a large sample of SDSS
galaxies.  The absolute $B$-band magnitudes, \mb, and $B-V$ colors in
Table~\ref{table:properties} have been corrected for foreground
Galactic reddening \citep{schlegel98a} and converted to the Vega
system using the AB-to-Vega conversions in \citet{blanton07a}.  We
adopt a minimum photometric error of $10\%$ and $5\%$ in \mb{} and
$B-V$, respectively.

One of the strengths of the SINGS sample is that it was designed to
span a broad range of galaxy properties.  We illustrate this point in
Figure~\ref{fig:cmd} where we plot $B-V$ versus \mb{} of the full
sample, divided into three bins of morphological type.  For
comparison, we show the corresponding color-magnitude diagram for a
flux-limited ($r_{\rm AB}<17.6$) sample of $\sim5\times10^{5}$ SDSS
galaxies at $0.01<z<0.25$ selected from the NYU Value-Added Galaxy
Catalog\footnote{http://sdss.physics.nyu.edu/vagc}
\citep[NYU-VAGC;][]{blanton05b} and the SDSS Data Release 7
\citep[DR7;][]{abazajian09a}.  The rest-frame luminosities of the SDSS
galaxies were estimated using {\sc k-correct}
\citep[v.~4.1.4;][]{blanton07a}.\footnote{http://howdy.physics.nyu.edu/index.php/Kcorrect}
Among bright ($\mb<-18$) galaxies, the two samples span a comparable
range of luminosity and color, whereas the SINGS sample includes a
significant number of faint blue galaxies.

\subsection{Nuclear, Circumnuclear, and Radial-Strip
  Spectroscopy}\label{sec:optspec}

\subsubsection{Observations \& Reductions}\label{sec:redux}

The overall strategy guiding our optical observing program was to
complement the mid- and far-infrared spectra of the SINGS sample
obtained as part of the principal \emph{Spitzer} observations
\citep{kenn03b, smith07a, dale09a}.  Briefly, the infrared
observations produce low-resolution ($\mathcal{R}\approx100$)
$5-15$~\micron{} spectral cubes of the central
$34\arcsec\times55\arcsec$ of each galaxy, and $10-37$~\micron{}
spectral cubes of the inner $18\arcsec\times24\arcsec$ at higher
resolution, $\mathcal{R}\approx600$, using the IRS spectrograph
\citep{houck04a}.\footnote{The center of NGC~3034=M~82 was a reserved
  target; therefore, it was not mapped by SINGS with the IRS
  instrument, although we did obtain optical spectra of its center
  (see Appendix~\ref{sec:thedata}).}  In addition, there are
$15-38$~\micron{} radial strip IRS spectra at $\mathcal{R}\approx100$
enclosing a $56\arcsec$ wide rectangular aperture that extend over
$30\%-100\%$ of the optical diameter of each galaxy, and complementary
low-resolution ($\mathcal{R}\approx15$) $55-95$~\micron{} radial strip
spectra of approximately the same region using the MIPS instrument in
SED mode \citep{rieke04a}.

The optical spectra were obtained between $2001$ November and $2006$
May at the Bok 2.3-meter telescope on Kitt Peak for galaxies in the
northern hemisphere, and at the CTIO 1.5-meter telescope for galaxies
in the southern hemisphere.  At the Bok telescope we used the B\&C
spectrograph and a $400$~mm$^{-1}$ grating, blazed at $\sim5200$~\AA,
to obtain spectral coverage from $3600-6900$~\AA{} at $\sim8$~\AA{}
FWHM ($\mathcal{R}\approx700$) resolution through a
$200\arcsec\times2\farcs5$ slit.  The CTIO observations were obtained
using the R-C spectrograph and a $300$~mm$^{-1}$ grating (\#9), blazed
at $\sim4000$~\AA, which provided spectral coverage from
$3450-6900$~\AA{} at $\sim8.5$~\AA{} FWHM ($\mathcal{R}\approx600$)
resolution through a $460\arcsec\times3\arcsec$ slit.  

We used the long-slit drift-scanning technique pioneered by
\citet{kenn92a} to obtain integrated optical spectra over the same
physical area targeted by the IRS and MIPS observations.
Drift-scanning consists of driving a long-slit back-and-forth
perpendicular to the slit during a single exposure in order to build a
luminosity-weighted integrated spectrum of a large spatially extended
area at the spectral resolution of a narrow slit \citep{kenn92a,
  jansen00a, gavazzi04a, moustakas06a}.  For each object we obtained a
circumnuclear spectrum targeting the central regions of each galaxy,
and a large rectangular radial-strip spectrum spatially coincident
with the IRS and MIPS observations (see below).  We also obtained a
pointed (spatially fixed) spectrum centered on the nucleus of each
object.  The total exposure time for the radial strip, circumnuclear,
and nuclear spectra ranged from $20-60$, $10-30$, and $5-10$ minutes,
respectively, usually split into two or more exposures to allow cosmic
rays to be identified.

The data reduction steps consisted of overscan- and bias-subtraction,
flat-fielding, a correction for the varying illumination pattern, and
sky subtraction (see \citealt{moustakas06a} for details).  We combined
multiple exposures using inverse variance weighting while iteratively
rejecting cosmic rays.  The galaxies in SINGS with the largest
projected angular diameters usually required multiple pointings to
achieve full spatial coverage of the corresponding infrared spectra,
or to ensure that the sky was adequately sampled.  For these objects
we obtained $2-6$ overlapping spectra, which we subsequently stitched
together by examining the spatial profiles in the overlap region.
Flux-calibration was facilitated by observations of standard stars
selected from the
\emph{HST}/CALSPEC\footnote{http://www.stsci.edu/hst/observatory/cdbs/calspec.html}
\citep{bohlin01a} and \citet{massey88a} star lists.  We estimate that
our \emph{relative} spectrophotometric accuracy is $\lesssim\pm5\%$
across the full spectral range \citep[see also][]{moustakas06a}.

We extracted one-dimensional spectra using a $2\farcs5\times2\farcs5$
(or $2\farcs5\times3\farcs0$ for the CTIO spectra) and
$20\arcsec\times20\arcsec$ aperture for each nuclear and circumnuclear
spectrum, respectively.  The radial strip spectra were extracted over
a $(0.55\times D_{25})\times\Delta_{\rm scan}$~arcsec$^{2}$
rectangular region, where $D_{25}$ is the major-axis diameter of each
galaxy (see Table~\ref{table:properties}) and $\Delta_{\rm
  scan}=56\arcsec$ (in some cases $55\arcsec$) is the drift-scan
length perpendicular to the slit.  The position angle and drift-scan
length of each radial-strip spectrum was chosen to maximize the
spatial overlap with the corresponding IRS spectra having at least
double coverage in both spectral orders (see
Appendix~\ref{sec:thedata} and \citealt{smith04a} for more details).
Finally, all the spectra were corrected for foreground Galactic
reddening \citep{odonnell94a, schlegel98a} assuming $R_{V}\equiv
A_{V}/E(B-V)=3.1$ (see Table~\ref{table:properties}).

Table~\ref{table:journal} summarizes our observations; we present the
data themselves in Appendix~\ref{sec:thedata}.  We obtained at least
one optical spectrum for $65$ galaxies, or $87\%$ of the SINGS sample.
Three of the ten objects without any spectra--- NGC~5408, IC~4710, and
NGC~7552--- are in the southern hemisphere and were not accessible
during the 2001 December CTIO observing run.  The remaining seven
galaxies lacking an optical spectrum--- Ho~II, M~81~Dw~A, Ho~I, Ho~IX,
DDO~154, DDO~165, and NGC~6822--- are faint, low surface-brightness
dwarfs for which we were unable to obtain usable data.  In total, we
obtained nuclear, circumnuclear, and radial-strip spectra for $57$,
$65$, and $62$ of the SINGS galaxies, respectively.

\begin{figure}[!ht]
\epsscale{1.2}
\plotone{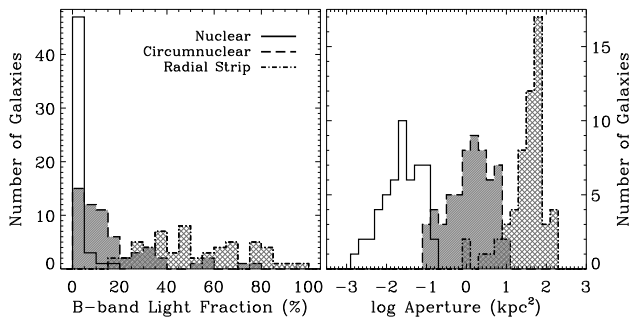}
\caption{Distribution of (\emph{left}) $B$-band light fraction and
  (\emph{right}) physical area (in kpc$^{2}$) subtended by our nuclear
  (\emph{solid}), circumnuclear (\emph{dashed}), and radial-strip
  (\emph{dot-dashed}) spectra.  The median light fraction enclosed by
  our nuclear, circumnuclear, and radial-strip spectra is $0.84\%$,
  $12\%$, and $49\%$, respectively, corresponding to a physical area
  of $0.17\times0.17$~kpc$^{2}$, $1.2\times1.2$~kpc$^{2}$, and
  $6.6\times6.6$~kpc$^{2}$.  Thus, our radial-strip spectra should be
  representative of the integrated (globally-averaged) properties of
  the sample for most objects, while our circumnuclear spectra should
  be representative of the central regions of most
  galaxies. \label{fig:bhist}} 
\end{figure}

In Figure~\ref{fig:bhist} we characterize the range of spatial scales
spanned by our observations by plotting the distribution of $B$-band
light fraction and the physical area (in kpc$^{2}$) subtended by our
nuclear, circumnuclear, and radial-strip spectra.  The light fractions
were measured from the $B$-band images of the SINGS galaxies by
projecting the rectangular apertures listed in
Table~\ref{table:journal} and computing the fraction of light within
each aperture relative to the integrated flux \citep{dale07a}.  Our
circumnuclear spectra enclose $0.54\%-80\%$ of the integrated optical
light, with a typical (median) value of $12\%$, while our radial-strip
spectra enclose $16\%-96\%$ of the integrated light, with a median
value of $49\%$.  Therefore, our circumnuclear and radial-strip
spectra should be representative of the central and integrated
(globally averaged) optical properties, respectively, of most of the
galaxies in the sample \citep{kewley05a}.

\subsubsection{Optical Emission-Line Measurements}\label{sec:ispec}

With the latest generation of high-resolution stellar population
synthesis models it has become possible to separate accurately the
stellar continuum spectrum of a galaxy from its integrated
emission-line spectrum, so that the emission-line strengths can be
studied free from the systematic effects of Balmer and metal-line
absorption \citep[e.g.,][]{brinchmann04a, cid05a, moustakas06a,
  asari07a}.

We fit each spectrum using the \citet[hereafter BC03]{bruzual03a}
stellar population synthesis models assuming the \citet{chabrier03a}
initial mass function between $0.1$ and $100$~\msun.  We model the
data as a non-negative linear combination of ten BC03 models with ages
ranging between $5$~Myr and $13$~Gyr and three different
metallicities: $Z=0.004$, $0.02$, $0.05$.  We derive the best-fitting
model, separately for each stellar metallicity, using a modified
version of the {\sc
  pPXF}\footnote{http://www-astro.physics.ox.ac.uk/$\sim$mxc/idl}
continuum-fitting code \citep{cappellari04a}.  First, we isolate the
$3700-4850$~\AA{} spectral range to derive the absorption-line
redshift of the galaxy, $z_{\rm abs}$, and an estimate of the velocity
dispersion, $\sigma_{\rm disp}$, deconvolved for the instrumental
resolution of our spectra and the BC03 models.  Next, we model the
full spectral range, fixing $z_{\rm abs}$ and $\sigma_{\rm disp}$ at
the derived values and aggressively mask pixels that may be affected
by emission lines, sky-subtraction residuals, or telluric absorption.
We treat reddening as a free parameter assuming the
\citet{calzetti00a} dust attenuation law.  Finally, we choose the
stellar metallicity that results in the best fit (i.e., lowest reduced
$\chi^{2}$).  

We emphasize that our principal objective in modeling the underlying
stellar continuum is to obtain a pure emission-line spectrum that has
been corrected self-consistently for stellar absorption, \emph{not} to
constrain star-formation history and stellar metallicity within each
spectroscopic aperture.  Star-formation history, age, metallicity, and
dust reddening are notoriously degenerate, which means that the
multidimensional parameter space is a very complex manifold that
likely contains numerous local minima.  Our simple $\chi^{2}$
minimization algorithm is ill-suited for solving this kind of problem,
although more general methods for inferring these quantities from
integrated optical spectra have been developed
\citep[e.g.,][]{panter03a, gallazzi05a, ocvirk06a, cid07a, tojeiro07a,
  blanton07a}.  For our purposes, we have verified that varying our
model assumptions does not significantly impact the measured
emission-line strengths.  Specifically, we checked that fixing the
stellar metallicity of the models to solar, adopting a different
extinction law \citep[e.g.,][]{odonnell94a, charlot00a, gordon03a},
assuming a different initial mass function \citep{salpeter55a}, or
changing the number of model ages yields emission-line strengths that
are within the statistical uncertainties for most objects.

Given the best-fitting stellar continuum, we subtract it from the data
to obtain a pure emission-line spectrum corrected for stellar
absorption.  Next, we remove any remaining residuals (typically of
order a few percent) due to imperfect sky-subtraction or template
mismatch using a $51$-pixel sliding median.  Finally, we model the
strong optical emission lines --- \oiilam, \hglam, \hblam,
\oiiidoublet, \niidoublet, \halam, and \siidoublet{} --- as individual
Gaussian line-profiles using a modified version of the {\sc
  gandalf}\footnote{http://www.strw.leidenuniv.nl/sauron}
emission-line fitting code \citep{sarzi06a, schawinski07a},
constraining the $\oiii~\lambda5007/\oiii~\lambda4959$ and
$\nii~\lambda6584/\nii~\lambda6548$ doublet ratios to be 3:1
\citep{osterbrock06a}.  We perform the emission-line fitting in two
iterations.  First, we tie the redshifts and intrinsic velocity widths
(deconvolved for the instrumental resolution) of the forbidden and
Balmer lines together to aid in the detection and deblending of weak
lines \citep{tremonti04a}.  On the second iteration we relax most of
the constraints on the line-profiles and use the best-fitting
parameters from the first iteration as initial guesses.  This second
step is necessary for two reasons: first, our knowledge of the
wavelength-dependent instrumental resolution is imperfect, which the
code can compensate for by allowing the velocity widths of the
emission lines to vary with wavelength; and second \oiilam{} is, in
fact, a doublet which is better represented at the spectral resolution
of our data as a single, slightly broader Gaussian line than two
closely-spaced Gaussian line-profiles.  Note that even on the second
iteration, however, we (separately) constrain the redshifts and
velocity widths of the \oiii, \nii, and \sii{} doublets to have the
same values.

For a handful of AGN in our sample we had to include a second, broad
component for the \ha, \hb, and \hg{} Balmer lines in order to obtain
a satisfactory fit to the observed line-profiles.  Specifically, our
nuclear spectra of the following objects required both broad and
narrow Balmer lines: NGC~1566, NGC~3031, NGC~4579, NGC~4594, and
NGC~5033.  The emission-line contribution from the broad-line region
obviously depends on the enclosed light-fraction of the spectrum.
Therefore, for our circumnuclear spectrum of NGC~4594 we did
\emph{not} need to include broad Balmer lines.  Finally, among our
radial-strip spectra the only galaxy that required broad Balmer
emission lines was NGC~1566.

\begin{figure}
\epsscale{1.2} 
\plotone{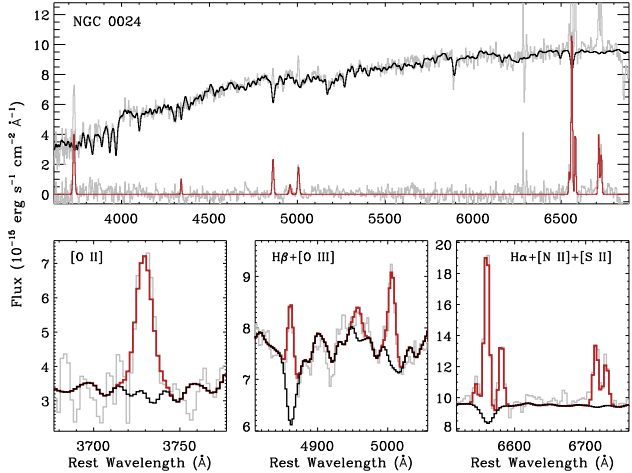}
\caption{Demonstration of our stellar continuum and emission-line
  fitting technique applied to the circumnuclear spectrum of NGC~0024.
  In the top panel we plot the observed spectrum (\emph{grey}) and the
  best-fitting continuum model (\emph{black}), where the limits of the
  plot have been chosen to emphasize the stellar continuum over the
  emission lines.  For reference, the best-fitting continuum spectrum
  at this spatial scale of NGC~0024 has $Z=0.02$, a
  luminosity-weighted age of $\sim10$~Gyr, and $E(B-V)=0.048$~mag of
  dust reddening, although as we emphasize in \S\ref{sec:ispec} these
  physical quantities should not be overinterpreted.  In the lower
  portion of the top panel we plot the residual (continuum-subtracted)
  spectrum ({\em grey, centered around zero flux}) and the
  best-fitting emission-line spectrum ({\em dark red}).  The bottom
  three panels show zoomed-in portions of the data and best-fitting
  model centered around \oiilam, \hb{} and \oiiidoublet, and \ha,
  \niidoublet, and \siidoublet, from left to
  right.  The amount of stellar absorption underlying the \hb{}
  emission line highlights the importance of proper
  continuum subtraction when studying the integrated emission-line
  properties of galaxies. \label{fig:example}} 
\end{figure}

Figure~\ref{fig:example} shows an example of our stellar continuum and
emission-line fitting technique applied to the circumnuclear spectrum
of NGC~0024, which is typical in terms of S/N.  Our best-fitting
absorption- and emission-model for the central region of NGC~0024 is
clearly a good representation of the data, and this result is typical
(see Appendix~\ref{sec:thedata}).  In the lower-middle panel of
Figure~\ref{fig:example} we focus on the wavelengths around \hb{} and
the \oiii{} doublet to highlight the amount of stellar absorption
affecting the \hb{} emission line.  In this example the amount of
\hb{} absorption is $\sim2.9$~\AA, a considerable fraction of the
absorption-corrected emission-line equivalent width, $\sim3.5$~\AA,
which emphasizes the importance of subtracting the underlying stellar
continuum when studying the integrated emission lines of galaxies.
For reference, the mean \hb{} stellar absorption for the full sample
is $2.5\pm1$~\AA, slightly larger than the amount of absorption found
by previous studies \citep{mccall85a, kobulnicky03a}, presumably
because of the lower instrumental resolution of our spectra.

Tables~\ref{table:fluxes} and \ref{table:ews} list the fluxes and
rest-frame equivalent widths (EWs) of the emission lines with
significant detections measured from our nuclear, circumnuclear, and
radial-strip spectra.  We construct the EW of each line by dividing
the integrated flux in each line by the mean surrounding continuum
\citep{moustakas06a}.  For the broad-line AGN identified above these
tables only give the flux and EW of the narrow-line component of the
Balmer lines.  For the detection of a line to be considered
significant we require the amplitude of the best-fitting Gaussian
model to be $>3\sigma_{c}$ above the residual (continuum-subtracted)
spectrum, where $\sigma_{c}$ is the standard deviation of the residual
spectrum near the line.  We further recommend that for most
applications a minimum signal-to-noise (S/N) ratio cut ${\rm S/N}>2$
be applied to the fluxes and EWs given in Tables~\ref{table:fluxes}
and \ref{table:ews}.

\subsection{Ancillary \ion{H}{2}-Region
  Spectroscopy}\label{sec:hiidata}  

We supplement our integrated and nuclear optical spectra with
previously published optical line-ratios of $561$ \hii{} regions in
$38$ of the SINGS galaxies (see Appendix~\ref{sec:hiiappendix}).
These data are complementary in several respects.  First, as discussed
in \S\ref{sec:redux}, we were unable to obtain an integrated optical
spectrum of most of the faint, low-surface brightness dwarfs in SINGS,
which are among the most metal-poor galaxies in the sample
\citep{kenn03b, walter07a}.  Fortunately, previously published optical
spectroscopy of least one \hii{} region in most of these objects
exists, enabling us to estimate their gas-phase oxygen abundances
self-consistently with respect to the rest of the sample (see
\S\ref{sec:hiioh}).  Second, unlike an integrated optical spectrum of
a galaxy, the spectrum of an individual \hii{} region typically
exhibits a much simpler underlying stellar continuum, mitigating
continuum-subtraction errors and enabling faint lines to be detected
and measured.  In addition, each integrated spectrum is a
surface-brightness weighted average of all the \hii-regions contained
within the spectroscopic aperture, and likely includes contributions
from diffuse- and shock-ionized gas emission \citep{kenn92b,
  lehnert94a, dopita06a}, although we do not expect these effects to
be a significant source of bias in our abundance estimates
\citep{kobulnicky99a, pilyugin04b, moustakas06c}.  Finally, the
emission-line spectrum of an \hii-region is impervious to AGN
contamination, which can be a significant source of systematic error
when deriving oxygen abundances from the strong nebular lines (see
\S\ref{sec:class}).  On the other hand, our database of \hii-region
line-ratios constitutes an heterogeneous compilation of data from the
literature.  Although we have made every effort to select only
high-fidelity observations, the data are not of uniform quality for
all galaxies, a point that should be kept in mind in the subsequent
analysis.  Appendix~\ref{sec:hiiappendix} contains details regarding
how the emission-line database was built and tabulates the properties
of each \hii{} region used in the subsequent analysis.

\section{Optical Spectral Classifications}\label{sec:class} 

Neglecting the AGN contribution to the optical emission-line spectrum
of a galaxy can lead to catastrophic errors in the derived oxygen
abundances, particularly those that rely on the \oiii/\hb{} ratio
\citep{kewley08a}.  Therefore, we classify the SINGS galaxies into
three broad categories: AGN, star-forming (SF) galaxies, and composite
systems, or SF/AGN, as defined below; further separating the AGN into
Seyferts \citep{seyfert43a} and LINERs \citep{heckman80a, kewley06b,
  ho08a} is beyond the scope of this paper.  A detailed investigation
of the nuclear properties of the SINGS galaxies using both optical and
infrared diagnostic diagrams has also been performed by
\citet{dale06a}.

Clearly, the degree to which the observed emission-line spectrum is
`contaminated' by an AGN depends both on the level of AGN activity,
and on the area (e.g., in kpc$^{2}$) subtended by the spectroscopic
aperture.  For example, a galaxy with an AGN-like nuclear spectrum may
have a circumnuclear or radial strip spectrum that is dominated by the
integrated emission from star-forming regions in the disk.  In these
objects, we can use the circumnuclear and radial-strip spectra to
obtain a reliable estimate of the oxygen abundance, but not the
nuclear spectrum.  Consequently, we classify each object in SINGS
using all three of our nuclear, circumnuclear, and radial strip
spectra, where available.

\begin{figure}
\epsscale{1.2}
\plotone{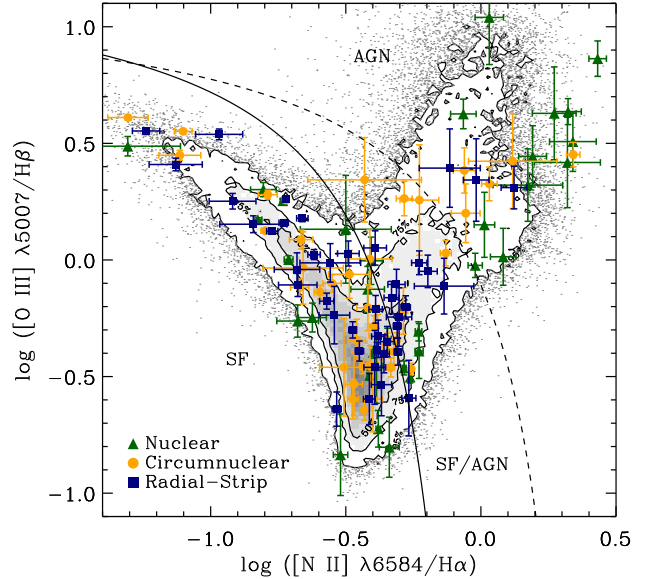}
\caption{Emission-line diagnostic diagram used to separate
  star-forming (SF) galaxies from active galactic nuclei (AGN).  We
  plot line-ratios based on our nuclear, circumnuclear, and
  radial-strip spectra using green triangles, orange points, and blue
  squares, respectively.  The contoured greyscale shows for comparison
  the distribution of SDSS emission-line galaxies in this diagram.
  The solid and dashed lines are the empirical and theoretical
  boundaries between SF galaxies and AGN defined by
  \citet{kauffmann03c} and \citet{kewley01b}, respectively.  We
  identify objects positioned above and to the right of the Kewley
  et~al. curve as AGN, objects located below and to the left of the
  Kauffmann et~al. curve as SF, and objects lying between the two
  curves as composite systems, or SF/AGN.  \label{fig:bpt}}
\end{figure}

In Figure~\ref{fig:bpt} we plot the \niilam/\ha{} versus
\oiiilam/\hb{} emission-line diagnostic diagram \citep{baldwin81a,
  veilleux87a}, which has been shown to be a sensitive diagnostic of
AGN activity \citep{kewley00a, kewley01a, kauffmann03c, stasinska06a}.
We plot line-ratios measured from our nuclear, circumnuclear, and
radial strip spectra as green triangles, orange points, and blue
squares, respectively.  In order to be included in this diagram all
four emission lines must have ${\rm S/N}>2$ (see \S\ref{sec:ispec}).
The contoured greyscale shows for comparison the distribution of SDSS
emission-line galaxies in this diagram.\footnote{The emission-line
  measurements for the SDSS sample described in \S\ref{sec:sample}
  were taken from the publically available MPA/JHU database
  (http://www.mpa-garching.mpg.de/SDSS/DR7).}  We use the
\citet{kewley01b} and \citet{kauffmann03c} classification curves to
separate the SINGS galaxies into SF, AGN, and SF/AGN: we classify
galaxies above and to the right of the Kewley et~al. curve as AGN;
galaxies below and to the left of the Kauffmann et~al. curve as SF;
and objects between the two curves as SF/AGN.

For some objects either \hb{} or \oiii{} were not detected; however, we
can still use the \nii/\ha{} ratio to discriminate between
star-forming galaxies and AGN because the star-forming galaxy sequence
asymptotes to a roughly constant \nii/\ha{} ratio as the \oiii/\hb{}
ratio diminishes \citep[see Fig.~\ref{fig:bpt};][]{kauffmann03c}.
Examining Figure~\ref{fig:bpt}, for these objects we adopt
$\log\,(\nii/\ha)=-0.25$ as the boundary between AGN and star-forming
galaxies.

Table~\ref{table:class} summarizes the results of classifying the
SINGS galaxies using our optical spectra.  A question mark in any of
these columns indicates that we failed to detect one or more of the
requisite emission lines.  As a consistency check we also classified
our sample using the nuclear emission-line fluxes published by
\citet{ho97a}, which are available for nearly half ($35/75$) the SINGS
sample.  We find that our nuclear classifications generally agree very
well with the classifications derived using the \citet{ho97a}
line-ratios. 

For many applications it is useful to have a single spectral class for
each galaxy, which we provide in the last column of
Table~\ref{table:class}.  This final classification was generally
adopted from our nuclear spectrum; therefore, the final classification
tends to favor AGN.  However, we emphasize that for many objects the
AGN contributes negligibly to the central or integrated spectrum of
the galaxy, as traced by our circumnuclear and radial-strip spectra,
respectively. 

\begin{figure}[h]
\epsscale{1.2}
\plotone{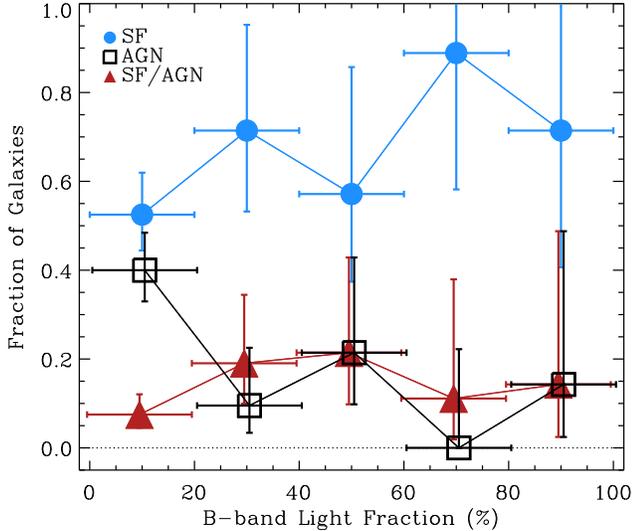}
\caption{Fraction of galaxies classified as SF ({\em filled blue
    points}), AGN ({\em open black squares}), and SF/AGN ({\em filled
    red triangles}) as a function of the $B$-band light-fraction.  At
  fixed light fraction the symbols have been offset slightly from one
  another for clarity.  The asymmetric $1\sigma$ error bars are based
  on the number of galaxies in each bin based on the formulae in
  \citet{gehrels86a}.  We find a positive (negative) correlation
  between the fraction of objects classified as SF (AGN) with
  increasing light fraction, while the proportion of objects
  classified as SF/AGN stays roughly constant.  We caution against
  over-interpreting these results, however, because they are sensitive
  to sample selection effects and incompleteness in the spectral
  classifications (e.g., missing classifications due to undetected
  emission lines.). \label{fig:classfrac}}
\end{figure}

We illustrate this last point explicitly in
Figure~\ref{fig:classfrac}, where we plot the proportion of galaxies
classified as SF, AGN, and SF/AGN as a function of the $B$-band light
fraction (see \S\ref{sec:redux}).  We find the anticipated trend that
the fraction of galaxies classified as SF increases with enclosed
light fraction at the expense of the proportion of objects classified
as AGN, while the fraction of galaxies classified as SF/AGN remains
roughly constant at $\sim15\%$.  However, we note that these results
are sensitive to incompleteness in our spectral classifications (e.g.,
due to undetected emission lines), and to sample selection effects
\citep[e.g., SINGS intentionally excluded powerful AGN;][]{kenn03b}.
These various effects should be considered carefully depending on the
specific application of the derived optical classifications.

\section{Oxygen Abundance Analysis}\label{sec:analysis}

Our objective in this section is to derive the gas-phase metallicities
of all $75$ SINGS galaxies.  There are two broad issues to consider.
First, given an optical spectrum, the observed emission-line ratios
must be converted into an estimate of the gas-phase metallicity, a
procedure that is subject to both random and systematic uncertainties.
Second, for each galaxy we often have multiple oxygen abundance
estimates sampling a wide range of spatial scales; these various
measurements must be combined in a sensible way to ensure a consistent
set of metallicities for the full sample.

We address the first question in \S\ref{sec:calib}, where we discuss
the general methodology of deriving the gas-phase metallicity of a
star-forming galaxy or \hii{} region from its observed emission-line
spectrum.  As numerous studies have pointed out \citep[and references
  therein]{kewley08a}, there exist large ($0.1-0.7$~dex) systematic
discrepancies among existing empirical and theoretical methods of
estimating oxygen abundances.  Consequently, the \emph{absolute}
uncertainty in the nebular abundance scale is a factor of $\sim5$.  To
explore this issue, we therefore compute oxygen abundances using two
different methods.  In \S\ref{sec:intnucoh} and \S\ref{sec:hiioh} we
apply these two methods to derive oxygen abundances from our nuclear,
circumnuclear, and radial-strip spectra, and from our ancillary \hii-
region spectroscopy, respectively.  We combine all the available
abundance measurements for each galaxy in \S\ref{sec:final} based on
the estimated fraction of the integrated optical light enclosed by
each spectroscopic aperture, resulting in an estimate of the
\emph{central} and \emph{characteristic}, or globally-averaged
metallicity of each galaxy.

\subsection{Strong-Line Abundance Calibration}\label{sec:calib} 

The most direct, physically motivated method of deriving the oxygen
abundance of an \hii{} region or star-forming galaxy is to measure the
electron temperature ($T_{e}$) of the ionized gas using the intensity
(relative to a hydrogen recombination line) of one or more
temperature-sensitive auroral line such as \oiii~$\lambda4363$,
\nii~$\lambda5755$, \siii~$\lambda6312$, and \oii~$\lambda7325$
\citep{dinerstein90a, skillman98a, garnett02b, stasinska07a}.  This
technique is frequently referred to as the \emph{direct}, or $T_{e}$,
method of deriving abundances.  Unfortunately, the requisite lines are
intrinsically faint, particularly in metal-rich galaxies and
star-forming regions, and in general are not detected in our spectra.
Therefore, we must rely on a so-called \emph{strong-line} abundance
calibration to estimate the metallicity of the ionized gas.
Strong-line abundance calibrations essentially relate the oxygen
abundance to one or more line-ratios involving the strongest
recombination and collisionally excited (forbidden) lines (e.g.,
\oiilam, \ha, \hb, \oiiidoublet, \niidoublet, and \siidoublet).
Although strong-line methods are \emph{indirect} and often
model-dependent, they are important because they can be used to infer
the physical conditions in star-forming galaxies across a significant
fraction of cosmic time \citep{kenn98c, pettini06a, delucia09a};
moreover, the direct method is not without its own limitations, as we
discuss in \S\ref{sec:discussion}.

Over the last three decades numerous strong-line calibrations have
been developed, but in general they fall into one of three categories:
semi-empirical, empirical, and theoretical.  The older,
\emph{semi-empirical} calibrations were generally tied to electron
temperature abundance measurements at low metallicity and
photoionization models at high metallicity \citep[e.g.,][]{alloin79a,
  pagel79a, edmunds84a, mccall85a, dopita86a, skillman89b}.  These
hybrid calibrations were born from the observational difficulty of
measuring the electron temperature of metal-rich \hii{} regions, which
remains challenging even with $6-10$-meter class telescopes (e.g.,
\citealt{castellanos02a, kenn03a, garnett04a, bresolin04a,
  bresolin05a}, but see \citealt{kinkel94a} for a heroic early
effort).  By contrast, the more recent \emph{empirical} methods are
calibrated against high-quality observations of individual \hii{}
regions with measured direct (i.e., $T_{e}$-based) oxygen abundances
\citep{pilyugin00a, pilyugin01a, denicolo02a, pettini04a, pilyugin05a,
  perez-montero05a, nagao06a, stasinska06b, yin07a,
  peimbert07a}.\footnote{Strictly speaking, the `empirical'
  calibrations presented by \citet{pettini04a} and \citet{denicolo02a}
  do include a handful of metal-rich \hii{} regions whose abundances
  were derived using photoionization models.}  One of the limitations
of the empirical calibrations, especially in the metal-rich regime, is
that they are based on observations of relatively small samples of
high-excitation \hii{} regions, whereas most integrated spectra of
galaxies exhibit softer ionizing radiation fields (see
\S\ref{sec:discussion}).  Finally, the class of \emph{theoretical}
abundance calibrations are based on {\em ab initio} photoionization
model calculations, in which various nebular emission-line ratios are
tracked as a function of the input metallicity and ionization
parameter \citep{mcgaugh91a, dopita00a, dopita06b, charlot01a,
  kewley02a}.\footnote{Confusingly, some of these theoretical
  calibrations are occasionally refered to as `semi-empirical'
  calibrations \citep[e.g.,][]{kenn96a, croxall09a}.}

Among published strong-line calibrations there exist large, poorly
understood systematic discrepancies, in the sense that empirical
calibrations generally yield oxygen abundances that are factors of
$1.5-5$ \emph{lower} than abundances derived using theoretical
calibrations \citep{kenn03a, garnett04a, bresolin04a, bresolin05a,
  shi06a, nagao06a, liang06b, yin07a, kewley08a}.  Unfortunately, the
physical origin of this systematic discrepancy remains unresolved.
Therefore, we have chosen to compute the oxygen abundances of the
SINGS galaxies using two different strong-line calibrations: the
theoretical calibration published by \citet[hereafter
  KK04]{kobulnicky04a}, and the empirical \citet[hereafter
  PT05]{pilyugin05a} calibration.  Our goal is not to conduct a
detailed intercomparison of all the available calibrations \citep[see,
  e.g.,][]{kewley08a}, but instead to bracket the range of oxygen
abundances one would derive using existing strong-line calibrations.
We discuss the various strengths and limitations of the abundances
derived using these two calibrations in \S\ref{sec:discussion}.

The KK04 and PT05 calibrations we have chosen both rely on the
metallicity-sensitive \pagel{} parameter \citep{pagel79a}:

\begin{equation}
\pagel \equiv \frac{\oiilam + \oiiidoublet}{\hblam}.
\label{eq:r23}
\end{equation}

\noindent The principal advantage of \pagel{} as an oxygen abundance
diagnostic is that it is directly proportional to both principal
ionization states of oxygen, unlike other diagnostics that have a
second-order dependence on the abundance of other elements like
nitrogen and sulfur.  Moreover, because \pagel{} depends on blue
rest-wavelength lines, it can be used to study the chemical history of
star-forming galaxies over a significant fraction of cosmic time
(\citealt{pettini01a, kobulnicky03b, kobulnicky04a, savaglio05a,
  maiolino08a}; J.~Moustakas et~al., in prep.).  The disadvantages of
\pagel{} are that it is sensitive to AGN contamination, and it must be
corrected for stellar absorption and dust attenuation (but see
\citealt{kobulnicky03a}; J.~Moustakas et~al., in prep.).  An
additional complication is that the relation between \pagel{} and
metallicity is famously double-valued (see, e.g.,
Fig.~\ref{fig:r23_vs_12oh}).  Metal-rich objects lie on the
\emph{upper} \pagel{} branch, while metal-poor galaxies and \hii{}
regions lie on the \emph{lower} branch; the transition between the
upper and lower \pagel{} branches is called the \emph{turn-around}
region.  The non-monotonic relation between \pagel{} and O/H arises
because in metal-rich star-forming regions \pagel{} decreases with
increasing O/H as the far-IR fine-structure lines (predominantly
\oiii~$\lambda52$~\micron{} and \oiii~$\lambda88$~\micron)
increasingly dominate the nebular cooling.  At low metallicity, on the
other hand, the optical transitions dominate the nebular cooling, and
so \pagel{} decreases in tandem with decreasing metallicity because
its strength is directly proportional to the abundance of oxygen
atoms.

For the KK04 calibration, we have

\begin{eqnarray}
12 & + & \log\,({\rm O/H})_{\rm lower} = 9.40 + 4.65x - 3.17x^2 \nonumber \\  
&-& \logq\,(0.272 + 0.547x -0.513x^2), 
\label{eq:kk04_lower}
\end{eqnarray}

\noindent and 

\begin{eqnarray}
12 & + & \log\,({\rm O/H})_{\rm upper} = 9.72 - 0.777x - 0.951x^2
-0.072x^3 \nonumber \\ &-& 0.811x^4 - \logq\,(0.0737 -0.0713x -
0.141x^2 \nonumber \\ &+& 0.0373x^3 - 0.058x^4),
\label{eq:kk04_upper}
\end{eqnarray}

\noindent for galaxies on the lower and upper branch, respectively,
where $x\equiv\log\,(\pagel)$.  The ionization parameter $q$ in
${\rm cm~s}^{-1}$ is given by

\begin{eqnarray}
\logq & = & 32.81 - 1.153y^2 \nonumber \\ 
      & + & z(-3.396 - 0.025y + 0.1444y^2) \nonumber \\ 
      & \times & [4.603 - 0.3119y - 0.163y^2 \nonumber \\ 
      & + & z(-0.48+0.0271y+0.02037y^2)]^{-1},
\label{eq:kk04_logu}
\end{eqnarray}

\noindent where $z\equiv\logoh$, $y\equiv\log\,(\ioniz)$, and

\begin{equation}
\ioniz \equiv \frac{\oiiidoublet}{\oiilam}
\label{eq:o32}
\end{equation}

\noindent characterizes the hardness of the ionizing radiation field
\citep{kewley02a}.  Hereafter we will use the dimensionless ionization
parameter $U\equiv q/c$ where $c=2.99\times10^{10}$~\cms{} is the
speed-of-light \citep{kewley02a}.\footnote{The ionization parameter
  also can be written as $U\propto (n_{\rm e}f^2 Q)^{1/3}$, where
  $n_{\rm e}$ is the electron density, $f$ is the volume filling
  factor, and $Q$ is the rate of photoionizing photons injected into
  the gas by massive stars \citep{shields90a}.}  Note that equations
(\ref{eq:kk04_lower})-(\ref{eq:kk04_logu}) must be solved iteratively
for both the ionization parameter and the oxygen abundance;
convergence is typically achieved in a handful of iterations.

For the PT05 calibration, we have 

\begin{eqnarray}
12 & + & \log\,({\rm O/H})_{\rm lower} = \nonumber \\ & & \frac{\pagel
  + 106.4 + 106.8P - 3.40P^2} {17.72 + 6.60P + 6.95P^2 - 0.302\pagel},
\label{eq:pt05_lower}
\end{eqnarray}

\noindent and 

\begin{eqnarray}
12 & + & \log\,({\rm O/H})_{\rm upper} = \nonumber \\ & & \frac{\pagel
  + 726.1 + 842.2P + 337.5P^2} {85.96 + 82.76P + 43.98P^2 +
  1.793\pagel}, 
\label{eq:pt05_upper}
\end{eqnarray}

\noindent for the lower and upper branch, respectively, where

\begin{equation}
P \equiv \frac{\oiiidoublet}{\oiilam+\oiiidoublet}
\label{eq:pt05_p}
\end{equation}

\noindent is an excitation parameter \citep{pilyugin01a} that is
analogous to equation~(\ref{eq:kk04_logu}) for the KK04 calibration.

The KK04 calibration is an updated parameterization of the
\citet{kewley02a} \pagel{} calibration, which is based on the
state-of-the-art photoionization model calculations carried out by
\citet{dopita00a} and \citet{kewley01b}.\footnote{Note that we do not
  use the calibration explicitly recommended by \citet{kobulnicky04a},
  which they obtain by averaging equations~(\ref{eq:kk04_lower}) and
  (\ref{eq:kk04_upper}) with the corresponding equations from
  \citet{mcgaugh91a}.  Although the differences between these two
  particular calibrations are not large ($<0.1$~dex), we generally
  advocate using one calibration derived from the same set of
  observations or theoretical models.}  The PT05 calibration, by
comparison, is based on an extensive compilation of \hii-regions from
the literature with well-determined electron temperatures, and is a
significant improvement over the original calibrations presented in
\citet{pilyugin00a, pilyugin01a}.  Note, however, that due to a
paucity of electron temperature measurements for low-excitation
regions, the PT05 calibration is only strictly applicable to
star-forming regions with $P\gtrsim0.4$ (see \S\ref{sec:discussion}).

We selected these two particular calibrations for several reasons.
First, as indicated above, they bracket the range of oxygen abundances
one would derive using other strong-line calibrations, thereby serving
as useful limiting cases.  Second, they are relatively recent, and
therefore they utilize the best available observations, atomic data,
and theoretical stellar atmospheres.  Third, they are both
two-parameter calibrations which account for variations in excitation
at fixed metallicity \citep{mcgaugh91a, pilyugin00a, pilyugin01a}.
Finally, they provide separate calibrations for objects on the lower
and upper \pagel{} branches, and therefore can be applied to a sample
such as SINGS which spans a wide range of luminosity and gas-phase
metallicity.  We refer the interested reader to
\citet{perez-montero05a}, \citet{liang06b}, and \citet{kewley08a} for
a detailed intercomparison of these and other popular strong-line
calibrations.

\subsection{Nuclear, Circumnuclear, and Radial-Strip
  Abundances}\label{sec:intnucoh}

In this section we derive oxygen abundances for the SINGS galaxies
from our nuclear, circumnuclear, and radial-strip spectra.  In order
for a spectrum to be included for chemical abundance analysis we
require \ha, \hb, \oiilam, and \oiiilam{} to be measured with a
minimum ${\rm S/N}>2$ (see \S\ref{sec:ispec}).  We also remove spectra
dominated by the central AGN, but retain those classified as SF/AGN
(see \S\ref{sec:class} and Table~\ref{table:class}).  A total of
$42/65$ galaxies ($65\%$) satisfy these criteria, of which $33/44$
($79\%$) have at least two optical spectra.

Following standard practice we correct the emission-line fluxes for
dust reddening using the observed Balmer decrement, $(\ha/\hb)_{\rm
  obs}$ assuming an intrinsic case~B recombination value of
$(\ha/\hb)_{\rm int} = 2.86^{+0.18}_{-0.11}$, where the uncertainty
reflects the variation in $(\ha/\hb)_{\rm int}$ with electron
temperature \citep{storey95a, osterbrock06a}.  Assuming a foreground
dust screen and the \citet{odonnell94a} Milky Way extinction curve,
the reddening is given by $E(B-V) =
2.184\times\log_{10}\,[(\ha/\hb)_{\rm obs}/(\ha/\hb)_{\rm int}]$
\citep{calzetti01a, moustakas06b}.  Table~\ref{table:reddening} lists
the observed \ha/\hb{} ratios and corresponding $E(B-V)$ values
inferred from our nuclear, circumnuclear, and radial-strip spectra.
In some cases the measured \ha/\hb{} ratio was less than, but
statistically consistent with, the adopted intrinsic ratio (within
$1\sigma$).  For these spectra we set $E(B-V)$ equal to zero and
propagate the statistical uncertainty in \ha/\hb{} into the reddening
error.

\begin{figure}[ht]
\centerline{\includegraphics[scale=0.45,angle=0]{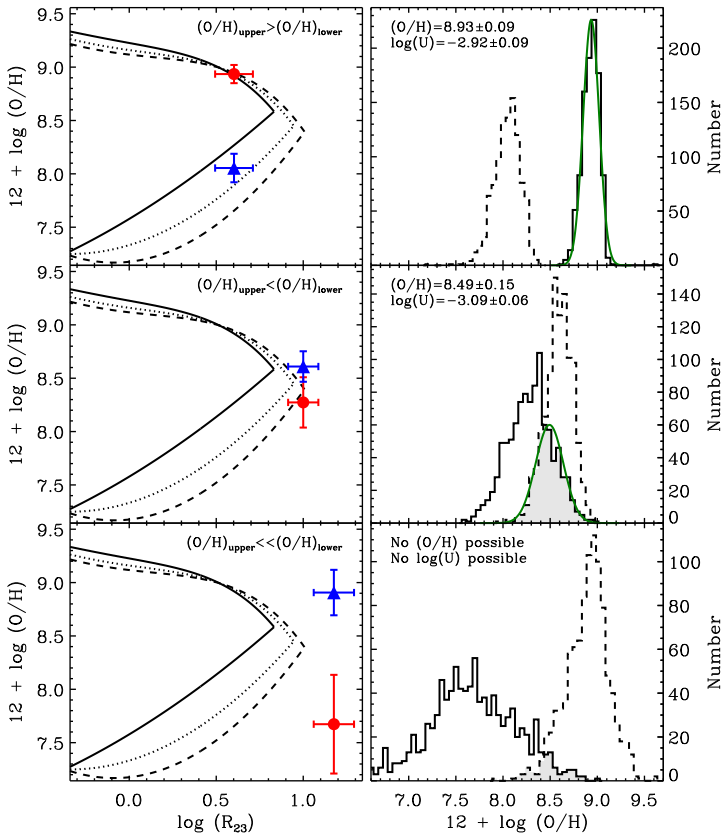}}
\caption{{\scriptsize Illustration of the methodology we use to derive
    oxygen abundances and ionization parameters.  The algorithm uses
    the measured errors on \pagel{} and \ioniz{} to derive a realistic
    abundance uncertainty for objects that are near the
    \emph{turn-around} region of the \pagel-O/H relation, and to
    ascertain quantitatively whether an object is formally ``off'' the
    \pagel-O/H calibration.  These particular examples use the KK04
    calibration, but the method can be applied to any \pagel-based
    abundance calibration.  Each row corresponds to a different
    hypothetical galaxy or \hii{} region.  In the left-hand panels we
    plot \pagel{} vs.~oxygen abundance for three representative
    ionization parameters, $\logu = -3.77$ (\emph{solid}), $-2.87$
    (\emph{dotted}), and $-2.30$ (\emph{dashed}).  The red point and
    blue triangle in each panel shows the resulting metallicity
    assuming the object lies on the upper and lower \pagel{} branch,
    respectively.  [Note that this method does \emph{not} decide
      whether an object is on the upper or lower branch; that choice
      is an \emph{input} into this algorithm (see \S\ref{sec:calib}).]
    The right-hand panels show, for each object, the Monte Carlo
    distribution of oxygen abundances corresponding to the solution on
    the upper ({\em solid histogram}) and lower branch ({\em dashed
      histogram}), based on the errors on the line-fluxes after $500$
    trials.  (\emph{Top}) For this galaxy the upper and lower-branch
    solutions are distinct and well-separated.  Assuming the object
    belongs on the upper branch, the final metallicity and error are
    given by the mean and standard deviation of the solid histogram,
    represented by the green Gaussian profile in the top-right panel.
    (\emph{Middle}) Here, the metallicity on the lower \pagel{} branch
    is formally larger than the upper-branch metallicity, which is not
    physical.  However, given the uncertainties on \pagel{} and
    \ioniz, a \emph{subset} of the possible abundance solutions have
    $({\rm O/H})_{\rm upper}>({\rm O/H})_{\rm lower}$, shown as the
    shaded histogram in the middle-right panel.  Consequently, we
    adopt the mean and standard deviation of the distribution of
    abundances in the shaded overlap region as the final metallicity
    and uncertainty in this situation, illustrated with the green
    Gaussian profile.  (\emph{Bottom}) The \pagel{} parameter measured
    for this galaxy places it well outside the region of parameter
    space defined by the KK04 calibration: $({\rm O/H})_{\rm
      upper}\ll({\rm O/H})_{\rm lower}$; therefore, no estimate of the
    oxygen abundance or ionization parameter is possible for this
    object.} \label{fig:r23_vs_12oh}}
\end{figure}

Next, we use the reddening-corrected line-fluxes and
equations~(\ref{eq:r23}), (\ref{eq:o32}), and (\ref{eq:pt05_p}) to
derive \pagel, \ioniz, and $P$, respectively.  Before computing oxygen
abundances, however, every galaxy must be assigned to either the lower
or upper branch of the KK04 and PT05 calibration (see
\S\ref{sec:calib}).  Following \citet{contini02a}, we assign galaxies
to the lower branch if $\log(\nii/\ha)<-1$ and
$\log(\nii/\oii)<-1.05$, while we identify upper-branch objects as
having $\log(\nii/\ha)>-1$ and $\log(\nii/\oii)>-0.8$ \citep[see
  also][]{kewley08a}.  In some cases these criteria are either
inconclusive or cannot be applied because of a poorly measured \nii{}
line, in which case we use the optical luminosity to choose the most
likely branch (see \S\ref{sec:final}).

Because the \pagel-O/H relation is double-valued, special care is
required when computing the oxygen abundances of objects near the
\emph{turn-around} region, where the upper and lower branches
intercept.  In particular, objects that are statistically consistent
with being on \emph{either} the upper branch \emph{or} the lower
branch must have correspondingly large abundance errors.  In addition,
measurement uncertainties or residual AGN contamination occasionally
result in an \pagel{} parameter that is larger than $\sim10$, the
approximate theoretical limit for photoionization by massive stars
\citep{kewley02a}.  Although these objects formally lie ``off'' the
\pagel{} calibration, rather than rejecting them outright, which is
typically what has been done in the literature, it is better to assess
whether they are statistically consistent with being \emph{on} the
\pagel{} calibration under consideration.  In
Figure~\ref{fig:r23_vs_12oh} we illustrate the quantitative procedure
we have developed to compute \pagel-based oxygen abundances that
addresses all these issues.  Here, we focus on the KK04 calibration,
but the same procedure applies to the PT05 calibration and, indeed, to
any \pagel-based abundance calibration.

In the left panels of Figure~\ref{fig:r23_vs_12oh} we plot \pagel{}
versus \logoh{} for three hypothetical galaxies.  The curves show the
KK04 calibration for three representative values of the ionization
parameter, $\logu = -3.77$ (\emph{solid}), $-2.87$~dex
(\emph{dotted}), and $-2.30$~dex (\emph{dashed}).  In each panel the
filled red point corresponds to the oxygen abundance on the upper
branch, while the filled blue triangle corresponds to the lower-branch
solution (for the same object).  In the top-left panel the upper- and
lower-branch solutions are distinct and well-separated: once the
appropriate branch has been chosen (e.g., using the criteria described
above), the corresponding metallicity follows.  The top-right panel
shows the resulting Monte Carlo distribution of \logoh{} values
corresponding to the solution on the upper ({\em solid histogram}) and
lower branch ({\em dashed histogram}), assuming Gaussian errors on the
\oii, \oiii, and \hb{} line-fluxes
after $500$ trials.  Assuming that this object belongs on the upper
branch, the $1\sigma$ uncertainty on \logoh{} is given by the width of
the solid histogram; a Gaussian profile ({\em green curve}) of the
appropriate width has been overplotted to guide the eye.  The
identical procedure leads to the uncertainty on the ionization
parameter, \logu.

The middle panels in Figure~\ref{fig:r23_vs_12oh} illustrate a more
ambiguous, albeit frequently encountered situation.  In this case the
formal solution on the lower branch is \emph{larger} than the solution
on the upper branch (note that the blue triangle is now above the red
circle).  In the middle-right panel the overlapping, shaded region
corresponds to values of \pagel{} and \ioniz{} that lie \emph{on} the
KK04 calibration, that is, where $({\rm O/H})_{\rm upper}>({\rm
  O/H})_{\rm lower}$.  If the central values of the two distributions
are within $1\sigma$ of one another, as measured by the width of the
shaded histogram ({\em green Gaussian profile}), we adopt the
\emph{average} of the two solutions as the oxygen abundance, and the
width of the shaded histogram as the $1\sigma$ uncertainty.  We
further characterize the \pagel{} branch as \emph{ambiguous}.  In
practice, these types of objects all have KK04 abundances equal to the
abundance around the turn-around region, $\approx8.5$~dex; however,
they also have large abundance errors, which reflects the \pagel{}
branch ambiguity.

Finally, the bottom panels in Figure~\ref{fig:r23_vs_12oh} illustrate
a situation in which no oxygen abundance measurement is possible using
the KK04 calibration.  Here, the upper- and lower-branch solutions are
statistically inconsistent with one another, given the measurement
uncertainties; therefore no solution exists, and these objects must be
rejected.

Applying the above procedure to the star-forming galaxies in our
sample we derive \pagel{} branches, ionization and excitation
parameters, and gas-phase oxygen abundances, as well as robust
uncertainties, using both the KK04 and PT05 calibrations.  For
reference, we are able to estimate oxygen abundances using the KK04
calibration for all the star-forming SINGS galaxies, and for all but
one object (the circumnuclear spectrum of NGC~5474) using the PT05
calibration.  We list the results in Table~\ref{table:intnucoh} and
discuss them in \S\ref{sec:final}.

\subsection{\ion{H}{2}-Region Abundances}\label{sec:hiioh}

In the previous section we derived oxygen abundances using our new
nuclear and integrated optical spectra.  Here, we analyze the
abundances of the SINGS galaxies using our \hii-region database (see
\S\ref{sec:hiidata}).  In \S\ref{sec:gradients} we use these data to
constrain the form of the radial abundance gradient in $21$ of the
SINGS galaxies, and in \S\ref{sec:hiiavg} we compute the average
oxygen abundances of all $38$ galaxies with observations of at least
one \hii{} region.

\subsubsection{Radial Gradients}\label{sec:gradients}

\begin{figure*}[h]
\centerline{\includegraphics[scale=0.75,angle=0]{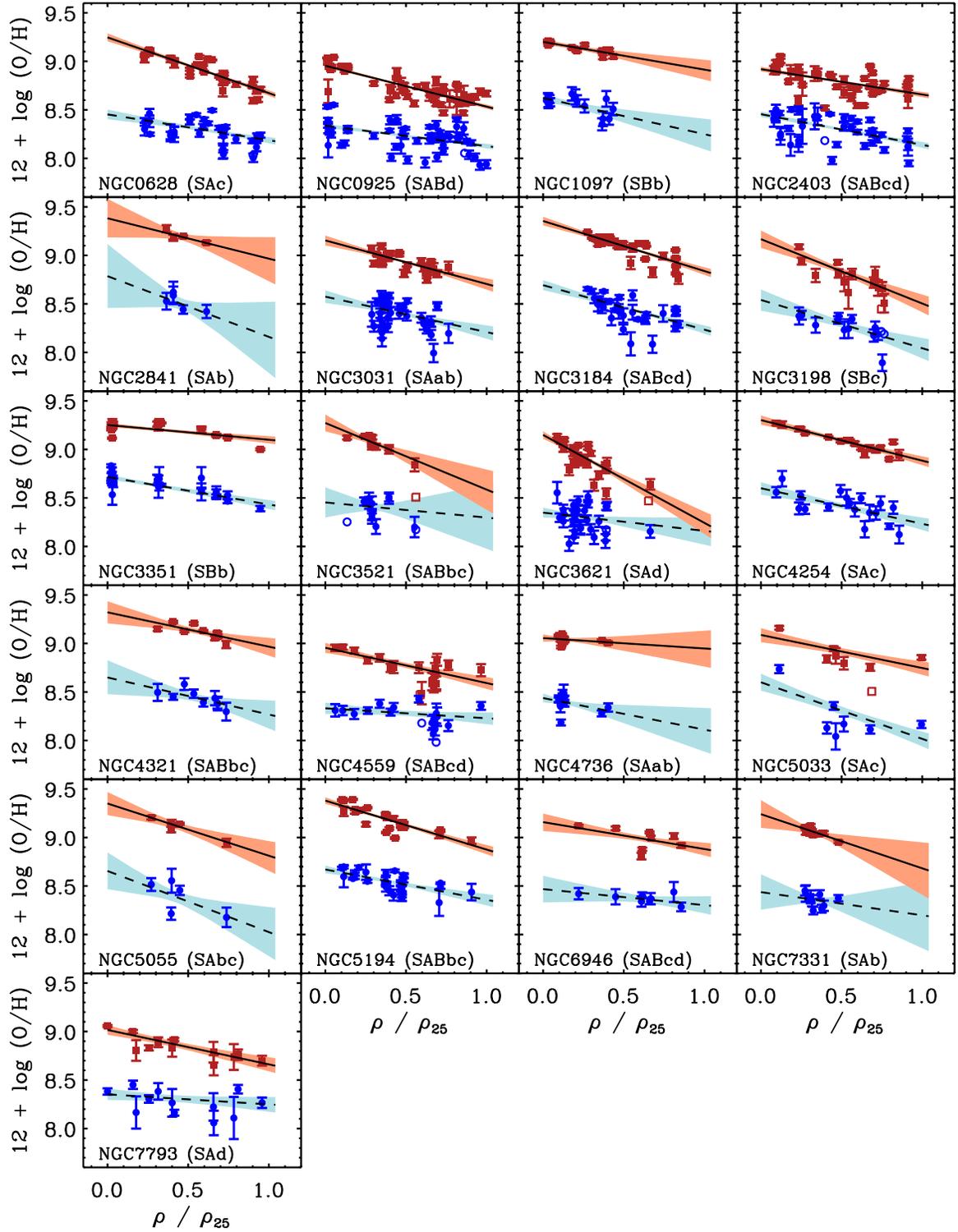}}
\caption{Oxygen abundance vs.~deprojected galactocentric radius for
  individual \hii{} regions in $21$ of the SINGS galaxies.  The dark
  red squares and blue points correspond to oxygen abundances computed
  using the KK04 and PT05 calibration, respectively; open symbols
  without error bars represent \hii{} regions with an ambiguous
  \pagel{} branch (see \S\ref{sec:intnucoh} and
  Fig.~\ref{fig:r23_vs_12oh}).  The solid and dashed lines show the
  best-fitting linear abundance gradients based on each calibration,
  and the light-red and light-blue shaded regions reflect the
  $1\sigma$ range of gradients that are consistent with the
  observations.  We provide the galaxy name and visual morphology of
  each object for reference in the lower-left corner of each panel.
  We find that the KK04 calibration yields abundances that are a
  factor of $\sim4$ higher than those based on the PT05 calibration,
  and that the abundance gradients derived using the KK04 calibration
  are systematically steeper.  \label{fig:gradients}}
\end{figure*}

Building on L.~H. Aller's pioneering study of \hii{} regions in M~33
\citep{aller42a}, \citet{searle71a} was the first to suggest that disk
galaxies might possess radial abundance gradients.  Subsequent work
confirmed his interpretation that the gas-phase metallicity of disk
galaxies, including the Milky Way, decreases from the center outward
\citep[e.g.,][]{webster83a, shaver83a, garnett87a, vila-costas92a,
  zaritsky94a, kenn96a, garnett97a, vanzee98a, dutil99a, pilyugin04a,
  rosolowsky08a, rosales09a}.  Abundance gradient measurements are
important because they provide crucial constraints on the
(time-dependent) inside-out gas accretion and star-formation histories
of galactic disks (e.g., \citealt{molla96a, boissier99a, prantzos00a,
  carigi05a, colavitti09a, marcon10a}).

An accurate measurement of the abundance gradient of a galaxy requires
observations of a minimum number of \hii{} regions spanning a large
enough fraction of the disk to constrain the slope \citep{zaritsky94a,
  dutil01a, bresolin09a}.  Among the $75$ SINGS galaxies, there are
$21$ disk galaxies with published spectroscopy for five or more \hii{}
regions spanning at least $10\%$ of the disk radius, $\rho_{25}$ (see
Table~\ref{table:properties}).  In Figure~\ref{fig:gradients} we plot
oxygen abundance versus normalized deprojected galactocentric radius,
$\rho/\rho_{25}$, for all the star-forming regions in these objects.
We derive oxygen abundances using both the KK04 ({\em filled dark red
  squares}) and PT05 ({\em filled dark blue points}) abundance
calibrations assuming the \pagel{} branches listed in
Table~\ref{table:hiioh}.  We plot \hii{} regions with an ambiguous
\pagel{} branch, according to the criteria defined in
\S\ref{sec:intnucoh}, using open symbols without error bars; these
regions are not used when fitting the abundance gradient.  Finally, we
add $0.05$~dex in quadrature to the statistical abundance uncertainty
of each \hii{} region to ensure that the fit is not dominated by a
small number of objects with the smallest statistical errors.  We
model the radial gradient in each galaxy, separately for the KK04 and
PT05 abundance calibration, using a weighted linear least-squares fit,
and plot the results using solid and dashed lines in
Figure~\ref{fig:gradients}, respectively.  The light-red and
light-blue shaded regions reflect the $1\sigma$ range of linear
gradients that are consistent with the observations, accounting for
the covariance matrix of the best-fitting parameters.  We tabulate the
derived abundance gradients and uncertainties in
Table~\ref{table:hiioh}, and discuss the results below and in
\S\ref{sec:final}.

Figure~\ref{fig:gradients} illustrates the well-known observation that
disk galaxies exhibit a wide range of abundance gradient slopes.
Using the KK04 calibration, the radial gradients in these objects
range from $-0.91$~dex~$\rho_{25}^{-1}$ in the late-type disk galaxy
NGC~3621, to $-0.11$~dex~$\rho_{25}^{-1}$ in the SAab galaxy NGC~4736,
with a mean slope of $-0.42\pm0.19$~dex~$\rho_{25}^{-1}$.  Using the
PT05 calibration the abundance gradients are generally shallower, as
we discuss below; the average slope is
$-0.33\pm0.16$~dex~$\rho_{25}^{-1}$, ranging from
$-0.63$~dex~$\rho_{25}^{-1}$ in NGC~5055, to
$-0.10$~dex~$\rho_{25}^{-1}$ in the barred galaxy NGC~4559.  These
gradients imply a factor of $\sim1.7-4$ decrease in the oxygen
abundances of galaxies from the center to the optical edge of the disk
using the KK04 calibration, or a factor of $\sim1.5-3$ using the PT05
calibration.  Using the KK04-based abundances we find a weak
correlation between abundance gradient slope and Hubble type in the
sense that early-type disk galaxies tend to exhibit shallower
abundance gradients, confirming previous studies that also relied on
theoretical strong-line calibrations \citep[e.g.,][]{oey93a,
  zaritsky94a, garnett97a}.  By contrast, the PT05-based abundance
gradients are independent of Hubble type \citep[see
  also][]{pilyugin04a}.  There is also no statistically significant
correlation between $B$-luminosity and slope using either calibration,
although a noisy, but significant correlation appears if the slope is
expressed in physical units, i.e., dex~kpc$^{-1}$, owing to the
tendency for luminous disk galaxies to be larger \citep{garnett97a,
  blanton09a}.  Finally, previous studies have suggested that barred
galaxies exhibit shallower abundance gradients
\citep[e.g.,][]{vila-costas92a, dutil99a}, presumably due to enhanced
mixing via bar-driven radial inflows of gas \citep[and references
  therein]{kormendy04a}; unfortunately, there are too few galaxies in
our sample to test this hypothesis.

\begin{figure}[h]
\centerline{\includegraphics[scale=0.3,angle=0]{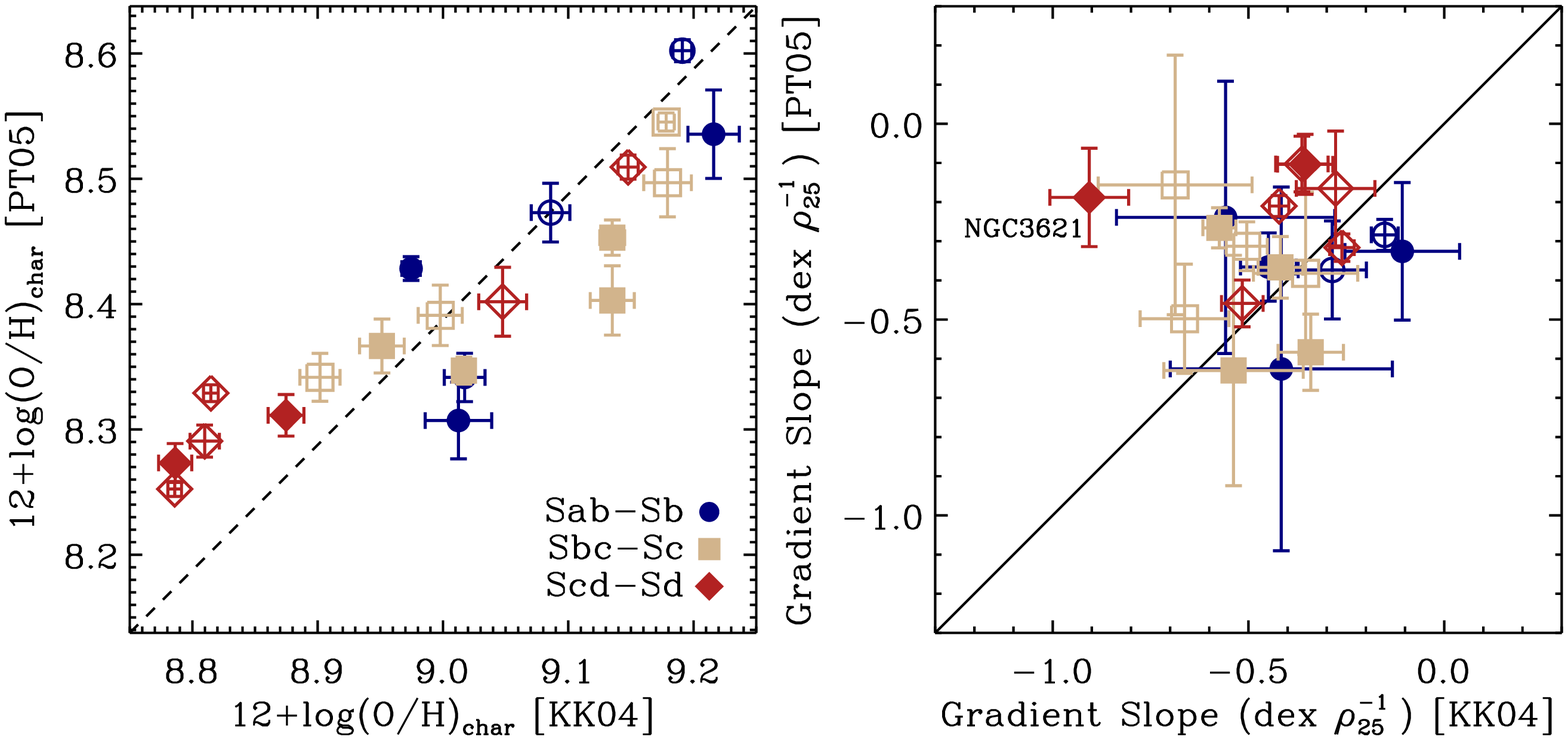}}
\caption{Comparison of the (\emph{left}) characteristic abundances and
  (\emph{right}) abundance gradient slopes derived using the KK04 and
  PT05 calibrations.  Open (filled) symbols indicate barred (unbarred)
  galaxies, and the different symbols and colors correspond to the
  following range of morphological types: Sab-Sb ({\em dark blue
    points}); Sbc-Sc ({\em tan squares}); and Scd-Sd ({\em dark red
    diamonds}).  We find that the characteristic abundances derived
  using the KK04 calibration are, on average, $\sim0.6$~dex higher
  than those based on the PT05 calibration ({\em dashed line}),
  although the difference varies from $0.55-0.65$~dex from low to high
  metallicity.  To first order, the abundance gradient slopes derived
  using either calibration are correlated; however, to second order
  the KK04 calibration results in moderately steeper abundance
  gradients.  NGC~3621, the galaxy with the most discrepant abundance
  gradient, has been labeled in the right panel and is discussed in
  \S\ref{sec:gradients}.  \label{fig:slope}}
\end{figure}

Perhaps the most striking result in Figure~\ref{fig:gradients} is the
considerable systematic offset between the two abundance scales: the
oxygen abundances derived using the theoretical KK04 calibration are,
on average, a factor of $\sim4$ higher than those based on the
empirical PT05 calibration.  We quantify this result in
Figure~\ref{fig:slope} (\emph{left}), where we compare the
\emph{characteristic} abundances of these galaxies derived using each
calibration.  The characteristic abundance is defined as the oxygen
abundance at $\rho=0.4\rho_{25}$ \citep{zaritsky94a, garnett02a},
which has been shown by \citet{moustakas06c} to be statistically
consistent with the luminosity-weighted mean metallicity of the whole
galaxy based on integrated spectroscopy (see also
\citealt{kobulnicky99a, pilyugin04b, rosales09a}).  Open and filled
symbols in this figure differentiate between barred and unbarred
galaxies, respectively, and the symbols types correspond to different
morphological classes: Sab-Sb ({\em blue points}); Sbc-Sc ({\em pink
  squares}); and Scd-Sd ({\em tan diamonds}).  The characteristic
abundances of these galaxies based on the KK04 calibration are
$0.61\pm0.06$~dex higher than the corresponding metallicities derived
using the PT05 calibration ({\em dashed line}).  The residuals are
also a weak function of metallicity: below $\logoh_{\rm
  KK04}\lesssim8.95$, the PT05 characteristic abundances are offset by
$\sim0.55$~dex from the KK04 abundances, while at higher metallicity
the offset is slightly larger, $\sim0.65$~dex.

Returning to Figure~\ref{fig:gradients}, we find that despite the
significant zeropoint offset in the two abundance scales, to first
order the \emph{slope} of the abundance gradients in these galaxies
are correlated (see also \citealt{moustakas06c} and
\citealt{rosales09a}).  We quantify this result in
Figure~\ref{fig:slope} (\emph{right}), where we find a well-defined,
albeit noisy correlation between the abundance gradient slopes derived
using the PT05 and KK04 calibrations.  The most significant outlier
from the one-to-one relation ({\em solid line}) is NGC~3621: according
to the KK04 calibration the abundance gradient slope is
$-0.91\pm0.10$~dex~$\rho_{25}^{-1}$, considerably steeper than the
slope derived using the PT05 calibration,
$-0.19\pm0.13$~dex~$\rho_{25}^{-1}$.  We emphasize that the steep
slope derived for NGC~3621 using the KK04 calibration is not being
driven by either of the \hii{} regions at $\rho/\rho_{25}\approx0.65$
(S3A1 and S3A2; see Appendix~\ref{sec:hiiappendix}); we obtain a slope
that is within the statistical error whether or not these regions are
included in the fit.

To second order, however, the abundance gradients derived using the
KK04 calibration are systematically steeper.  Excluding the three
galaxies with the least well-determined abundance gradients (NGC~2841,
NGC~3521, and NGC~7331; see Table~\ref{table:hiioh} and
Fig.~\ref{fig:gradients}), the weighted mean difference in slope is
$0.10\pm0.02$~dex~$\rho_{25}^{-1}$.  \citet{bresolin09a} report a
similar result for NGC~0300; they find that various theoretical
strong-line abundance diagnostics yield steeper abundance gradients
than the gradient inferred from electron-temperature abundance
estimates.  Indeed, for NGC~5194 the abundance gradient slope we
derive using the empirical PT05 calibration,
$-0.31\pm0.06$~dex~$\rho_{25}^{-1}$, is statistically consistent with
the slope derived by \citet{bresolin04a},
$-0.28\pm0.14$~dex~$\rho_{25}^{-1}$, using a sample of $10$ \hii{}
regions with high-quality electron temperature measurements; by
comparison, the theoretical KK04 calibration yields a slope that is
$\sim50\%$ steeper, $-0.50\pm0.05$~dex~$\rho_{25}^{-1}$.  In the
context of galactic chemical evolution models
\citep[e.g.,][]{prantzos00a, marcon10a}, the slope of the abundance
gradient in a disk galaxy places tight constraints on its radially and
time-dependent gas-accretion and star-formation history, making it
important to determine whether the steeper or shallower gradients
predicted by the KK04 or PT05 calibration, respectively, are more
correct.  On the other hand, the abundance differences at $\rho_{25}$
due to the slightly different gradients are typically $<0.1$~dex,
negligible compared to the zeroth-order systematic difference in the
two abundance scales, $\sim0.6$~dex.  Nevertheless, these results
emphasize the need for a resolution to the nebular abundance scale
discrepancy (see \S\ref{sec:discussion}).

The final point we raise regarding Figure~\ref{fig:gradients} is the
amount and possible physical origin of the dispersion in metallicity
at fixed galactocentric radius.  Relative to the KK04-based abundance
gradients, the dispersion ranges from $0.03-0.11$~dex, with a mean
value of $\pm0.06$~dex.  The scatter around the best-fitting
PT05-based abundance gradients is $0.05-0.18$~dex, or $\pm0.10$~dex,
on average.  It is interesting to determine why the dispersion in
metallicity when using the PT05 calibration is $\sim50\%$ larger than
when adopting the KK04 calibration.  Correlating the abundance
residuals against various properties of the \hii{} regions, we find
that the larger dispersion is being driven primarly by low-excitation
($P\lesssim0.2$) \hii{} regions.  As emphasized in \S\ref{sec:calib},
the paucity of low-excitation \hii{} regions with electron temperature
abundance measurements means that the PT05 calibration is not
well-constrained in this regime (see also the discussion in
\S\ref{sec:discussion}); therefore, PT05-based abundances of
low-excitation \hii{} regions may be susceptible to additional
systematic errors (i.e., due to extrapolation).  Consequently, in the
subsequent discussion we focus exclusively on the dispersion around
the KK04-based abundance gradients.

The dispersion we measure is comparable to or smaller than the scatter
reported by previous studies, $\pm0.1-0.2$~dex
\citep[e.g.,][]{mccall85a, zaritsky94a, vanzee98a}.  One reason we
measure a smaller dispersion in metallicity at fixed galactocentric
radius compared to these older studies, even though we are using data
compiled from the very same surveys, is because the KK04 calibration
accounts for variations in metallicity \emph{and} ionization for a
given emission-line spectrum.  Neglecting the tendency for metal-poor
(i.e., outer) \hii{} regions to have harder ionizing radiation fields
would introduce additional scatter due to the variation in physical
conditions among different \hii{} regions.  \citet{kenn96a} speculate
that in M101 the dispersion is metallicity may be due to large-scale
deviations from azimuthal symmetry in the gas disk, perhaps due to
tidal interactions with its nearby companions.  One way to test this
hypothesis is to build two-dimensional abundance maps of statistically
significant samples of nearby galaxies, which has become possible
recently with the latest generation of integral-field-unit
spectrographs \citep[e.g.,][]{rosales10a, blanc10a}.

\subsubsection{Average \ion{H}{2}-Region Oxygen 
  Abundances}\label{sec:hiiavg}

\begin{figure}[!ht]
\epsscale{1.2}
\plotone{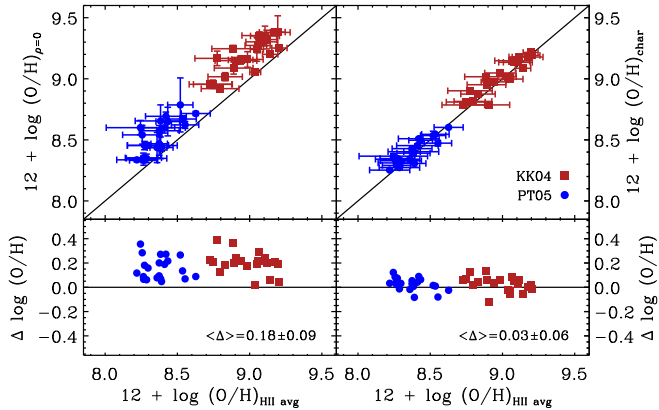}
\caption{Comparison of the average \hii-region oxygen abundance
  vs.~(\emph{left}) the metallicity at $\rho=0$, and (\emph{right})
  the characteristic oxygen abundance (defined at
  $\rho/\rho_{25}=0.4$) for the $21$ SINGS galaxies with measured
  radial gradients (see Fig.~\ref{fig:gradients}).  We plot abundances
  computed using the KK04 and PT05 calibration as dark red squares and
  blue points, respectively, and the solid line in each panel
  represents the line-of-equality.  The lower panels show the
  abundance residuals without error bars for clarity, and give the
  mean and standard deviation of the residuals for both abundance
  calibrations.  This comparison shows that the extrapolated central
  metallicity is, on average, a factor of $\sim1.5$ higher than the
  average \hii-region metallicity in this sample, while the average
  and characteristic abundances are statistically consistent with one
  another with a $1\sigma$ scatter of $\lesssim\pm15\%$ and no
  significant systematic offset.  We conclude that the average
  \hii-region abundance is a reliable proxy for the characteristic
  metallicity when the radial gradient cannot be
  constrained.  \label{fig:hiiavg_hiichar}}
\end{figure}

In this section we consider all $38$ SINGS galaxies with observations
of one or more \hii{} regions (see Appendix~\ref{sec:hiiappendix}),
including the $21$ objects studied in \S\ref{sec:gradients}, and
compute the average \hii-region metallicity of each galaxy assuming
the \pagel{} branches listed in Table~\ref{table:hiioh}.
Specifically, we compute the average metallicity as the unweighted
mean of all the individual \hii-region abundances in each object.  To
ensure a reliable estimate we exclude \hii{} regions with an ambiguous
\pagel{} branch assignment (see \S\ref{sec:intnucoh}), as well as
regions beyond the optical diameter of the galaxy (i.e., those with
$\rho/\rho_{25}>1$).  Estimating the uncertainty in the average
metallicity is not straightforward because galaxies exhibit genuine
abundance inhomogeneities, although the amplitude of these variations
are expected to be relatively small in dwarf galaxies
\citep[e.g.,][]{vanzee06b}.  Consequently, we compute the error in the
average metallicity as the unweighted standard deviation of the
distribution of oxygen abundances, but require the resulting
uncertainty to be greater than or equal to the mean statistical
uncertainty of the individual measurements.  This method ensures that
we do not underestimate the metallicity error in galaxies with only a
handful of \hii{} regions.  We list the final average abundances and
uncertainties for all $38$ galaxies in Table~\ref{table:hiioh}.

In Figure~\ref{fig:hiiavg_hiichar} we compare the average abundances
we derive against the (\emph{left}) central ($\rho=0$) and
(\emph{right}) characteristic ($\rho/\rho_{25}=0.4$) oxygen abundances
for the $21$ SINGS galaxies with measured radial gradients
(\S\ref{sec:gradients}).  The dark red squares and blue points
correspond to the KK04 and PT05 abundance calibration, respectively,
and the solid line is the line-of-equality.  The upper panels compare
the abundances directly and include error bars, while the lower panels
plot the residuals without error bars, for clarity.  The mean and
standard deviation of the residuals are given in the lower panels.  We
find that the central oxygen abundances in these galaxies are
$0.18\pm0.09$~dex, or a factor of $1.2-1.9$ higher than the average
metallicity, while the characteristic and average oxygen abundances
agree to within $\lesssim\pm15\%$) with no statistically significant
systematic offset, independent of the adopted abundance calibration.
This result indicates that, to first order, the average \hii-region
metallicity is a reliable surrogate for the characteristic abundance
in galaxies where we are unable to constrain the form of the radial
abundance gradient.

\subsection{Synthesis: Central and Characteristic Oxygen
  Abundances of the SINGS Galaxies}\label{sec:final}

In the previous two sections we derived several different estimates of
the gas-phase oxygen abundances for each SINGS galaxy spanning a wide
range of spatial scales (see Tables~\ref{table:intnucoh} and
\ref{table:hiioh}).  For most applications, however, a single,
characteristic oxygen abundance that is representative of the whole
galaxy may be desired; in other applications, a nuclear or central
metallicity may be needed.  The goal of this section, therefore, is to
combine all the various abundance measurements to derive a uniform set
of characteristic (i.e., globally averaged) and central oxygen
abundances for the full sample.

\begin{figure}[!hb]
\centerline{\includegraphics[scale=0.35,angle=0]{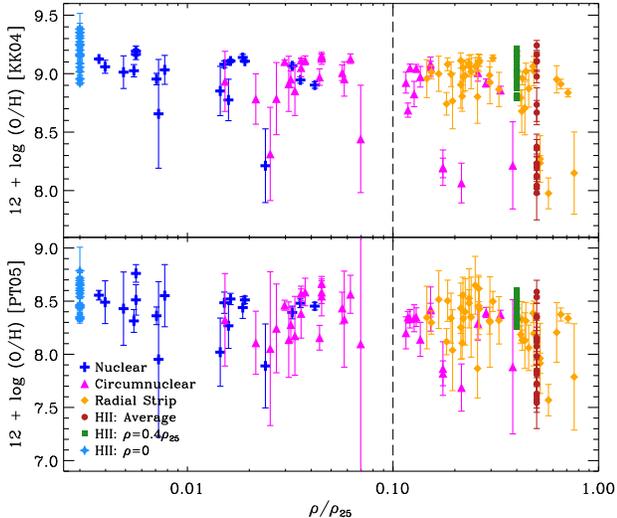}}
\caption{Oxygen abundance vs.~normalized galactocentric radius based
  on all the available data and the (\emph{top}) KK04 and
  (\emph{bottom}) PT05 abundance calibration.  We plot abundances
  inferred from our nuclear, circumnuclear, and radial-strip spectra
  using dark blue crosses, magenta triangles, and orange diamonds,
  respectively, and the average, characteristic, and central
  \hii-region abundances as dark red points, green squares, and light
  blue stars, respectively.  Note that we only plot the average
  \hii-region abundances of galaxies without measured radial gradients
  (\S\ref{sec:gradients}) so that the same \hii{} regions are not
  counted twice.  We estimate $\rho/\rho_{25}$ for our nuclear,
  circumnuclear, and radial-strip spectra using the effective radius
  of the spectroscopic aperture relative to $\rho_{25}$, as described
  in \S\ref{sec:final}.  By definition the characteristic abundances
  correspond to $\rho/\rho_{25}=0.4$, and for simplicity we assign the
  average \hii-region abundances a relative radius of
  $\rho/\rho_{25}=0.5$ (see Fig.~\ref{fig:hiiavg_hiichar},
  \emph{right}).  Finally, we plot the central ($\rho=0$) \hii-region
  abundances based on the measured abundance gradients at
  $\rho/\rho_{25}=0.003$ so that they appear on this logarithmic plot.
  We derive the central abundance of each galaxy by averaging all the
  available abundances with $\rho/\rho_{25}<0.1$, and the
  characteristic abundance of each galaxy as the weighted average of
  the individual metallicities with $\rho/\rho_{25}>0.1$, as indicated
  by the vertical dashed line.  \label{fig:ohall}}
\end{figure}

We begin our analysis with Figure~\ref{fig:ohall} by plotting
\emph{all} the oxygen abundances listed in Tables~\ref{table:intnucoh}
and \ref{table:hiioh} versus $\rho/\rho_{25}$.  We plot abundances
determined from our nuclear, circumnuclear, and radial-strip spectra
using dark blue crosses, magenta triangles, and orange diamonds,
respectively.  We derive an approximate $\rho/\rho_{25}$ value for
each spectrum by computing the mean radius of the spectroscopic
aperture, $\rho\approx \sqrt{\Delta_{\parallel}\Delta_{\perp}}/2$,
where $\Delta_{\parallel}$ and $\Delta_{\perp}$ is the diameter of the
extraction aperture along and perpendicular to the slit, respectively
(see Table~\ref{table:journal}).  Our results are not sensitive to the
details of this calculation, as our goal is simply to distinguish
between, for example, a nuclear spectrum that encloses a small
fraction of the light of the galaxy (e.g., $\rho/\rho_{25}\approx0.02$
for NGC~1482) and a radial-strip spectrum that extends over a much
larger area relative to the size of the galaxy (e.g.,
$\rho/\rho_{25}\approx0.52$ for NGC~1705).  We plot the characteristic
($\rho/\rho_{25}=0.4$) and central ($\rho=0$) abundances determined
from our radial metallicity gradients (see \S\ref{sec:gradients}) as
green squares and light blue stars, respectively.  Finally, we plot
the average \hii-region abundances derived in \S\ref{sec:hiiavg} as
dark red points at $\rho/\rho_{25}=0.5$.  Note that we only plot the
average \hii-region abundances of objects without measured radial
gradients so that the same \hii{} regions are not counted twice.

Next, we divide Figure~\ref{fig:ohall} into central and characteristic
metallicity regimes at $\rho/\rho_{25}=0.1$ ({\em vertical dashed
  line}).  For each galaxy, we define the central abundance as the
weighted average of all the available abundances at
$\rho/\rho_{25}<0.1$; similarly, we define the characteristic
metallicity of each galaxy as the weighted average of all the
abundances at $\rho/\rho_{25}>0.1$.  Although this division is
somewhat arbitrary, once again our goal is to ensure that we are not
averaging metallicities originating from widely disparate parts of the
galaxy.  Note that by using the weighted average we tend to favor
metallicities derived from the \hii{} regions, which typically have
smaller uncertainties than the metallicities derived from our nuclear
and integrated spectra.  Occasionally the average central metallicity
is formally \emph{lower} than the characteristic metallicity, although
the difference is never statistically significant (i.e., never
$>1\sigma$).  In other cases, either the central or characteristic
metallicity carries a considerably larger uncertainty because it is
based on a single abundance estimate.  For these objects we adopt the
weighted average of all the available metallicities at
$0<\rho/\rho_{25}<1$ as indicative of both the central and
characteristic abundance.  We list the final set of oxygen abundances,
using both the KK04 and PT05 abundance calibrations, in
Table~\ref{table:final}.

Summarizing, Table~\ref{table:final} contains central and
characteristic abundances for $55$ galaxies, or $\sim73\%$ of the
SINGS sample.  The characteristic abundances range from $\logoh_{\rm
  KK04}=8.00-9.21$ based on the KK04 calibration, or $\logoh_{\rm
  PT05}=7.54-8.60$ using the PT05 calibration.  Among the $20$ objects
with no metallicity estimate, $16$ ($80\%$) are early-type,
bulge-dominated galaxies lacking prominent emission lines in their
integrated spectrum (e.g., NGC~0584, NGC~5866), or their optical
emission-line spectrum is dominated by the central AGN (e.g.,
NGC~1266, NGC~4579).  Unfortunately, \hii{} regions either are not
present in these galaxies, or have not been observed.  The remaining
four objects are late-type galaxies (e.g., NGC~3938), including three
dwarfs (M~81~Dw~A, NGC~4236, and IC~4710), for which we were unable to
obtain a useful optical spectrum (see \S\ref{sec:redux}), and for
which no \hii{} regions have been observed in these objects based on
our search of the literature (see Appendix~\ref{sec:hiiappendix}).

In an effort to make our compilation of oxygen abundances for the
SINGS galaxies as comprehensive as possible, we derive an
\emph{approximate} metallicity for the remaining $20$ galaxies using
the $B$-band luminosity-metallicity (\lz) relation.  The statistical
correlation between optical luminosity (or stellar mass) and gas-phase
metallicity has been known for several decades \citep[e.g.,][and
  references therein]{lee06b}, and provides a useful tool for deriving
a rough estimate of the metallicities of the SINGS galaxies without
spectroscopic abundances.  We emphasize that the metallicities we
derive using the \lz{} relation may be susceptible to additional
systematic biases.  In particular, the bulk of the objects without
spectroscopic abundances are early-type galaxies which may not obey
the same underlying \lz{} relation defined by late-type galaxies.  In
fact, early-type disk galaxies exhibit shallower radial abundance
gradients and higher overall metallicities than late-type galaxies
\citep[\S\ref{sec:gradients};][]{garnett87a, oey93a, zaritsky94a,
  dutil99a}, although the latter correlation is likely driven by the
fact that early-type galaxies tend to be more luminous, and therefore
more metal rich.  On the other hand, we derive the $B$-band \lz{}
relation using all the galaxies in SINGS with spectroscopic
abundances, including a number of early-type disk galaxies, which
should mitigate some of these systematic effects.

\begin{figure}
\centerline{\includegraphics[scale=0.35,angle=0]{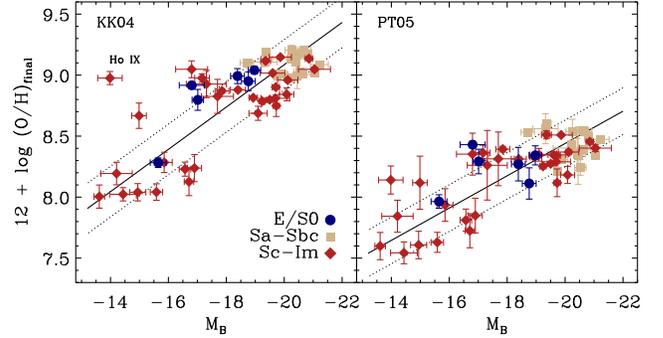}}
\caption{$B$-band luminosity-metallicity relation using the
  \emph{final} characteristic (\emph{left}) KK04 and (\emph{right})
  PT05 oxygen abundances of the SINGS galaxies (see
  Table~\ref{table:final}).  The symbols correspond to broad bins of
  morphological type: E-Sa ({\em dark blue points}), Sab-Sbc ({\em tan
    squares}), and Sc-Im ({\em dark red diamonds}).  The solid lines
  indicate the linear ordinary least-squares bisector fits to each
  luminosity-metallicity relation, and the dotted lines correspond to
  the $1\sigma$ scatter around the best fit, $\pm0.2$~dex.  In the
  left panel we label Ho~IX, a tidal-dwarf galaxy in the M~81 group
  that is considerably more metal-rich than comparably luminous
  galaxies.  \label{fig:lz}}
\end{figure}

With these caveats in mind, in Figure~\ref{fig:lz} we plot the
$B$-band \lz{} relation for the SINGS galaxies using the (\emph{left})
KK04 and (\emph{right}) PT05 abundance calibrations and the
characteristic oxygen abundances listed in Table~\ref{table:final}.
The symbols correspond to three broad bins of morphological type: E-Sa
({\em dark blue points}), Sab-Sbc ({\em tan squares}), and Sc-Im ({\em
  dark red diamonds}).  Ho~IX, a metal-rich tidal-dwarf galaxy
\citep{makarova02a, croxall09a}, deviates significantly from the
KK04-based luminosity-metallicity relation, and has been labeled in
this figure.  The solid lines show the linear ordinary least-squares
bisector fits to each set of abundances \citep{isobe90a}, while the
dotted lines in each panel indicate the $1\sigma$ residual scatter in
either \lz{} relation, $\pm0.2$~dex.  Using these \lz{} relations we
estimate the oxygen abundances of all the SINGS galaxies and list the
results in Table~\ref{table:final}.

The \lz{} relations we obtain are:

\begin{equation}
\logoh_{\rm KK04} = (5.62\pm0.24) + (-0.173\pm0.012)\mb
\label{eq:lz_kk04}
\end{equation}

\noindent and 

\begin{equation}
\logoh_{\rm PT05} = (5.79\pm0.21) + (-0.132\pm0.011)\mb
\label{eq:lz_pt05}
\end{equation}

\noindent using the KK04 and PT05 abundance calibration, respectively.
Comparing these results with previously determined $B$-band \lz{}
relations is difficult because the slope and intercept are sensitive
to sample selection effects (e.g., the distribution of absolute
magnitudes) and, obviously, the adopted abundance calibration.  For
example, \citet{tremonti04a} obtained a steeper \lz{} relation,
$\logoh=5.24-0.185\mb$ based on a large sample of SDSS galaxies with
$-18\gtrsim\mb\gtrsim-23$ and a theoretical abundance calibration.  By
comparison, \citet{lee06b} studied a sample of $25$ nearby galaxies
with $-11\gtrsim\mb\gtrsim-18$ and oxygen abundances derived using the
direct method and obtained $\logoh=5.94-0.128\mb$.  Other studies have
obtained \lz{} relation slopes ranging from $-0.149$~dex~mag$^{-1}$ to
$-0.280$~dex~mag$^{-1}$ (\citealt{skillman89a, hidalgo98a, jlee04a,
  lama04a, salzer05a, vanzee06a}; J.~Moustakas et~al., in prep.).
Given the impact of the assumed abundance calibration and sample
selection effects we conclude that equations~(\ref{eq:lz_kk04}) and
(\ref{eq:lz_pt05}) are both reasonable descriptions of the $B$-band
\lz{} relation for the SINGS galaxies.

\section{Discussion of the Nebular Abundance
  Scale}\label{sec:discussion}

The factor of $\sim5$ absolute uncertainty in the nebular abundance
scale poses one of the most important outstanding problems in
observational astrophysics.  In this paper we have sidestepped this
issue by computing the gas-phase metallicities of the SINGS galaxies
using two independent strong-line calibrations: PT05, which was
empirically calibrated against electron-temperature abundance
measurements of individual \hii{} regions; and KK04, a purely
theoretical calibration based on a large grid of state-of-the-art
photoionization model calculations (see \S\ref{sec:calib}).  We have
seen that the KK04 calibration yields abundances that are
$\sim0.6$~dex higher than metallicities derived using the PT05
calibration, {\em based on the same input emission-line ratios.}  The
question we explore in this section is: ``Which set of oxygen
abundances should one use?''

As discussed in \S\ref{sec:calib}, the absolute uncertainty in the
nebular abundance scale is largely due to the systematic difference
between abundances computed using empirical versus theoretical
calibrations (see, e.g., Fig.~2 in \citealt{kewley08a}).  Therefore,
we begin our discussion by exploring the various strengths and
limitations of the empirical and theoretical strong-line methods (see
\citealt{stasinska10a} for a complementary discussion).  Our principal
conclusion is that the empirical calibrations likely underestimate the
`true' metallicity by $\sim0.2-0.3$~dex, while the theoretical
calibrations yield abundances that may be too high by the same amount;
a compromise procedure, therefore, would be to \emph{average} the two
abundance estimates presented in \S\ref{sec:final} and
Table~\ref{table:final}.  We conclude with a brief discussion of other
strong-line abundance calibrations.

One indirect argument in favor of empirical abundance calibrations is
that they generally yield abundances for $L^{*}$ (i.e., typical)
galaxies that are more consistent with the oxygen abundance of the
Sun, $\logsunoh=8.69\pm0.05$ \citep{asplund09a}.  For example,
luminous, metal-rich galaxies on the empirical abundance scale have
$\logoh\approx8.5-8.9$, or $(0.7-1.6)\times Z_{\sun}$, while the
metallicities of the same galaxies using the theoretical strong-line
calibrations are $8.8-9.2$~dex, or $(1.3-3.2)\times Z_{\sun}$
\citep[Fig.~\ref{fig:lz};][]{tremonti04a, salzer05a, pilyugin07a,
  kewley08a}.  Consequently, on the empirical abundance scale $L^{*}$
galaxies like the Milky Way have roughly solar metallicity, whereas
theoretical abundance methods suggest that the overwhelming majority
of star-forming galaxies in the local universe are more metal-rich
than the Sun (see, e.g., Fig.~4 in
\citealt{tremonti04a}).\footnote{Note, however, that the metallicities
  derived from the SDSS $3\arcsec$ diameter fiber spectra may be too
  high by $\sim0.1$~dex due to aperture bias \citep{tremonti04a,
    kewley05a}.}  Application of the Copernican principal suggests
that the theoretical abundance scale is likely too high given the
currently accepted solar oxygen abundance \citep[but
  see][]{serenelli09a}.

Another powerful way to test the absolute zeropoint in the nebular
abundance scale is to compare the gas-phase oxygen abundances of
\hii{} regions in the Milky way and other nearby galaxies against the
\emph{stellar} abundances of young O-, B-, and A-type stars in the
same galaxies.  \citet{bresolin09a} performed this experiment by
obtaining high-quality electron temperature measurements of $28$
\hii{} regions in the nearby late-type galaxy NGC~0300.  They compared
the inferred gas-phase abundance gradient with the stellar gradient
derived from observations of blue supergiants in the same galaxy
\citep{urbaneja05b, kudritzki08a} and found excellent statistical
agreement.  \citet{pilyugin06a} reported a similar result for the
Milky Way; they applied an empirical strong-line method
\citep{pilyugin05b} to observations of Galactic \hii{} regions and
found excellent agreement with the stellar oxygen abundance gradient
from \citet{daflon04a} over a similar range of Galactocentric radii.
By comparison, three widely-used theoretical strong-line calibrations
\citep{mcgaugh91a, kewley02a, tremonti04a} applied to the \hii{}
regions in NGC~0300 yielded abundance gradients that were offset from
the stellar gradient toward higher metallicity by $0.3-0.5$~dex
\citep{bresolin09a}.

Despite these successes, direct and empirical abundance methods may
systematically \emph{underestimate} the oxygen abundances of
star-forming regions, especially in the metal-rich regime.  For
example, using the direct method \citet{esteban04a} measured the
oxygen abundance of the Orion nebula to be $\logoh=8.51\pm0.03$.
However, this metallicity is $\sim0.2-0.3$~dex lower than the oxygen
abundances of B-type stars in the solar neighborhood and in the Orion
nebula \citep[i.e., young stars within $\sim1$~kpc;][]{cunha06a,
  przybilla08a, simon-diaz10a}.  One way to (partially) resolve this
discrepancy would be if oxygen atoms are depleted by $0.1-0.2$~dex
onto dust grains \citep{jenkins04a, cartledge06a}, although presumably
this correction would affect the abundances derived from the
theoretical methods in the same way.  Alternatively, if \hii{} regions
exhibit significant spatial temperature fluctuations then the
abundances derived from collisionally excited lines such
\oiii~$\lambda4363$ could be biased low.

\citet{peimbert67a} was the first to point out that temperature
inhomogeneities in \hii{} regions could cause the electron temperature
inferred from the collisionally excited forbidden lines to be
overestimated, and therefore the abundance to be underestimated
\citep{stasinska05a, bresolin06a}.  He defined a quantity, $t^{2}$,
equal to the root-mean-square deviation of the temperature from the
mean value.  Temperature inhomogenieties are expected to be more
severe in metal-rich \hii{} regions because the higher efficiency of
metal-line cooling leads to strong temperature gradients as a function
of distance from the ionizing star or star cluster \citep{garnett92a,
  stasinska02a, stasinska05a}.  Typically, direct abundances, and the
empirical methods that are calibrated against them, assume $t^{2}=0$.
Unfortunately, deriving $t^{2}$ is very challenging observationally;
it has been measured in a relatively small number of metal-rich
[$\logoh_{T_{e}}\gtrsim8.1$] Galactic and extragalactic \hii{} regions
\citep[and references therein]{garcia07a, esteban09a}.  Using the
intensity of the faint, temperature-insensitive metal recombination
lines, the He~I recombination line spectrum, and the electron
temperature implied by the Balmer discontinuity, these studies suggest
$t^{2}=0.03-0.07$ for metal-rich \hii{} regions, corresponding to an
upward revision of the abundances derived from empirical methods of
$0.2-0.3$~dex \citep[the so-called abundance discrepancy
  factor;][]{garcia07a, esteban09a}.  Indeed, applying a correction
for $t^{2}\neq0$ to the oxygen abundance of Orion implied by the
direct method, \citet{esteban04a} obtain $\logoh=8.67\pm0.04$, in much
better agreement with the mean abundance of B-type stars in the same
star-forming region, $\logoh=8.74\pm0.04$ \citep{simon-diaz10a},
assuming that $\sim0.1$~dex of oxygen atoms are locked in dust grains.

These observations suggest that it might be possible to develop an
empirical strong-line calibration using abundances measured from metal
recombination lines, which are not susceptible to temperature
fluctuations \citep{peimbert05b, peimbert07a, bresolin07a}.
\citet{peimbert07a} derive such a calibration for \pagel{} using a
small number ($\sim20$) of metal-rich, high-excitation emission-line
galaxies and \hii{} regions.  For a given value of \pagel, this
calibration yields oxygen abundances that are $\sim0.25$~dex, or a
factor of $\sim1.8$ higher than the abundances implied by the direct
and empirical methods assuming $t^{2}=0$.  However, this calibration
should be used with caution because it has been tested on a relatively
small, biased (metal-rich, high-excitation) sample of Galactic and
extragalactic \hii{} regions.  In addition, recent theoretical work
suggests that abundances derived from recombination lines may not be
as unbiased as once believed \citep{stasinska07b, ercolano07a,
  ercolano09a}.  Additional observations of the Balmer discontinuities
and \ion{O}{2} recombination-line intensities of \hii{} regions
spanning a wider range of metallicity and physical conditions would be
of considerable value.

\begin{figure}
\epsscale{1.2}
\plotone{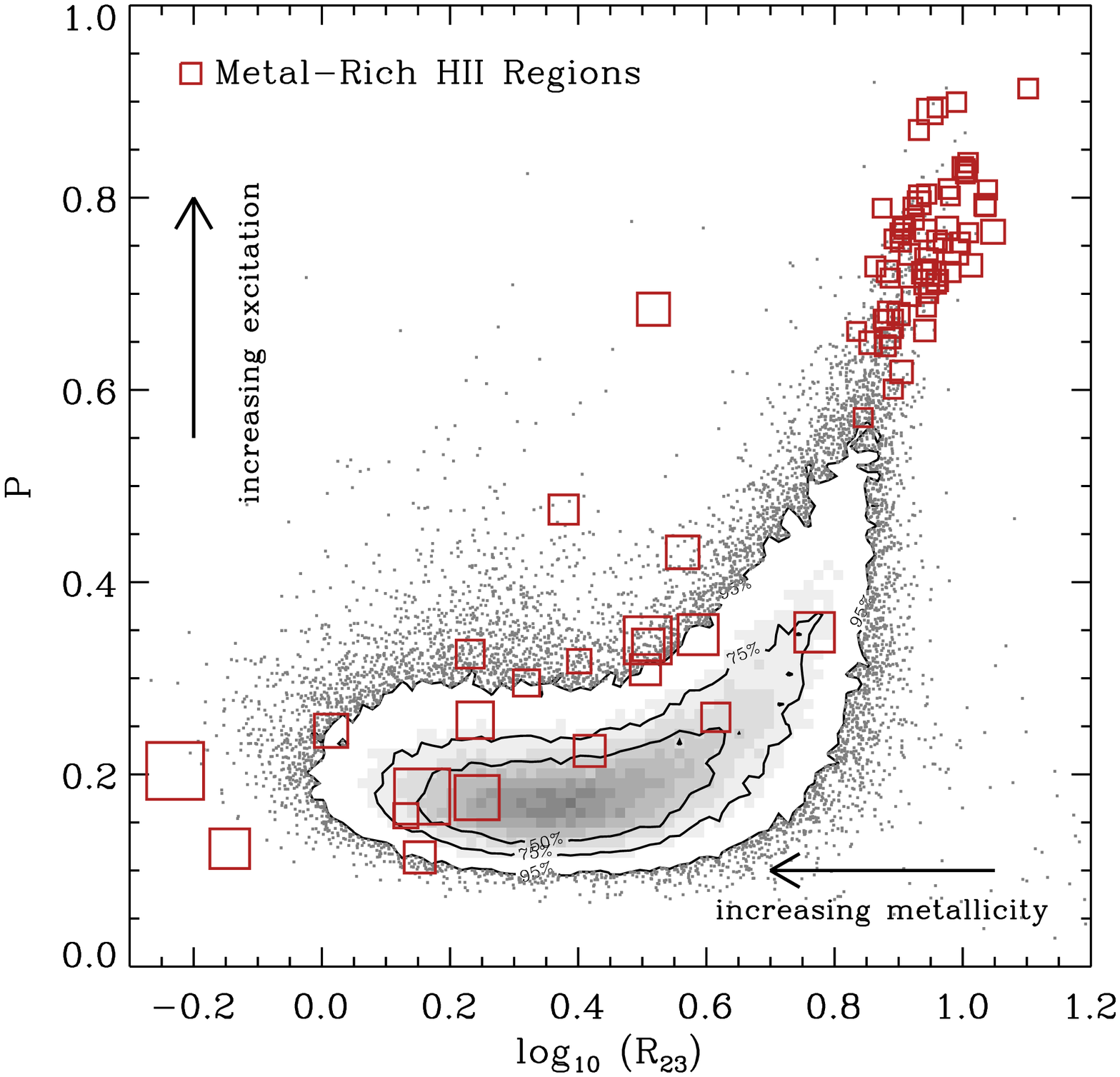}
\caption{Metallicity-sensitive \pagel{} parameter vs.~excitation, $P$,
  for star-forming galaxies in the SDSS ({\em contoured greyscale})
  and a representative sample of metal-rich \hii{} regions with
  well-measured electron temperatures collected from the literature
  \citep[{\em red squares};][]{garnett97a, vanzee98a, kenn03a,
    kniazev04a, bresolin04a, bresolin05a, izotov04a, izotov06a,
    rosolowsky08a}.  The symbol size for each \hii{} region is
  proportional to its metallicity, ranging from $\logoh_{T_{e}}=8.21$
  \citep[H143 in M101;][]{kenn03a} to $8.94$~dex \citep[H11 in
    M83;][]{bresolin05a}.  This scatterplot demonstrates that the bulk
  of the galaxies in the SDSS are low-excitation, whereas most
  metal-rich \hii{} regions with well-measured electron temperatures
  are high-excitation, which is a well-known observational bias (see
  also Fig.~12 in \citealt{pilyugin05a}).  The practical implication
  is that empirical abundance calibrations such as PT05 in general
  should \emph{not} be applied to large emission-line galaxy surveys
  such as the SDSS because the calibrations would be extrapolated to a
  part of physical parameter space that is not well-constrained by
  current observations.  \label{fig:sdss_r23_p}}
\end{figure}

As alluded to above, one of the principal limitations of the empirical
strong-line abundance methods is that they are only strictly
applicable to \hii{} regions and star-forming galaxies spanning the
same range of excitation and metallicity as the \hii{} regions that
were used to build the calibration.  To illustrate this point, in
Figure~\ref{fig:sdss_r23_p} we plot the (reddening-corrected) \pagel{}
parameter versus excitation, $P$, for star-forming galaxies in the
SDSS and a \emph{representative} sample of metal-rich \hii{} regions
with well-measured electron temperatures \citep[{\em red
    squares};][]{garnett97a, vanzee98a, kenn03a, kniazev04a,
  bresolin04a, bresolin05a, izotov04a, izotov06a, rosolowsky08a}.  The
symbol size for each \hii{} region is proportional to its oxygen
abundance in the range $\logoh_{T_{e}}\approx8.2-8.95$ assuming
$t^{2}=0$.  Because all these objects are on the upper \pagel{}
branch, $P$ increases in tandem with nebular excitation, while a
decrease in \pagel{} corresponds to an increase in metallicity, as
indicated by the arrows in Figure~\ref{fig:sdss_r23_p}.  This figure
demonstrates that star-forming galaxies in the SDSS are
lower-excitation and more metal-rich than the bulk of the \hii{}
regions with direct abundance estimates.\footnote{As indicated in the
  previous footnote, at least some of the tendency for the SDSS
  galaxies to appear metal-rich and low-excitation may be due to
  aperture bias.}  Note that a similar conclusion would be reached if
\ewhb{} had been used as a measure of excitation \citep[see, e.g.,
  Fig.~5 in][]{bresolin07a}: the median \ewhb{} value of the SDSS
galaxies is $\sim6$~\AA, nearly a factor of $15$ smaller than the
median \ewhb{} of the \hii{} regions, $\sim85$~\AA.  The practical
implication of this result is that empirical abundance calibrations
such as PT05 should \emph{not} in general be applied to large
emission-line galaxy surveys such as the SDSS because the calibrations
would be extrapolated to a part of physical parameter space that is
not well-constrained by current observations.

Meanwhile, theoretical strong-line abundance calibrations have the
advantage that they are based on models that, by construction, span a
wide range of ionization parameter ($-3.5<\logu<-1.9$), metallicity
($0.0001<Z<0.05$), and stellar effective temperature
\citep{mcgaugh91a, dopita00a, charlot01a, kewley01b, moy01a,
  dopita06b, martin-manjon10a, levesque10a}.  In these methods the UV
(i.e., photoionizing) spectral energy distribution of a young star
cluster is coupled to a photoionization code such as {\sc cloudy}
\citep{ferland98a} or {\sc mappings} \citep{sutherland93a} and the
output emission-line spectrum is studied as a function of input
parameters.  In recent years there have been significant improvements
in the evolutionary tracks for young stars, including the effects of
mass loss and rotation \citep{vazquez05a}, the calculation of non-LTE,
line-blanketed stellar atmospheres \citep{simon-diaz08a}, and the
inclusion of dust grains in the photoionization models
\citep{shields95a, groves04a}.  Despite these improvements, however,
some problems persist.  For example \citet{levesque10a} find that the
latest generation of photoionizing spectra are too soft to reproduce
the observed \sii/\ha{} ratios of nearby star-forming galaxies.
Photoionization codes also have difficulty reproducing the observed
line-strengths of the auroral lines \citep{stasinska99a, jamet05a},
and predict insignificant temperature fluctuations within \hii{}
regions \citep{baldwin91a}, in contradiction with the (albeit limited)
observations.  The theoretical methods also rely on a number of
simplifying assumptions regarding the electron density, volume filling
factor, and nebular geometry (e.g., spherically symmetric,
plane-parallel); the metal-abundance pattern, including the proportion
of each metal depleted onto dust grains; the initial mass function,
star-formation history (e.g., instantaneous, continuous, stochastic),
and age of the central cluster; and chemical evolution effects (e.g.,
the relationship between primary and secondary nitrogen), among
others.  In addition to the discrepancies alluded to above, a
breakdown in one or more of these assumptions may be responsible for
the seemingly ``high'' oxygen abundances predicted by theoretical
strong-line calibrations.

So what are the ways forward?  From the observational standpoint,
high-quality optical spectrophotometry of larger samples of \hii{}
regions in nearby galaxies is clearly warranted.  These observations
should span a wide wavelength range in order to include a large number
of independent electron temperature and density diagnostics, and
should target \hii{} regions spanning a wide range of metallicity and
excitation \citep[e.g.,][]{kenn03a, bresolin05a, bresolin07a,
  rosolowsky08a, esteban09a}.  In particular, a concerted effort to
measure the Balmer discontinuities and metal recombination lines in
metal-rich \hii{} regions will elucidate the prevalence and relative
importance of temperature-fluctuations, and will help constrain
empirical strong-line calibrations in the metal-rich regime.  Although
these observations are extraordinarily challenging, they are possible
with the latest generation of multi-object spectrographs on $8-$ and
$10-$meter class telescopes (e.g., Keck/{\sc deimos}, LBT/{\sc mods},
VLT/{\sc xshooter}, etc.).  These observations also will help
establish the radial and azimuthal dependence of abundances in nearby
galaxies; many existing observations are more than twenty-five years
old, and for many systems data are available for only a handful of
\hii{} regions \citep[Fig.~\ref{fig:gradients};][]{pilyugin04a}.
Finally, observations of the mid- and far-infrared metal cooling lines
(e.g., \siii~$\lambda19$~\micron, \oiii~$\lambda52$~\micron,
\niii~$\lambda57$~\micron, \oiii~$\lambda88$~\micron,
\nii~$\lambda122$~\micron), which are insensitive to variations in
electron temperature, will be crucial in establishing the abundances
of metal-rich \hii{} regions \citep{martin-hernandez02a, garnett04c,
  rudolph06a, rubin08a}.  Spectroscopy from $60-210$~\micron{} with
the PACS instrument on-board the {\em Herschel Space Telescope} (e.g.,
from the {\sc kingfish} project; \citealt{poglitsch08a}) will expand
significantly the number of star-forming regions observed in the
critical far-infrared wavelength regime.

Improvements in the grid-based theoretical strong-line abundance
calibrations are also warranted.  In particular, a systematic
exploration of the input parameters and simplifying assumptions in the
models \citep[e.g.,][]{moy01a, mathis05a, ercolano07a, ercolano10a,
  levesque10a} may elucidate the origin of the discrepancy between
theoretical and empirical abundances.  For example, \citet{yin07a}
show that the difference in the oxygen abundances derived using the
direct method and the \citet{charlot01a} photoionization models
\citep{brinchmann04a, tremonti04a} correlates strongly with the N/O
abundance ratio.  Because nitrogen is both a primary and secondary
nucleosynthetic product, the N/O ratio varies systematically with
oxygen abundance, as well as the time since the most recent episode of
star formation \citep{edmunds78a, contini02a, mouhcine02a,
  pilyugin03a, izotov06a}, a complexity that is not reflected in the
grid-based theoretical calibrations.  Finally, we note that
\emph{tailored} photoionization models of individual \hii{} regions
frequently \emph{can} be tuned to reproduce most or all the observed
emission-line ratios (e.g., \citealt{castellanos02a, garnett04c}, but
see \citealt{jamet05a}).  This result again suggests the possibility
of an as-yet unidentified shortcoming in the \emph{grid-based}
theoretical calibrations.

Summarizing the preceding discussion, we find that empirical abundance
calibrations may underestimate the oxygen abundances of \hii{} regions
and star-forming galaxies because of systematic biases (e.g.,
temperature fluctuations) in the $T_{e}$-based abundances that are
used to calibrate the empirical methods.  Empirical methods also
suffer from a shortage of metal-rich \hii{} regions with measured
electron temperatures, which means that the calibrations are being
extrapolated to an unconstrained region of parameter space when
applied to large galaxy samples such as the SDSS.  On the other hand,
indirect evidence suggests that theoretical calibrations yield
abundances that are too high, possibly due to one or more simplifying
assumptions in the photoionization model grids.  Therefore, until
these issues are resolved, we recommend that any study utilizing the
oxygen abundances provided in Table~\ref{table:final} be carried out
either using the \emph{relative} (KK04 or PT05) abundances of the
sample, or using \emph{both} the KK04 and PT05 abundances separately,
to ensure that the conclusions are not affected by the error in the
absolute zeropoint of the nebular abundance scale.  Alternatively, if
a single set of oxygen abundances for the SINGS galaxies is required,
a compromise procedure would be to \emph{average} the KK04 and PT05
metallicities and conservatively adopt the \emph{difference} in the
two metallicity estimates as the metallicity error for each galaxy.

Although we have focused exclusively on the \pagel{} parameter to
compute oxygen abundances, over the past three decades many other
strong-line calibrations have been developed for estimating the
abundances of \hii{} regions and star-forming galaxies.  Among the
most popular diagnostics are those based on the \niilam{} emission
line \citep[e.g., \nii/\ha, \nii/\oii, and \oiii/\nii;][]{alloin79a,
  storchi94a, vanzee98a, kewley02a, pettini04a, liang06b}, although
calibrations based on an assortment of other forbidden lines have been
proposed \citep{dopita86a, vilchez96a, diaz00b, perez-montero05a,
  nagao06a, stasinska06b, wu08a, okada08a}, as well as techniques
designed to model all the emission lines simultaneously
\citep{charlot01a, brinchmann04a, maier06a}.  Most \nii-based
calibrations have the advantage that, unlike \pagel, they vary
monotonically with metallicity.  On the other hand, for most galaxies
\nii{} is usually weaker than \oii, \oiii, and \hb, and it is more
sensitive to contributions from shock-ionized gas and a non-thermal
(i.e., AGN) photoionizing continuum.\footnote{A notable exception is
  the abundance-sensitive \nii/\oii{} ratio, which is insensitive to
  variations in excitation because \nii{} and \oii{} have similar
  ionization potentials \citep{bresolin07a, kewley08a}; on the other
  hand, \nii/\oii{} is sensitive to dust reddening.}  As discussed
above, oxygen abundances inferred from \nii{} also depend implicitly
on the recent star formation history of the galaxy, which may
introduce non-negligible systematic errors \citep[e.g.,][]{yin07a}.
Nevertheless, because \nii{} is observable from the ground to
$z\sim2$, strong-line calibrations that rely on this line provide an
important window into the gas-phase abundances of high-redshift
galaxies \citep{shapley04a, shapley05a, maier05a, erb06a, erb10a,
  hainline09a}.

One final cautionary note is that both empirical and theoretical
strong-line methods have been calibrated against observations of
\hii{} regions and star-forming galaxies in the nearby universe, and
may not be applicable at higher redshift where the physical conditions
in galaxies may differ dramatically from physical conditions in local
galaxies \citep{shapley05a, liu08a, brinchmann08a, hainline09a,
  lehnert09a}.  Testing the validity of locally calibrated abundance
diagnostics in higher-redshift galaxies, therefore, will be an
important objective of the first generation of highly multiplexed
near-infrared spectrographs and, eventually, with the NIRSpec
instrument onboard the {\em James Webb Space Telescope}, through
observations of the full complement of rest-frame optical
emission-line diagnostics at high redshift.

\section{Summary}\label{sec:summary}

We have obtained intermediate-resolution ($\sim8$~\AA{} FWHM), high
signal-to-noise ratio (${\rm S/N}=5-100$~pixel$^{-1}$) optical
($\sim3600-6900$~\AA) spectrophotometry of the SINGS galaxies on three
spatial scales to complement and enhance the legacy value of existing
mid- and far-infrared observations obtained as part of the
\emph{Spitzer}/SINGS legacy program.  Our integrated circumnuclear and
radial-strip spectra characterize the central and globally-averaged
optical properties of each galaxy, respectively, and are spatially
coincident with mid- ($\sim5-38$~\micron) and far-infrared
($\sim55-95$~\micron) spectral cubes from the IRS spectrograph and the
MIPS instrument in SED-mode, respectively.  A third optical spectrum
targeting the nucleus of each galaxy provides a reliable means of
identifying galaxies hosting an AGN, among other applications.  We
make the fully reduced, spectrophotometrically calibrated
one-dimensional spectra available to the community.  In addition, we
use state-of-the art stellar population synthesis models to separate
the underlying stellar continuum from the optical emission-line
spectrum, and generate tables of the fluxes and equivalent widths of
the strongest optical emission lines.  Together with existing
ancillary multi-wavelength observations of the sample, these data
should facilitate a broad range of studies on the physical properties
of nearby galaxies.

As a first effort demonstrating the utility of these data, we classify
the sample into star-forming (SF), AGN, and SF/AGN on three spatial
scales, and carry out a detailed analysis of the gas-phase oxygen
abundances of the SINGS sample using our optical spectra and
observations of more than $550$ \hii{} regions.  Our principal results
are:

\begin{enumerate}
\item{We find that the proportion of galaxies classified as SF and AGN
  is a strong function of the fraction of the integrated $B$-band
  light enclosed by the spectroscopic aperture.  The fraction of the
  sample classified as AGN decreases from $\approx40\%$ to
  $\lesssim15\%$ as the $B$-band light fraction increases from
  $\sim10\%$ to $\gtrsim60\%$, with a corresponding increase in the
  fraction of galaxies classified as SF.  The $\sim15\%$ incidence
  rate of objects classified as SF/AGN, on the other hand, is roughly
  independent of enclosed light fraction.}
\item{Because of significant systematic differences among oxygen
  abundances computed using different methods, we compute oxygen
  abundances using both a theoretical (KK04) and an empirical (PT05)
  strong-line abundance calibration, both based on the popular
  \pagel{} parameter.  We measure a systematic offset of $\sim0.6$~dex
  between the two nebular abundances, in the sense that the
  theoretical KK04 calibration yields significantly higher
  metallicities.  These two calibrations were chosen because they
  bracket the full range of metallicities one would obtain using other
  popular strong-line abundance calibrations, and because they could
  be applied to such a diverse sample such as SINGS.}
\item{We carry out a detailed analysis of the radial abundance
  gradients in $21$ of the SINGS galaxies using both calibrations.  We
  find that the slope of the abundance gradients inferred from these
  two calibrations are generally correlated; however, to second order
  the theoretical KK04 calibration results in systematically steeper
  abundance gradients, which has significant implications for
  constraining disk-galaxy formation models.  In addition, the
  dispersion in metallicity at fixed galactocentric radius is
  $\sim50\%$ larger when using the empirical PT05 calibration,
  presumably because the PT05 calibration is not well-constrained for
  low-excitation \hii{} regions.}
\item{We combine all the available abundance measurements, including
  abundances inferred from our nuclear, circumnuclear, and
  radial-strip spectra, \hii{} regions, and the $B$-band
  luminosity-metallicity relation, to compute the mean characteristic
  and central oxygen abundances of whole SINGS sample using both the
  KK04 and PT05 strong-line calibrations.  These abundances should
  facilitate a wide range of studies between the metal content and
  infrared emission properties of nearby galaxies.}
\item{Finally, we discuss some of the possible origins of the
  systematic differences in nebular abundance computed using direct or
  empirical versus theoretical strong-line abundance calibrations.  We
  conclude that additional observations of \hii{} regions spanning a
  wide range of physical conditions, as well as improvements in the
  theoretical models, are imperative for resolving this important
  outstanding problem.}
\end{enumerate}

Preliminary values of the oxygen abundances presented in this paper
already have been used in several publications by the SINGS team.
\citet{smith07a} and \citet{draine07a} found a remarkable decrease in
the PAH emission-line strengths and the proportion of dust mass locked
up in PAH grains below a characteristic oxygen abundance $\logoh_{\rm
  PT05}\approx8.1$.  \citet{calzetti07a, calzetti10a} investigated the
metallicity dependence of the $8$, $24$, $70$, and $160$~\micron{}
monochromatic luminosities as quantitative SFR diagnostics; they found
a strong metallicity dependence for $L(8~\micron)$, hampering its
suitability for high-redshift studies.  \citet{bendo08a} compared the
radial variations in the $8~\micron/24~\micron$ and
$8~\micron/160~\micron$ ratios against the radial abundance gradients
in a subset of the face-on disk galaxies in SINGS and concluded that
metallicity was \emph{not} the dominant parameter driving the observed
trends.  And finally \citet{munoz09a} found that both the
total-infrared to far-UV luminosity ratio (i.e., $L_{\rm TIR}/L_{\rm
  FUV}$) and the UV spectral slope (i.e., $\beta$) depend
significantly on gas-phase oxygen abundance, owing to the tendency for
metal-rich, star-forming galaxies to be dustier.  

These applications illustrate the broad utility of the measured
optical emission-line strengths, spectral classifications, and derived
oxygen abundances of the SINGS galaxies, and should facilitate a
diversity of future studies into the dust emission and star formation
properties of nearby galaxies.

\acknowledgements

We gratefully acknowledge Moire Prescott for assistance with some of
the observations, Stuart Ryder and Henry Lee for providing their
published \hii-region data in electronic format, Juan-Carlos
Mu{\~n}oz-Mateos for assistance assembling the optical photometry of
the sample, the Bok $2.3$-meter telescope operators, Dennis Means and
Vic Hansen, for their expertise during the course of this survey, and
illuminating conversations with James Aird, Michael Blanton, Alison
Coil, David Hogg, Janice Lee, Leonidas Moustakas, and Amelia Stutz.
The authors would also like to thank Bruce Draine for insightful
comments on an early draft of the manuscript, and the referee, Fabio
Bresolin, for suggestions that improved the clarity of the manuscript.
This work has relied extensively on the invaluable {\sc idlutils}
software library developed by M.~R. Blanton, S.~Burles,
D.~P. Finkbeiner, D.~W. Hogg, and D.~J. Schlegel, and on the Goddard
{\sc idl} library maintained by W.~Landsman.  We also thank
M.~Cappellari and M.~Sarzi for generously making their continuum- and
emission-line fitting software, {\sc pPXF} and {\sc gandalf},
respectively, publically available.

This research has made extensive use of NASA's Astrophysics Data
System Bibliographic Services, the VizieR catalog access tool, and the
NASA/IPAC Extragalactic Database, which is operated by the Jet
Propulsion Laboratory, California Institute of Technology, under
contract with the National Aeronautics and Space Administration.  The
Digitized Sky Surveys were produced at the Space Telescope Science
Institute under U.S. Government grant NAG W-2166. The images of these
surveys are based on photographic data obtained using the Oschin
Schmidt Telescope on Palomar Mountain and the UK Schmidt
Telescope. The plates were processed into the present compressed
digital form with the permission of these institutions.

Funding for the Sloan Digital Sky Survey (SDSS) and SDSS-II has been
provided by the Alfred P. Sloan Foundation, the Participating
Institutions, the National Science Foundation, the U.S. Department of
Energy, the National Aeronautics and Space Administration, the
Japanese Monbukagakusho, and the Max Planck Society, and the Higher
Education Funding Council for England. The SDSS Web site is
http://www.sdss.org/.

The SDSS is managed by the Astrophysical Research Consortium (ARC) for
the Participating Institutions. The Participating Institutions are the
American Museum of Natural History, Astrophysical Institute Potsdam,
University of Basel, University of Cambridge, Case Western Reserve
University, The University of Chicago, Drexel University, Fermilab,
the Institute for Advanced Study, the Japan Participation Group, The
Johns Hopkins University, the Joint Institute for Nuclear
Astrophysics, the Kavli Institute for Particle Astrophysics and
Cosmology, the Korean Scientist Group, the Chinese Academy of Sciences
(LAMOST), Los Alamos National Laboratory, the Max-Planck-Institute for
Astronomy (MPIA), the Max-Planck-Institute for Astrophysics (MPA), New
Mexico State University, Ohio State University, University of
Pittsburgh, University of Portsmouth, Princeton University, the United
States Naval Observatory, and the University of Washington.

{\it Facilities:} 
\facility{Bok (Boller \& Chivens spectrograph)}; \facility{CTIO:1.5m
  (R-C spectrograph)} 

\begin{appendix}

\section{Presentation of the Spectra}\label{sec:thedata}

The SINGS galaxies were selected to span a broad range of
morphological types, optical luminosities, and infrared properties,
and this diversity is clearly reflected in their optical spectra.
Figures~\ref{fig:spectra_first}-\ref{fig:spectra_last} present the
reduced one-dimensional spectra\footnote{Available electronically at
  http://sings.stsci.edu and http://ssc.spitzer.caltech.edu/legacy.}
in \flunits{} versus rest-frame wavelength in \AA.  We plot the data
in grey and overplot the best-fitting stellar continuum model (see
\S\ref{sec:ispec}) in black.  For each galaxy, we also show an optical
Digitized Sky Survey image to illustrate how our spectroscopic
apertures compare to the optical extent of the galaxy ({\em dashed
  purple ellipse}; see Table~\ref{table:properties}).  The size of the
spectroscopic apertures corresponding to our radial-strip and
circumnuclear spectra are shown as a red rectangle and blue square,
respectively.  The dotted yellow rectangle shows the spatial coverage
of the corresponding \emph{Spitzer}/IRS long-low spectrum obtained as
part of SINGS where there was at least double coverage in both
spectral orders \citep{smith04a, smith07a}.  In some cases there is a
mismatch between the orientation angle of the IRS and optical
radial-strip spectra.  This difference occurs because many of the
optical spectra were obtained before the IRS spectra had been
observed, or even scheduled.  For these objects when we were planning
the optical observations we made the best effort to select a position
angle for the optical spectrum based on the predicted/constrained
position angle of the IRS observations.

\section{\ion{H}{2}-Region Database}\label{sec:hiiappendix}  

We performed a search of the literature to find published observations
of \hii{} regions in the SINGS sample, resulting in data on $561$
star-forming regions in $\sim50\%$ of the sample ($38/75$ galaxies)
drawn from $36$ distinct papers published between 1984 and 2009.  In
order to be included in our catalog, the \oiilam, \hb, and \oiiilam{}
line-strengths must have been measured to enable an estimate of the
gas-phase abundance (see \S\ref{sec:hiioh}).  While this catalog is
not exhaustive, it does include all the major spectroscopic surveys of
\hii{} regions in nearby galaxies conducted to date
\citep[e.g.,][]{mccall85a, zaritsky94a, vanzee98a}.  In general, we
gave preference to the most recent observations over older data
obtained using non-linear detectors, which may be biased
\citep{torres89a}.  We also only tabulated data based on traditional
spectrophotometry, not narrowband imaging \citep[e.g.,][]{belley92a,
  dutil99a}, which may suffer from a variety of systematic errors
\citep{dutil01a}.  Finally, observations of the same \hii{} region by
different authors were retained as independent observations.

For each \hii{} region we tabulated the fluxes and uncertainties of
the strong nebular emission lines, as well as the coordinates of each
star-forming region relative to the galactic nucleus.  In most cases
the published emission-line strengths had been corrected for both
underlying stellar absorption and dust reddening.  Where necessary,
however, we applied a statistical $\ewhb=1.9$~\AA{} correction
\citep{mccall85a} to the Balmer lines and dereddened the
line-strengths using the observed \ha/\hb{} Balmer decrement as
described in \S\ref{sec:intnucoh}.  In all cases we assumed that
$\oiii~\lambda4959$ was $0.34$ times the intensity of
$\oiii~\lambda5007$ \citep{osterbrock06a}.  Finally, the coordinates
were used to derive the deprojected galactocentric radius of each
\hii-region using the position and inclination angles listed in
Table~\ref{table:properties}.

Table~\ref{table:hiidata} provides the derived properties of the
\hii{} regions used, including the \hii{} region name and reference to
the literature, the deprojected galactocentric radius, the \pagel,
\ioniz, and $P$ parameters (see \S\ref{sec:calib}), and the oxygen
abundances derived using the KK04 and PT05 strong-line calibrations,
assuming the \pagel{} branches listed in Table~\ref{table:intnucoh}.
\hii{ }regions whose oxygen abundance could not be derived based on
the quantitative criteria described in \S\ref{sec:intnucoh}, or which
had ambiguous \pagel{} branches using either abundance calibration,
are identified in the table using footnotes.  More detailed
information on each \hii{} region (e.g., relative coordinates,
emission-line ratios, etc.) are available from the first author as a
binary {\sc fits} table upon request.

\end{appendix}



\setcounter{figure}{0}
\renewcommand\thefigure{A\arabic{figure}}

\begin{figure*}
\begin{center}
\includegraphics[scale=0.32,angle=0]{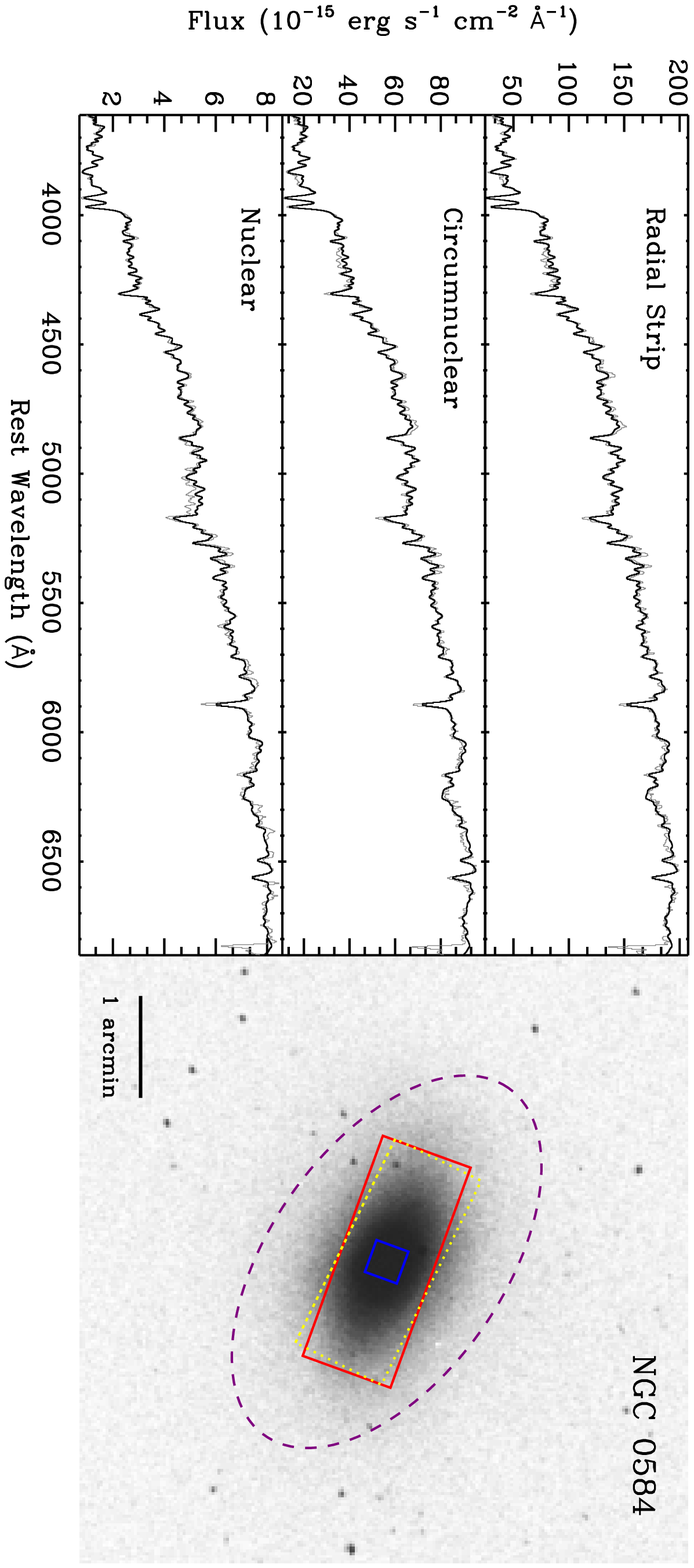}
\includegraphics[scale=0.32,angle=0]{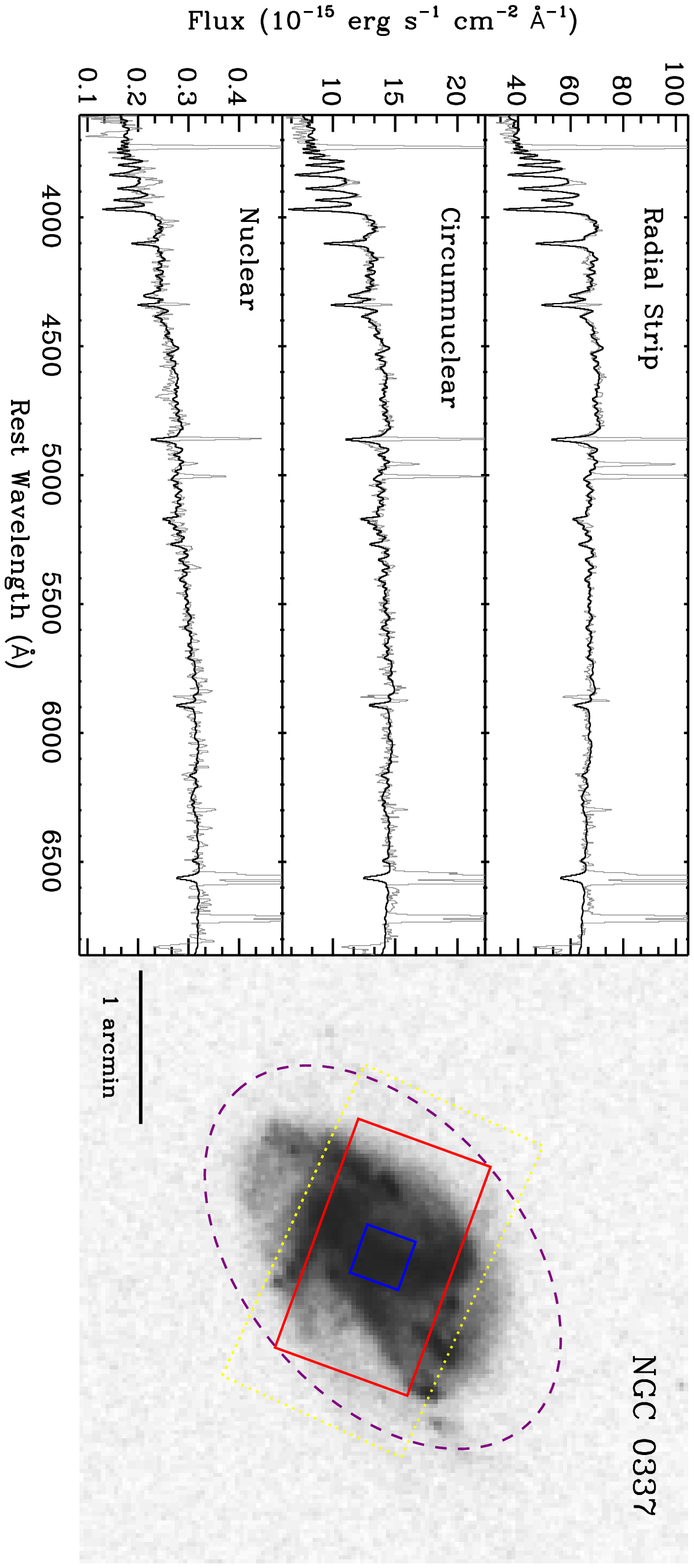}
\includegraphics[scale=0.32,angle=0]{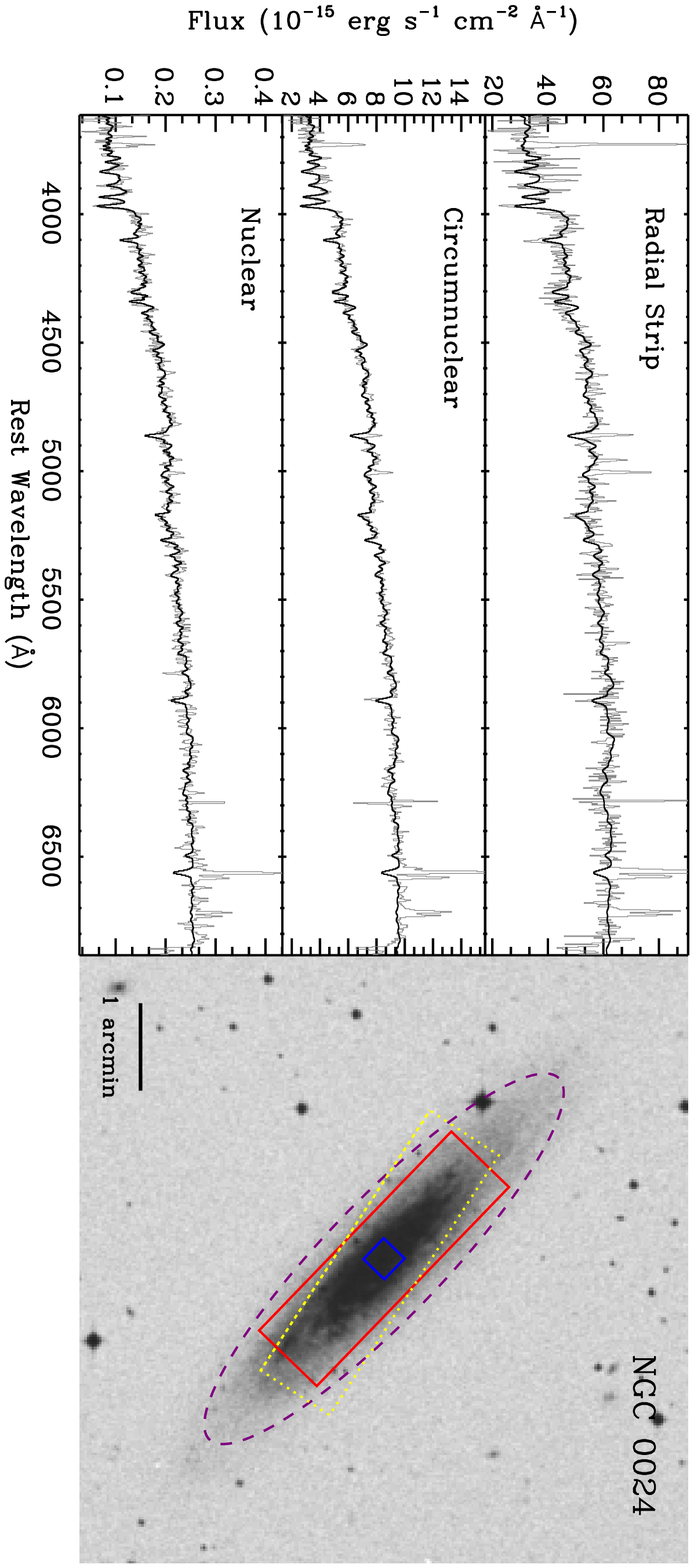} \\
\includegraphics[scale=0.32,angle=0]{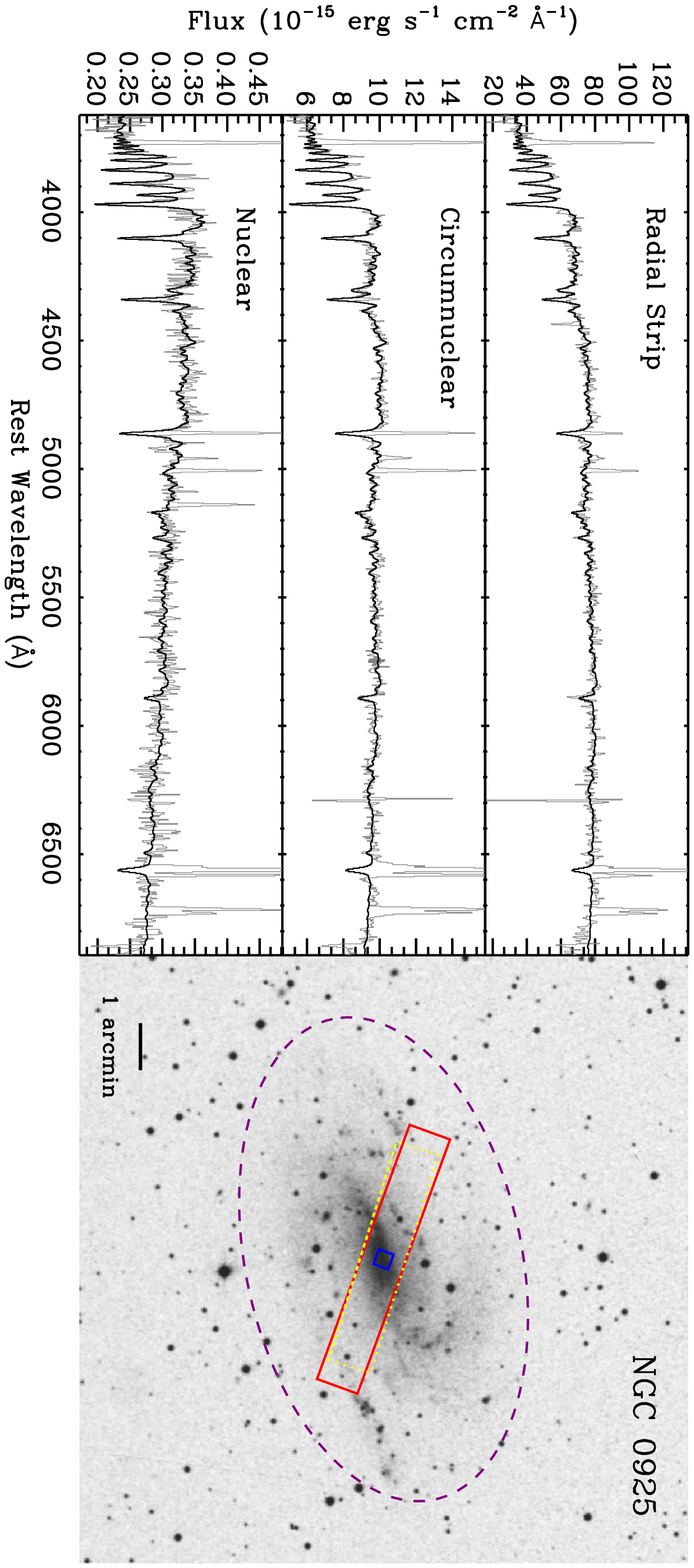}
\includegraphics[scale=0.32,angle=0]{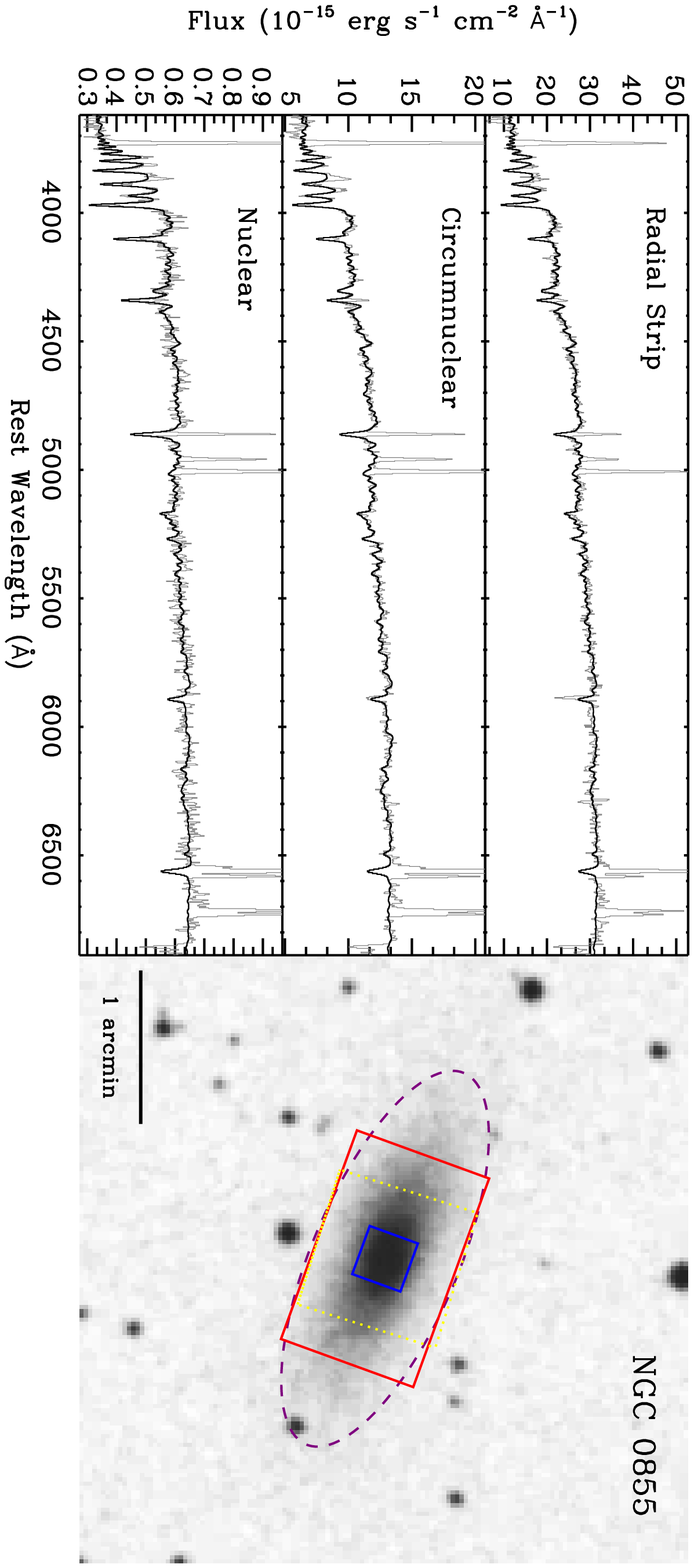}
\includegraphics[scale=0.32,angle=0]{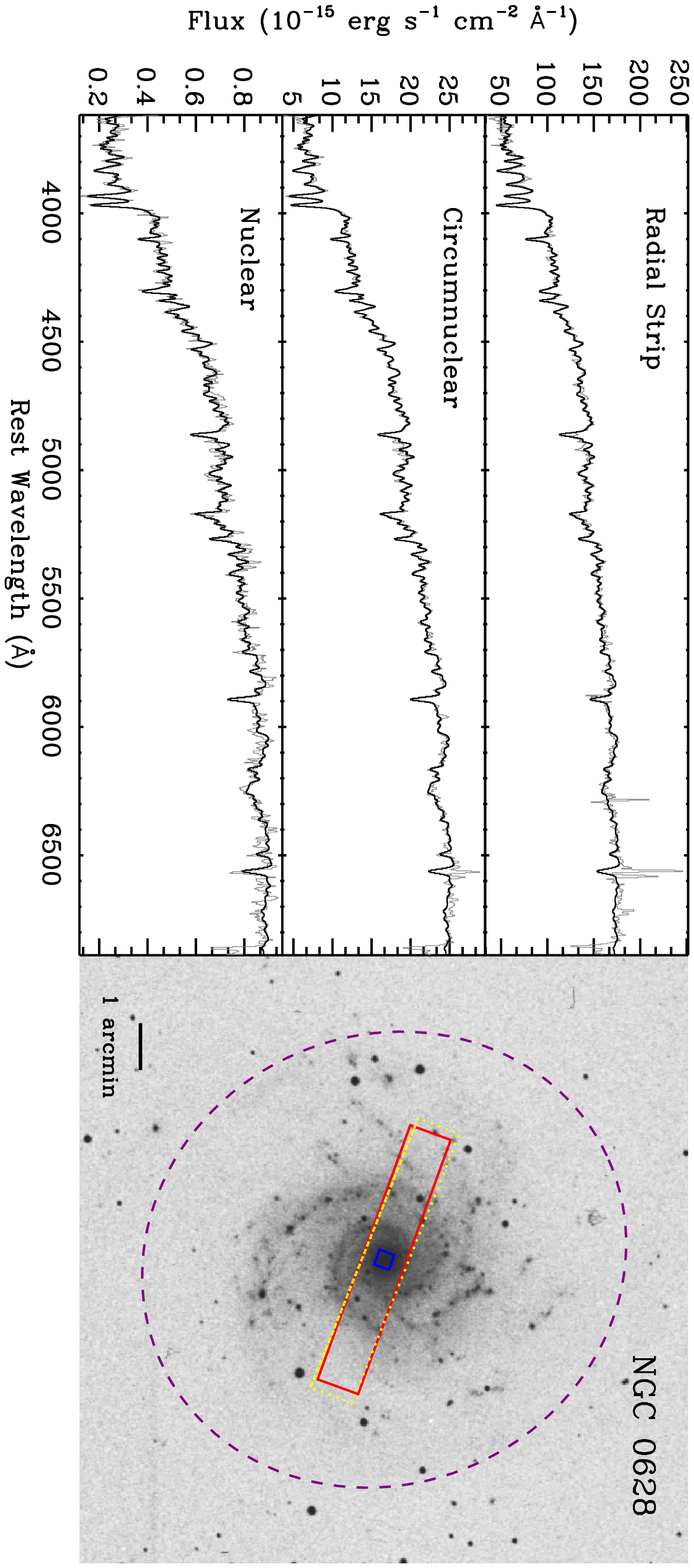}
\caption{The left panels plot the nuclear, circumnuclear, and
  radial-strip spectra optical spectra of the first six SINGS
  galaxies; the observed spectra are shown in grey and the
  best-fitting stellar continuum fit is overplotted in black (see
  \S\ref{sec:ispec}).  The right panels show the size of the
  circumnuclear ({\em blue square}) and radial-strip ({\em red
    rectangle}) spectroscopic apertures, overlaid on an optical DSS
  image of each galaxy.  The dashed purple ellipse indicates the
  optical extent of the galaxy (see Table~\ref{table:properties}), and
  the dotted yellow rectangle shows the spatial coverage of the
  \emph{Spitzer}/IRS radial-strip long-low spectrum obtained as part
  of SINGS \citep{smith04a, smith07a}. \label{fig:spectra_first}}
\end{center}
\end{figure*}

\begin{figure}[!h]
\begin{center}
\includegraphics[scale=0.4,angle=0]{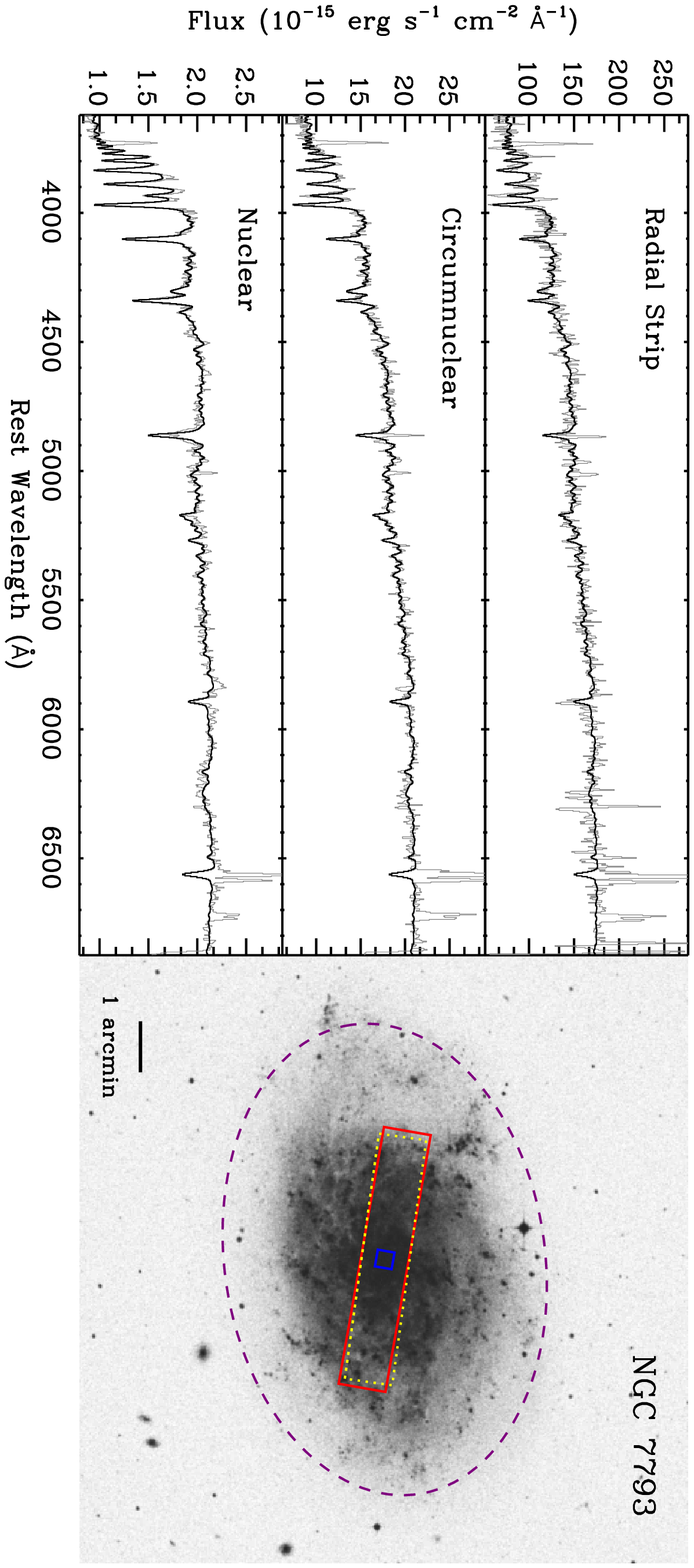}
\includegraphics[scale=0.4,angle=0]{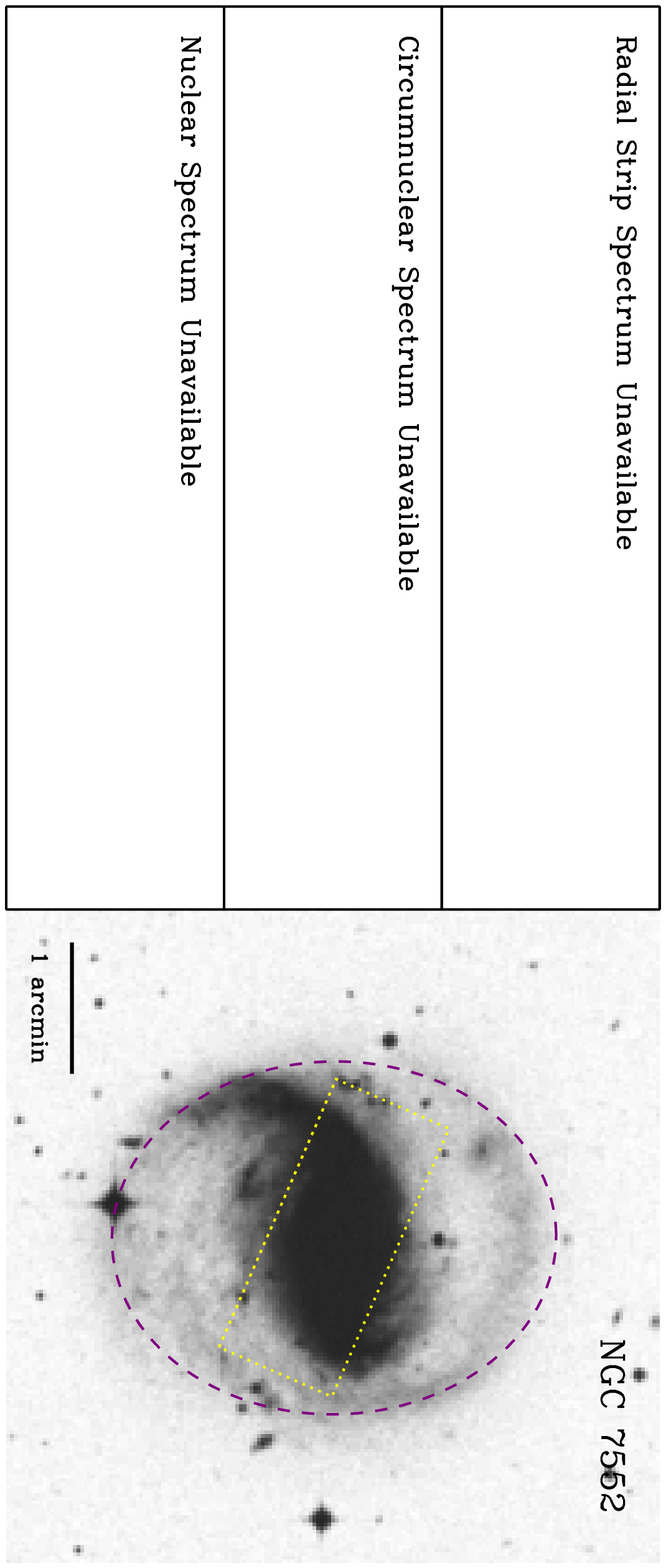}
\includegraphics[scale=0.4,angle=0]{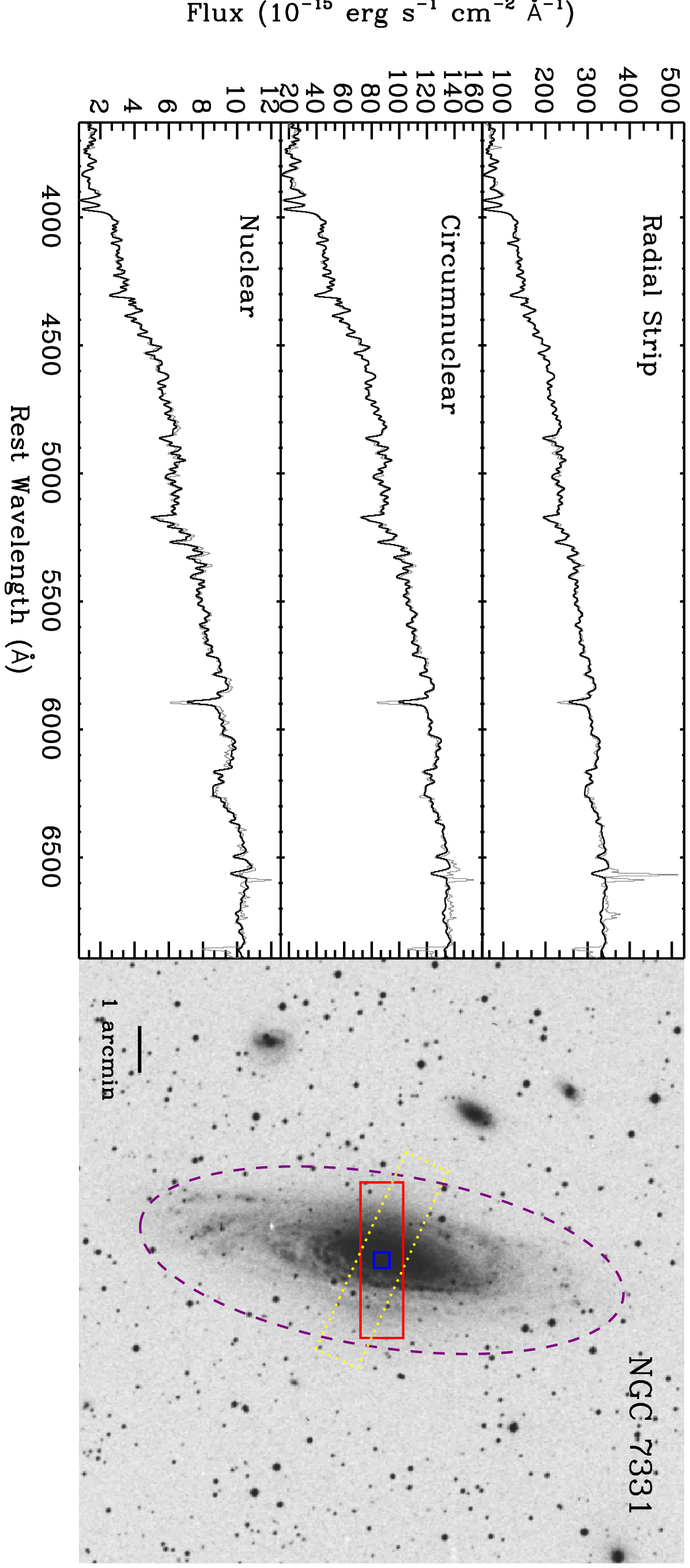} \\
 \caption{Same of Fig.~\ref{fig:spectra_first}, but for the final
   three SINGS galaxies. Note that the visualizations for the
   intervening galaxies can be found in the online edition of the
   published ApJS paper. \label{fig:spectra_last}}
\end{center}
\end{figure}
 

\clearpage
\input{tab1}

\clearpage
\input{tab2}

\clearpage
\input{tab3}

\clearpage
\input{tab4}

\clearpage
\input{tab5}

\clearpage
\input{tab6}

\clearpage
\input{tab7}

\clearpage
\input{tab8}

\clearpage
\input{tab9}

\clearpage
\input{tab1_appendix}

\end{document}

%% file: tab1.tex
\LongTables
\begin{deluxetable}{llccccccccc}
\tabletypesize{\small}
\tablecaption{Properties of the SINGS Galaxies\label{table:properties}}
\tablewidth{0pt}
\tablehead{
\colhead{} & 
\colhead{} & 
\colhead{$E(B-V)$} & 
\colhead{$\rho_{25}$} & 
\colhead{$i$} & 
\colhead{$\theta$} & 
\colhead{$M_{B}$} & 
\colhead{$B-V$} & 
\colhead{$D$} & 
\colhead{Distance} & 
\colhead{Distance} \\ 
\colhead{Galaxy} & 
\colhead{Type} & 
\colhead{(mag)} & 
\colhead{(arcmin)} & 
\colhead{(deg)} & 
\colhead{(deg)} & 
\colhead{(mag)} & 
\colhead{(mag)} & 
\colhead{(Mpc)} & 
\colhead{Method} & 
\colhead{Ref.} \\
\colhead{(1)} & 
\colhead{(2)} & 
\colhead{(3)} & 
\colhead{(4)} & 
\colhead{(5)} & 
\colhead{(6)} & 
\colhead{(7)} & 
\colhead{(8)} & 
\colhead{(9)} & 
\colhead{(10)} & 
\colhead{(11)} 
}
\startdata
NGC 0024 &  SAc &  0.020 &  2.88 &  82 &  46 &  $ -17.31\pm 0.69$ & 
$   0.68\pm   0.05$ &  $         7.3\pm         2.3$ &  Flow &  1 \\ 
NGC 0337 &  SBd &  0.112 &  1.44 &  52 &  130 &  $ -20.09\pm 0.24$ & 
$   0.38\pm   0.05$ &  $        22.4\pm         2.3$ &  Flow &  1 \\ 
NGC 0584 &  E4 &  0.042 &  2.08 &  58 &  55 &  $ -19.92\pm 0.22$ & 
$   1.13\pm   0.05$ &  $        20.1\pm         1.9$ &  SBF &  2 \\ 
NGC 0628 &  SAc &  0.070 &  5.24 &  24 &  25 &  $ -19.60\pm 0.43$ & 
$   0.50\pm   0.05$ &  $         7.3\pm         1.4$ &  BS &  3 \\ 
NGC 0855 &  E &  0.071 &  1.32 &  71 &  67 &  $ -17.01\pm 0.20$ & 
$   0.58\pm   0.05$ &  $        9.73\pm        0.76$ &  SBF &  2 \\ 
NGC 0925 &  SABd &  0.076 &  5.24 &  57 &  102 &  $ -19.24\pm 0.11$ & 
$   0.61\pm   0.05$ &  $        9.12\pm        0.17$\tablenotemark{a} &  Ceph & 
4 \\ 
NGC 1097 &  SBb &  0.027 &  4.67 &  48 &  130 &  $ -21.21\pm 0.31$ & 
$   0.67\pm   0.05$ &  $        17.1\pm         2.3$ &  Flow &  1 \\ 
NGC 1266 &  SB0~pec &  0.098 &  0.77 &  53 &  108 &  $ -18.89\pm 0.19$ & 
$   0.87\pm   0.05$ &  $        30.0\pm         2.3$ &  Flow &  1 \\
\enddata
\end{deluxetable}

%% file: tab2.tex
\LongTables
\begin{deluxetable}{lccccccccccccccccc}
\tabletypesize{\scriptsize}
\tablecaption{Journal of Observations\label{table:journal}}
\tablewidth{0pt}
\tablehead{
\colhead{} & 
\multicolumn{5}{c}{Nuclear} & 
\colhead{} & 
\multicolumn{5}{c}{Circumnuclear} & 
\colhead{} & 
\multicolumn{5}{c}{Radial Strip} \\
\cline{2-6}\cline{8-12}\cline{14-18}
\colhead{Galaxy} & 
\colhead{Aperture\tablenotemark{a}} & 
\colhead{$\theta_{\rm slit}$\tablenotemark{b}} & 
\colhead{$t_{\rm exp}$\tablenotemark{c}} & 
\colhead{Phot\tablenotemark{d}} & 
\colhead{Notes\tablenotemark{e}} & 
\colhead{} & 
\colhead{Aperture\tablenotemark{a}} & 
\colhead{$\theta_{\rm slit}$\tablenotemark{b}} & 
\colhead{$t_{\rm exp}$\tablenotemark{c}} & 
\colhead{Phot\tablenotemark{d}} & 
\colhead{Notes\tablenotemark{e}} & 
\colhead{} & 
\colhead{Aperture\tablenotemark{a}} & 
\colhead{$\theta_{\rm slit}$\tablenotemark{b}} & 
\colhead{$t_{\rm exp}$\tablenotemark{c}} & 
\colhead{Phot\tablenotemark{d}} & 
\colhead{Notes\tablenotemark{e}}
}
\startdata
NGC 0024 &  $2\farcs5\times2\farcs5$ &  177 &  300 &  Y &  \nodata &   & 
$20\arcsec\times20\arcsec$ &  46 &  75 &  Y &  \nodata &   & 
$190\arcsec\times55\arcsec$ &  46 &  109 &  Y &  \nodata \\ 
NGC 0337 &  $2\farcs5\times2\farcs5$ &  25 &  600 &  Y &  1 &   & 
$20\arcsec\times20\arcsec$ &  70 &  75 &  Y &  \nodata &   & 
$95\arcsec\times55\arcsec$ &  70 &  109 &  Y &  \nodata \\ 
NGC 0584 &  $2\farcs5\times2\farcs5$ &  25 &  300 &  Y &  \nodata &   & 
$20\arcsec\times20\arcsec$ &  70 &  75 &  Y &  \nodata &   & 
$138\arcsec\times55\arcsec$ &  70 &  109 &  Y &  \nodata \\ 
NGC 0628 &  $2\farcs5\times2\farcs5$ &  70 &  300 &  Y &  \nodata &   & 
$20\arcsec\times20\arcsec$ &  70 &  150 &  Y &  \nodata &   & 
$346\arcsec\times55\arcsec$ &  70 &  109 &  Y &  2,3 \\ 
NGC 0855 &  $2\farcs5\times2\farcs5$ &  70 &  300 &  Y &  \nodata &   & 
$20\arcsec\times20\arcsec$ &  70 &  75 &  Y &  \nodata &   & 
$86\arcsec\times55\arcsec$ &  70 &  109 &  N &  \nodata \\ 
NGC 0925 &  $2\farcs5\times2\farcs5$ &  70 &  300 &  Y &  \nodata &   & 
$20\arcsec\times20\arcsec$ &  70 &  112 &  Y &  \nodata &   & 
$346\arcsec\times55\arcsec$ &  70 &  109 &  Y &  2,3 \\ 
NGC 1097 &  $2\farcs5\times3\farcs0$ &  110 &  600 &  Y &  \nodata &   & 
$20\arcsec\times20\arcsec$ &  110 &  192 &  Y &  \nodata &   & 
$308\arcsec\times56\arcsec$ &  70 &  192 &  Y &  \nodata \\ 
NGC 1266 &  $2\farcs5\times2\farcs5$ &  70 &  600 &  Y &  \nodata &   & 
$20\arcsec\times20\arcsec$ &  70 &  75 &  Y &  \nodata &   & 
$51\arcsec\times55\arcsec$ &  70 &  109 &  Y &  \nodata \\
\enddata
\tablecomments{Summary of our nuclear, circumnuclear, and radial-strip observations of the SINGS galaxies.}
\tablenotetext{a}{Spectrophotometric extraction aperture along and perpendicular to the slit.}
\tablenotetext{b}{Long-slit position angle, measured positive from North to East, in degrees.}
\tablenotetext{c}{Effective exposure time, as defined in \citet{moustakas06a}, in seconds.}
\tablenotetext{d}{Flag indicating whether the spectra were obtained during photometric (Y) or non-photometric conditions (N).}
\tablenotetext{e}{Extraction notes: (1) Extended spatial profile; (2) One or more foreground stars subtracted; (3) Multiple pointings stitched together; (4) Extraction centered on bright \ion{H}{2} region; (5) Residual foreground stellar contamination.}
\end{deluxetable}

%% file: tab3.tex
\LongTables
\begin{deluxetable}{lcccccccc}
\tabletypesize{\small}
\tablecaption{Optical Emission-Line Fluxes\label{table:fluxes}}
\tablewidth{0pt}
\tablehead{
\colhead{Galaxy} & 
\colhead{[O~{\sc ii}]$~\lambda3727$} & 
\colhead{H$\gamma~\lambda4340$} & 
\colhead{H$\beta~\lambda4861$} & 
\colhead{[O~{\sc iii}]$~\lambda5007$} & 
\colhead{H$\alpha~\lambda6563$} & 
\colhead{[N~{\sc ii}]$~\lambda6584$} & 
\colhead{[S~{\sc ii}]$~\lambda6716$} & 
\colhead{[S~{\sc ii}]$~\lambda6731$}
}
\startdata
\cline{1-9}
\multicolumn{9}{c}{Nuclear} \\
\cline{1-9}
NGC 0024 &  $        1.02\pm        0.36$ &  \nodata & 
$        0.42\pm        0.19$ &  $        0.57\pm        0.17$ & 
$        2.73\pm        0.46$ &  $        0.86\pm        0.30$ & 
$        0.68\pm        0.26$ &  $        0.46\pm        0.18$ \\ 
NGC 0337 &  $        6.02\pm        0.39$ &  $        0.68\pm        0.17$ & 
$        2.20\pm        0.17$ &  $        1.24\pm        0.15$ & 
$        9.69\pm        0.34$ &  $        2.31\pm        0.26$ & 
$        2.78\pm        0.22$ &  $        2.15\pm        0.24$ \\ 
NGC 0584 &  $         4.0\pm         1.4$ &  \nodata &  \nodata & 
$         4.7\pm         2.2$ &  \nodata &  \nodata &  \nodata &  \nodata \\ 
NGC 0628 &  \nodata &  \nodata &  \nodata &  \nodata &  \nodata &  \nodata & 
\nodata &  \nodata \\ 
NGC 0855 &  $        8.51\pm        0.56$ &  $        1.16\pm        0.29$ & 
$        4.49\pm        0.27$ &  $        8.99\pm        0.31$ & 
$       23.53\pm        0.70$ &  $        3.69\pm        0.39$ & 
$        5.29\pm        0.61$ &  $        3.67\pm        0.39$ \\ 
NGC 0925 &  $        3.75\pm        0.40$ &  $        0.62\pm        0.20$ & 
$        2.57\pm        0.21$ &  $        1.41\pm        0.19$ & 
$        8.76\pm        0.36$ &  $        1.84\pm        0.32$ & 
$        2.25\pm        0.42$ &  $        1.22\pm        0.19$ \\ 
NGC 1097 &  $        6.69\pm        0.89$ &  \nodata & 
$        2.88\pm        0.82$ &  $        8.00\pm        0.77$ & 
$        19.7\pm         2.1$ &  $        30.6\pm         1.5$ & 
$         9.0\pm         1.7$ &  $         6.1\pm         1.2$ \\
\enddata
\tablecomments{Optical emission-line fluxes in units of $10^{-15}$~erg~s$^{-1}$~cm$^{-2}$, corrected for foreground Galactic extinction \citep[$R_{V}=3.1$;][]{odonnell94a, schlegel98a}. Note that the uncertainties include only statistical measurement uncertainties, and do not include systematic errors due to, for example, imperfect continuum subtraction.  For most applications we recommend that a minimum ${\rm S/N}>2$ signal-to-noise ratio cut be applied to the fluxes listed in this table.}
\end{deluxetable}

%% file: tab4.tex
\LongTables
\begin{deluxetable}{lcccccccc}
\tabletypesize{\small}
\tablecaption{Optical Emission-Line Equivalent Widths\label{table:ews}}
\tablewidth{0pt}
\tablehead{
\colhead{Galaxy} & 
\colhead{[O~{\sc ii}]$~\lambda3727$} & 
\colhead{H$\gamma~\lambda4340$} & 
\colhead{H$\beta~\lambda4861$} & 
\colhead{[O~{\sc iii}]$~\lambda5007$} & 
\colhead{H$\alpha~\lambda6563$} & 
\colhead{[N~{\sc ii}]$~\lambda6584$} & 
\colhead{[S~{\sc ii}]$~\lambda6716$} & 
\colhead{[S~{\sc ii}]$~\lambda6731$}
}
\startdata
\cline{1-9}
\multicolumn{9}{c}{Nuclear} \\
\cline{1-9}
NGC 0024 &  $        12.1\pm         4.3$ &  \nodata & 
$        2.08\pm        0.91$ &  $        2.82\pm        0.84$ & 
$        10.6\pm         1.8$ &  $         3.4\pm         1.2$ & 
$         2.6\pm         1.0$ &  $        1.78\pm        0.69$ \\ 
NGC 0337 &  $        36.1\pm         2.5$ &  $        2.79\pm        0.68$ & 
$        8.13\pm        0.63$ &  $        4.48\pm        0.54$ & 
$        30.3\pm         1.1$ &  $        7.34\pm        0.84$ & 
$        8.56\pm        0.68$ &  $        6.67\pm        0.73$ \\ 
NGC 0584 &  $         3.0\pm         1.1$ &  \nodata &  \nodata & 
$        0.93\pm        0.44$ &  \nodata &  \nodata &  \nodata &  \nodata \\ 
NGC 0628 &  \nodata &  \nodata &  \nodata &  \nodata &  \nodata &  \nodata & 
\nodata &  \nodata \\ 
NGC 0855 &  $        23.1\pm         1.6$ &  $        2.08\pm        0.51$ & 
$        7.42\pm        0.45$ &  $       14.87\pm        0.52$ & 
$        36.3\pm         1.1$ &  $        5.63\pm        0.59$ & 
$        7.88\pm        0.91$ &  $        5.49\pm        0.58$ \\ 
NGC 0925 &  $        15.6\pm         1.7$ &  $        1.89\pm        0.60$ & 
$        7.97\pm        0.65$ &  $        4.48\pm        0.61$ & 
$        31.8\pm         1.4$ &  $         6.8\pm         1.2$ & 
$         8.1\pm         1.5$ &  $        4.48\pm        0.72$ \\ 
NGC 1097 &  $         8.7\pm         1.2$ &  \nodata & 
$        1.00\pm        0.29$ &  $        2.71\pm        0.26$ & 
$        3.69\pm        0.40$ &  $        5.79\pm        0.28$ & 
$        1.71\pm        0.32$ &  $        1.15\pm        0.23$ \\
\enddata
\tablecomments{Rest-frame emission-line equivalent widths in \AA.  Note that the uncertainties include only statistical measurement uncertainties; they do not include systematic errors due to, for example, imperfect continuum subtraction.  For most applications we recommend that a minimum ${\rm S/N}>2$ signal-to-noise ratio cut be applied to the EWs listed in this table.}
\end{deluxetable}

%% file: tab5.tex
\begin{deluxetable}{lccccc}
\tabletypesize{\small}
\tablecaption{Optical Spectral Classifications\label{table:class}}
\tablewidth{0pt}
\tablehead{
\colhead{Galaxy} & 
\colhead{Nuclear} & 
\colhead{Circumnuclear} & 
\colhead{Radial Strip} & 
\colhead{Ho97} & 
\colhead{Adopted} \\ 
\colhead{(1)} & 
\colhead{(2)} & 
\colhead{(3)} & 
\colhead{(4)} & 
\colhead{(5)} & 
\colhead{(6)} 
}
\startdata
NGC 0024 &  SF &  SF &  SF &  \nodata &  SF \\ 
NGC 0337 &  SF &  SF &  SF &  \nodata &  SF \\ 
NGC 0584 &  ? &  ? &  ? &  \nodata &  AGN\tablenotemark{a} \\ 
NGC 0628 &  ? &  SF &  SF\tablenotemark{b} &  \nodata &  SF \\ 
NGC 0855 &  SF &  SF &  SF &  \nodata &  SF \\ 
NGC 0925 &  SF &  SF &  SF &  SF &  SF \\ 
NGC 1097 &  AGN &  SF &  SF\tablenotemark{e} &  \nodata &  AGN \\ 
NGC 1266 &  AGN\tablenotemark{b} &  AGN\tablenotemark{b} & 
AGN\tablenotemark{b} &  \nodata &  AGN \\
\enddata
\end{deluxetable}

%% file: tab6.tex
\begin{deluxetable}{lcccccccc}
\tabletypesize{\small}
\tablecaption{Nebular Dust Reddening\label{table:reddening}}
\tablewidth{0pt}
\tablehead{
\colhead{} & \multicolumn{2}{c}{Nuclear} & \colhead{} & \multicolumn{2}{c}{Circumnuclear} & \colhead{} & \multicolumn{2}{c}{Radial Strip} \\
\cline{2-3} \cline{5-6} \cline{8-9}
\colhead{Galaxy} & \colhead{H$\alpha$/H$\beta$} & \colhead{$E(B-V)$} & \colhead{} & \colhead{H$\alpha$/H$\beta$} & \colhead{$E(B-V)$} & \colhead{} & \colhead{H$\alpha$/H$\beta$} & \colhead{$E(B-V)$}
}
\startdata
NGC 0024 &  $6.48\pm3.05$ &  $0.78\pm0.45$ &    &  $4.09\pm0.69$ & 
$0.34\pm0.16$ &    &  $4.00\pm0.67$ &  $0.32\pm0.16$ \\ 
NGC 0337 &  $4.41\pm0.38$ &  $0.41\pm0.08$ &    &  $5.43\pm0.24$ & 
$0.61\pm0.04$ &    &  $4.90\pm0.12$ &  $0.51\pm0.02$ \\ 
NGC 0855 &  $5.24\pm0.35$ &  $0.58\pm0.06$ &    &  $4.44\pm0.19$ & 
$0.42\pm0.04$ &    &  $4.64\pm0.34$ &  $0.46\pm0.07$ \\ 
NGC 0925 &  $3.41\pm0.31$ &  $0.17\pm0.09$ &    &  $4.78\pm0.25$ & 
$0.49\pm0.05$ &    &  $3.84\pm0.33$ &  $0.28\pm0.08$ \\ 
NGC 1097 &  \nodata &  \nodata &    &  $6.52\pm0.23$ &  $0.78\pm0.03$ &    & 
$6.51\pm0.38$ &  $0.78\pm0.06$ \\ 
NGC 1482 &  $12.05\pm0.49$ &  $1.36\pm0.04$ &    &  $11.10\pm0.51$ & 
$1.29\pm0.04$ &    &  $7.85\pm0.49$ &  $0.96\pm0.06$ \\ 
NGC 1512 &  \nodata &  \nodata &    &  $5.01\pm0.37$ &  $0.53\pm0.07$ &    & 
$5.11\pm0.83$ &  $0.55\pm0.15$ \\ 
NGC 1705 &  \nodata &  \nodata &    &  $3.80\pm0.10$ &  $0.27\pm0.02$ &    & 
$3.82\pm0.08$ &  $0.28\pm0.02$ \\
\enddata
\tablecomments{Balmer decrement, H$\alpha$/H$\beta$, and corresponding color excess, $E(B-V)$, in magnitudes, asuming the \citet{odonnell94a} Milky Way extinction curve and an intrinsic (unreddened) decrement of $2.86^{+0.18}_{-0.11}$.  Note that the tabulated uncertainties only include the statistical measurement errors of the emission lines and the uncertainty in the adopted intrinsic Balmer decrement; they do not include possible systematic errors arising from, for example, a breakdown of our assumption of a foreground dust screen geometry (e.g., \citealt{witt00a}, but see \citealt{kenn09a}).}
\end{deluxetable}

%% file: tab7.tex
\LongTables
\begin{deluxetable}{lccccccc}
\tabletypesize{\small}
\tablecaption{Nuclear, Circumnuclear, \& Radial Strip Oxygen Abundances\label{table:intnucoh}}
\tablewidth{0pt}
\tablehead{
\colhead{Galaxy} & \colhead{$R_{23}$ Branch} & \colhead{$R_{23}$} & \colhead{$O_{32}$} & \colhead{$P$} & \colhead{$\log\,(U)_{\rm KK04}$} & \colhead{$12+\log\,({\rm O/H})_{\rm KK04}$} & \colhead{$12+\log\,({\rm O/H})_{\rm PT05}$} \\
\colhead{(1)} & 
\colhead{(2)} & 
\colhead{(3)} & 
\colhead{(4)} & 
\colhead{(5)} & 
\colhead{(6)} & 
\colhead{(7)} & 
\colhead{(8)} 
}
\startdata
\cline{1-8}
\multicolumn{8}{c}{Nuclear} \\
\cline{1-8}
NGC 0024 &  U &  $6.59\pm16.16$ &  $0.33\pm0.81$ &  $0.25\pm0.53$ & 
$-3.23\pm0.57$ &  $8.66\pm0.47$ &  $7.95\pm0.73$ \\ 
NGC 0337 &  U &  $4.74\pm2.18$ &  $0.18\pm0.08$ &  $0.15\pm0.06$ & 
$-3.33\pm0.22$ &  $8.85\pm0.21$ &  $8.02\pm0.32$ \\ 
NGC 0855 &  U &  $5.72\pm1.88$ &  $0.77\pm0.27$ &  $0.43\pm0.12$ & 
$-2.88\pm0.20$ &  $8.77\pm0.18$ &  $8.27\pm0.21$ \\ 
NGC 0925 &  U &  $2.42\pm1.13$ &  $0.42\pm0.21$ &  $0.30\pm0.13$ & 
$-2.86\pm0.27$ &  $9.06\pm0.06$ &  $8.49\pm0.20$ \\ 
NGC 1482 &  U &  $1.46\pm0.39$ &  $0.17\pm0.05$ &  $0.14\pm0.04$ & 
$-3.15\pm0.13$ &  $9.14\pm0.03$ &  $8.44\pm0.10$ \\ 
NGC 2798 &  U &  $1.72\pm0.27$ &  $0.32\pm0.05$ &  $0.24\pm0.04$ & 
$-2.93\pm0.08$ &  $9.11\pm0.02$ &  $8.52\pm0.05$ \\ 
NGC 2915 &  L &  $6.41\pm3.02$ &  $1.56\pm0.81$ &  $0.61\pm0.24$ & 
$-2.90\pm0.19$ &  $8.21\pm0.32$ &  $7.89\pm0.39$ \\
\enddata
\tablecomments{(1) Galaxy name;  (2-5) Adopted $R_{23}$ branch (U=upper, L=lower, A=ambiguous), \pagel{} parameter, and excitation parameters \ioniz{} and $P$ (see \S\ref{sec:calib} and \S\ref{sec:intnucoh});  (6-8) Ionization parameter and oxygen abundances based on the KK04 and PT05 strong-line calibrations.  Note that the uncertainties on the ionization parameter and oxygen abundances include statistical measurement errors only.}
\tablenotetext{a}{The \pagel{} branch of NGC~5474 based on its circumnuclear spectrum is ambiguous, according to the criteria defined in \S\ref{sec:intnucoh}.}
\end{deluxetable}

%% file: tab8.tex
\LongTables
\begin{deluxetable}{lccccccc}
\tabletypesize{\small}
\tablecaption{H~{\sc ii}-Region Oxygen Abundances\label{table:hiioh}}
\tablewidth{0pt}
\tablehead{
\colhead{} & \colhead{} & \colhead{Gradient} & \colhead{12+log(O/H)} & \colhead{12+log(O/H)} & \colhead{12+log(O/H)} & \colhead{} & \colhead{} \\
\colhead{Galaxy} & \colhead{$R_{23}$ Branch} & \colhead{(dex~$\rho_{25}^{-1}$)} & \colhead{at $\rho=0$} & \colhead{at $\rho=0.4\,\rho_{25}$} & \colhead{Average} & \colhead{N(H~{\sc ii})} & \colhead{Refs.} \\
\colhead{(1)} & 
\colhead{(2)} & 
\colhead{(3)} & 
\colhead{(4)} & 
\colhead{(5)} & 
\colhead{(6)} & 
\colhead{(7)} & 
\colhead{(8)} 
}
\startdata
\cline{1-8}
\multicolumn{8}{c}{KK04} \\
\cline{1-8}
NGC 0628 &  U &  $-0.57\pm0.04$ &  $9.19\pm0.02$ &  $9.02\pm0.01$ & 
$8.88\pm0.15$ &  34 &  1,2,3,4,5 \\ 
NGC 0925 &  U &  $-0.42\pm0.02$ &  $8.91\pm0.01$ &  $8.79\pm0.01$ & 
$8.73\pm0.15$ &  55 &  2,5,6 \\ 
NGC 1097 &  U &  $-0.29\pm0.09$ &  $9.17\pm0.01$ &  $9.09\pm0.02$ & 
$9.14\pm0.05$ &  15 &  7 \\ 
NGC 1705 &  L &  \nodata &  \nodata &  \nodata &  $8.37\pm0.06$ &  9 &  8 \\ 
NGC 2403 &  U &  $-0.26\pm0.03$ &  $8.89\pm0.01$ &  $8.81\pm0.01$ & 
$8.80\pm0.13$ &  46 &  1,4,5,9,10 \\ 
Ho II &  L &  \nodata &  \nodata &  \nodata &  $8.13\pm0.11$ &  21 & 
11,12,13 \\ 
DDO 053 &  L &  \nodata &  \nodata &  \nodata &  $7.98\pm0.23$ &  4 &  11,14 \\
\enddata
\tablecomments{(1) Galaxy name;  (2) Adopted $R_{23}$ branch (U=upper, L=lower, A=ambiguous);  (3) Slope of the radial abundance gradient;  (4) Central oxygen abundance based on the derived abundance gradient;  (5) Characteristic oxygen abundance, defined as the metallicity at $\rho=0.4\rho_{25}$, based on the derived abundance gradient;  (6) Unweighted average and standard deviation of all the H~{\sc ii}-region abundances (see \S\ref{sec:hiiavg});  (7) Number of H~{\sc ii} regions with measured abundances;  (8) References to the literature from which the \hii-region line-ratios were taken (see Appendix~\ref{sec:hiiappendix}).}
\tablenotetext{a}{Although Ho~IX is among the faintest galaxies in our sample, the \nii/\ha{} and \nii/\oii{} line-ratios clearly place it on the upper \pagel{} branch \citep{croxall09a}.}
\tablerefs{(1) \citet{bresolin99a}; (2) \citet{castellanos02a}; (3) \citet{ferguson98a}; (4) \citet{mccall85a}; (5) \citet{vanzee98a}; (6) \citet{zaritsky94a}; (7) \citet{storchi96a}; (8) \citet{lee04a}; (9) \citet{garnett97a}; (10) \citet{garnett99a}; (11) \citet{croxall09a}; (12) \citet{lee03a}; (13) \citet{masegosa91a}; (14) \citet{skillman89a}; (15) \citet{oey93a}; (16) \citet{lee03c}; (17) \citet{miller96a}; (18) \citet{guseva00a}; (19) \citet{garnett87a}; (20) \citet{stauffer84a}; (21) \citet{kong02a}; (22) \citet{bresolin02a}; (23) \citet{diaz07a}; (24) \citet{ryder95a}; (25) \citet{henry94a}; (26) \citet{shields91a}; (27) \citet{kenn01a}; (28) \citet{vanzee97a}; (29) \citet{bresolin04a}; (30) \citet{diaz91a}; (31) \citet{garnett04a}; (32) \citet{stasinska86a}; (33) \citet{lee06a}; (34) \citet{peimbert05a}; (35) \citet{storchi95a}; (36) \citet{edmunds84a}.}
\end{deluxetable}

%% file: tab9.tex
\begin{deluxetable}{lccccccc}
\tablecaption{Final Oxygen Abundances\label{table:final}}
\tablewidth{0pt}
\tablehead{
\colhead{} & \multicolumn{3}{c}{12+log(O/H)$_{\rm KK04}$} & \colhead{} & \multicolumn{3}{c}{12+log(O/H)$_{\rm PT05}$} \\
\cline{2-4} \cline{6-8}
\colhead{Galaxy} & 
\colhead{Central} & 
\colhead{Characteristic} & 
\colhead{$L-Z$} & 
\colhead{} & 
\colhead{Central} & 
\colhead{Characteristic} & 
\colhead{$L-Z$} 
}
\startdata
NGC 0024 &  $8.93\pm0.11$ &  $8.93\pm0.11$ &  $8.62$ &    &  $8.26\pm0.29$ & 
$8.26\pm0.29$ &  $8.08$ \\ 
NGC 0337 &  $8.84\pm0.05$ &  $8.84\pm0.05$ &  $9.10$ &    &  $8.18\pm0.07$ & 
$8.18\pm0.07$ &  $8.45$ \\ 
NGC 0584 &  \nodata &  \nodata &  $9.07$ &    &  \nodata &  \nodata & 
$8.43$ \\ 
NGC 0628 &  $9.25\pm0.03$ &  $9.02\pm0.01$ &  $9.02$ &    &  $8.45\pm0.03$ & 
$8.35\pm0.01$ &  $8.39$ \\ 
NGC 0855 &  $8.80\pm0.09$ &  $8.80\pm0.09$ &  $8.57$ &    &  $8.29\pm0.10$ & 
$8.29\pm0.10$ &  $8.04$ \\ 
NGC 0925 &  $8.96\pm0.01$ &  $8.79\pm0.01$ &  $8.95$ &    &  $8.34\pm0.02$ & 
$8.25\pm0.01$ &  $8.34$ \\ 
NGC 1097 &  $9.15\pm0.01$ &  $9.09\pm0.01$ &  $9.30$ &    &  $8.62\pm0.03$ & 
$8.47\pm0.02$ &  $8.60$ \\ 
NGC 1266 &  \nodata &  \nodata &  $8.89$ &    &  \nodata &  \nodata & 
$8.29$ \\
\enddata
\tablecomments{Final central and characteristic nebular oxygen abundances for the SINGS galaxies based on the KK04 and PT05 abundance calibrations.  The central abundance characterizes the inner, or nuclear metallicity of each galaxy, while the characteristic abundance is a globally-averaged metallicity.  Note that the uncertainties on the oxygen abundances include statistical measurement errors only.  We also provide the oxygen abundances inferred from the $B$-band luminosity-metallicity ($L-Z$) relation for every object; the uncertainty in the $L-Z$ abundance is at least $0.2$~dex (see \S\ref{sec:final} and Fig.~\ref{fig:lz}).}
\end{deluxetable}

%% file: tab1_appendix.tex
\LongTables
\begin{deluxetable}{cllccccccc}
\tabletypesize{\small}
\tablecaption{Oxygen Abundances of the H~{\sc ii} Regions in the SINGS Galaxies\label{table:hiidata}}
\tablewidth{0pt}
\tablehead{
\colhead{No.} & 
\colhead{Galaxy} & 
\colhead{Region} & 
\colhead{$\rho/\rho_{25}$} & 
\colhead{$R_{23}$} & 
\colhead{$O_{32}$} & 
\colhead{$P$} & 
\colhead{$12+\log\,({\rm O/H})_{\rm KK04}$} & 
\colhead{$12+\log\,({\rm O/H})_{\rm PT05}$} & 
\colhead{Ref.} \\ 
\colhead{(1)} & 
\colhead{(2)} & 
\colhead{(3)} & 
\colhead{(4)} & 
\colhead{(5)} & 
\colhead{(6)} & 
\colhead{(7)} & 
\colhead{(8)} & 
\colhead{(9)} & 
\colhead{(10)} 
}
\startdata
001 &  NGC 0628 &  H292 &  0.23 &  $2.44\pm0.17$ &  $0.24\pm0.04$ & 
$0.20\pm0.03$ &  $9.06\pm0.01$ &  $8.38\pm0.04$ &  1 \\ 
002 &  NGC 0628 &  H451 &  0.26 &  $1.75\pm0.18$ &  $0.15\pm0.06$ & 
$0.13\pm0.05$ &  $9.12\pm0.02$ &  $8.38\pm0.06$ &  1 \\ 
003 &  NGC 0628 &  H572 &  0.39 &  $2.82\pm0.18$ &  $0.15\pm0.02$ & 
$0.13\pm0.02$ &  $9.04\pm0.02$ &  $8.24\pm0.04$ &  1 \\ 
004 &  NGC 0628 &  H598 &  0.41 &  $3.13\pm0.25$ &  $0.23\pm0.05$ & 
$0.19\pm0.04$ &  $9.01\pm0.02$ &  $8.27\pm0.06$ &  1 \\ 
005 &  NGC 0628 &  H627 &  0.51 &  $4.24\pm0.18$ &  $0.80\pm0.07$ & 
$0.45\pm0.03$ &  $8.92\pm0.01$ &  $8.43\pm0.03$ &  1 \\ 
006 &  NGC 0628 &  H154-155 &  0.72 &  $5.29\pm0.24$ &  $0.27\pm0.02$ & 
$0.21\pm0.01$ &  $8.80\pm0.03$ &  $8.04\pm0.04$ &  1 \\ 
007 &  NGC 0628 &  H13 &  0.71 &  $4.82\pm0.18$ &  $0.48\pm0.03$ & 
$0.33\pm0.02$ &  $8.86\pm0.02$ &  $8.25\pm0.03$ &  1 \\ 
008 &  NGC 0628 &  FGW628A &  0.26 &  $2.28\pm0.08$ &  $0.10\pm0.01$ & 
$0.09\pm0.01$ &  $9.08\pm0.01$ &  $8.25\pm0.02$ &  2 \\
\enddata
\tablecomments{(1-2) Unique \hii-region identification number and SINGS galaxy name; (3) H~{\sc ii}-region name as listed in the originating reference, if available; (4) Deprojected galactocentric radius normalized by the $\rho_{25}$ radius of the galaxy (see Table~\ref{table:properties});  (5-7) Abundance-sensitive \pagel{} parameter and excitation parameters \ioniz{} and $P$ (see \S\ref{sec:calib}) (8-9) Oxygen abundances derived using the KK04 and PT05 strong-line calibrations assuming the \pagel{} branch listed in Table~\ref{table:intnucoh}, unless otherwise noted;  (10) References to the literature from which the \hii-region line-ratios were taken (see Appendix~\ref{sec:hiiappendix}).}
\tablenotetext{a}{The \pagel{} branch of this \hii{} region is ambiguous according to the criteria described in \S\ref{sec:intnucoh}.}
\tablenotetext{b}{The oxygen abundance of this \hii{} region is not defined according to the criteria described in \S\ref{sec:intnucoh}.}
\tablerefs{(1) \citet{bresolin99a}; (2) \citet{ferguson98a}; (3) \citet{vanzee98a}; (4) \citet{mccall85a}; (5) \citet{castellanos02a}; (6) \citet{zaritsky94a}; (7) \citet{storchi96a}; (8) \citet{lee04a}; (9) \citet{garnett99a}; (10) \citet{garnett97a}; (11) \citet{croxall09a}; (12) \citet{lee03a}; (13) \citet{masegosa91a}; (14) \citet{skillman89a}; (15) \citet{oey93a}; (16) \citet{lee03c}; (17) \citet{miller96a}; (18) \citet{guseva00a}; (19) \citet{garnett87a}; (20) \citet{stauffer84a}; (21) \citet{kong02a}; (22) \citet{diaz07a}; (23) \citet{bresolin02a}; (24) \citet{ryder95a}; (25) \citet{henry94a}; (26) \citet{shields91a}; (27) \citet{kenn01a}; (28) \citet{vanzee97a}; (29) \citet{bresolin04a}; (30) \citet{garnett04a}; (31) \citet{diaz91a}; (32) \citet{stasinska86a}; (33) \citet{lee06a}; (34) \citet{peimbert05a}; (35) \citet{storchi95a}; (36) \citet{edmunds84a}.}
\end{deluxetable}